\documentclass[
              reprint,
              superscriptaddress,
              nofootinbib,
              showpacs,
              amsmath,amssymb,aps,prd,
              ]{revtex4-1}
\pdfoutput=1 			

\usepackage{graphics}
\usepackage{epsfig}
\usepackage{graphicx}
\usepackage{dcolumn}
\usepackage{bm}
\usepackage{longtable}
\usepackage{ifpdf}
\ifpdf
\usepackage[pdfencoding=auto,psdextra]{hyperref}
\usepackage{bookmark}
\fi

\begin{document}


\title{\boldmath
	Measurement of the charge asymmetry of electrons from the decays of
	$W$ bosons produced in $p\bar{p}$ collisions at $\sqrt{s}=1.96$ TeV
       \unboldmath}

\input{cdf_auth_050621.itex}

\date{\today}

\begin{abstract}
At the Fermilab Tevatron proton-antiproton ($p\bar{p}$) collider,
high-mass electron-neutrino ($e\nu$) pairs are produced predominantly
in the process $p \bar{p} \rightarrow W(\rightarrow e\nu) + X$.
The asymmetry of the electron and positron yield as a function of
their pseudorapidity constrain the slope of the ratio of the
$u$- to $d$-quark parton distributions versus the fraction of the
proton momentum carried by the quarks. This paper reports on the
measurement of the electron-charge asymmetry using the full data
set recorded by the Collider Detector at Fermilab in 2001--2011 and
corresponding to 9.1~fb$^{-1}$ of integrated luminosity.
The measurement significantly improves the precision of the Tevatron
constraints on the parton-distribution functions of the proton.
Numerical tables of the measurement are provided.
\end{abstract}

\maketitle

\section{Introduction}

At the Fermilab Tevatron proton-antiproton ($p\bar{p}$) collider,
massive lepton pairs consisting
of charged leptons $(\ell)$ and their partner neutrinos $(\nu)$ are
produced in $p\bar{p}$ collisions at the center-of-momentum energy
($\sqrt{s}$) 1.96~TeV~\cite{DrellYan,*DrellYanE}.
In the standard model, the $\ell\nu$ pair is produced through an
intermediate $W$ boson whose production
occurs primarily through the quark-antiquark annihilation process,
\begin{displaymath}
  q+\bar{q}^\prime \rightarrow W \rightarrow \ell\nu \:,
\end{displaymath}
where the $q$ and $\bar{q}^\prime$ denote the incoming quark and
antiquark, respectively, from the colliding hadrons. In leading-order
quantum chromodynamics (QCD) calculations, 90\% of $W^+$ bosons are
produced via $u+\bar{d}$ collisions and a similar fraction of $W^-$
bosons via $d+\bar{u}$ collisions.
\par
The production rates of $W^+$ and $W^-$ bosons exhibit differences
as functions of their kinematic properties over their kinematic range
of production.
The momentum distributions of the $u$ ($\bar{u}$) and
$d$ ($\bar{d}$) quarks from the incoming proton (antiproton) differ,
affecting the $W^+$ and $W^-$ differential-production rates.
Momentum distributions of quarks and gluons are determined by the
parton-distribution functions (PDFs)
of the proton, which must be experimentally derived. Measurements of
production-rate differences can be used to constrain the PDFs. A highly
constraining measurement is the charged-lepton yield asymmetry as a
function of pseudorapidity
\begin{equation}
 A_\ell = \frac
	{d\sigma^+_\ell/d\eta - d\sigma^-_\ell/d\eta} 
	{d\sigma^+_\ell/d\eta + d\sigma^-_\ell/d\eta} \; ,
\label{eqnAlepDef}
\end{equation}
where $d\sigma^\pm_\ell/d\eta$ denotes the differential cross section
with respect to the
pseudorapidity $\eta$ of charged leptons $\ell^\pm$ from the production
of $W^\pm$ bosons and their subsequent decay via $W \rightarrow \ell\nu$.
The pseudorapidity is $-\!\ln \tan(\theta/2)$, where $\theta$ is the
polar angle of the lepton relative to the proton direction.
Effects from the interference between the axial and vector currents of
the electroweak interaction, and from the initial-state interactions of
the colliding partons alter the boson asymmetries.
While the leptonic asymmetry $A_\ell$ can be measured
well, its interpretation in terms of the underlying PDFs must include
these effects. The Tevatron measurement of $A_\ell$ constrains the slope
of the ratio of the $d$- to $u$-quark distribution functions as a function
of the Bjorken scaling parameter, the fraction $x$ of the proton momentum
taken by the colliding quark~\cite{BjorkenX}.
\par
The leptonic asymmetry $A_\ell$ has been measured at the Tevatron collider
with $p\bar{p}$ collisions and at the Large Hadron Collider (LHC)
with $pp$ collisions.
Tevatron measurements have been reported 
at $\sqrt{s}=1.8$~TeV by CDF~\cite{CDF180Aem},
and at $\sqrt{s}=1.96$~TeV by CDF~\cite{CDF196Ae1} and
D0~\cite{D0196Am1,D0196Ae1,D0196Am2,D0196Ae2,*D0196Ae2E}.
The boson-level asymmetry has also been inferred
at $\sqrt{s}=1.96$~TeV by
CDF~\cite{CDF196Awe1} and D0~\cite{D0196Awe1,*D0196Awe1E}
using a neutrino-weighting technique~\cite{WasyNuWeights}.
The LHC measurements at $\sqrt{s}=7$~(8)~TeV have been reported by
ATLAS~\cite{ATLAS7Am1,ATLAS7Aem1},
CMS~\cite{CMS7Aem1,CMS7Ae1,CMS7Am1} (CMS~\cite{CMS8Am1}), and
LHCb~\cite{LHCbAmu7} (LHCb~\cite{LHCbAmel8,LHCbAmujet8}).
Measurements from colliders of different types and energies provide
important constraints for global fits of PDFs
because the compositions of input parton fluxes that produce
$W$ bosons differ, and because the increased precision attainable
significantly improves the accuracy of the fitted PDFs.
At the LHC, $W$ bosons are mostly produced through quark-antiquark
collisions as they are at the Tevatron. However, at the Tevatron, the
collisions are primarily between valence quarks while at the LHC,
collisions are primarily between valence and sea quarks.

\par
In this paper, the final CDF measurement
of the asymmetry $A_\ell$ in the $W \rightarrow e\nu$ channel\footnote
{
The $W \rightarrow \mu\nu$ channel is not considered due to the
limited $\eta$ coverage and complexity of the muon measurement.
}
is presented,
using a data sample corresponding to an integrated $p\bar{p}$
luminosity of 9.1~fb$^{-1}$ collected at the Tevatron collider.
This measurement supersedes the previous CDF
measurements~\cite{CDF196Ae1,CDF196Awe1}, that were based on
subsamples at least nine times smaller.

\par
Section~\ref{AsymmetryDistr} of the paper provides an overview
of the formal definition of the asymmetries and of the existing
theoretical calculations.
Section~\ref{asymMeasurement} introduces the asymmetry measured
in this paper.
Section~\ref{CDFdetector} describes the experimental apparatus.
Section~\ref{DataSelection} reports on the selection of data.
Section~\ref{ExpDatSim} describes the simulation of the
reconstructed data.
Section~\ref{CorrOverview} presents an overview of the corrections
to the data and simulation, and Sec.~\ref{CorrDatSim} the details
of those corrections.
Section~\ref{AsymMeas} presents the measurement of the asymmetry,
Sec.~\ref{systUncerts} the systematic uncertainties,
and Sec.~\ref{finalResults} the results. Finally, 
Sec.~\ref{theEndSummary} presents a summary.

\section{\label{AsymmetryDistr}
Asymmetry distributions}

In the laboratory frame, the $p\bar{p}$-collision axis is
the $z$ axis, with the positive direction oriented along the
direction of the proton. The transverse component of any vector
quantity is defined relative to that
axis. The rapidity, transverse momentum, and mass of a particle
are denoted as $y$, $\vec{P}_{\rm T}$, and $M$, respectively. The
energy and momentum of particles are denoted as $E$ and $\vec{P}$,
respectively. The rapidity is
$y = \frac{1}{2} \, \ln[\,(E + P_{\rm z})/(E - P_{\rm z})\,]$,
where $P_{\rm z}$ is the component of the momentum vector along
the $z$ axis. For massless particles, the rapidity reduces to the
pseudorapidity $\eta$.
\par
The cross section for the production of $W$ bosons in hadronic
collisions, differential in the rapidity, squared mass, and squared
transverse momentum, is denoted by $d^3\sigma_W / dy dP_{\rm T}^2 dM^2$.
The charge asymmetry at a given $y$ value is defined as 
\begin{equation}
 A_W = \frac
        {d\sigma^+_W/dy - d\sigma^-_W/dy}
        {d\sigma^+_W/dy + d\sigma^-_W/dy} \; ,
\label{eqnAwDef}
\end{equation}
where $d\sigma^\pm_W/dy$ denotes the cross section for
$W^\pm$ production integrated over $P_{\rm T}^2$ and $M^2$.
\par
Since the $P_{\rm z}$ component of the neutrino momentum cannot
be measured on an event-by-event basis,
the charge asymmetry of the lepton
$A_\ell(\eta)$ is measured. The cross-section input
to $A_\ell(\eta)$ is a combination of the $W$-boson cross section
and the angular distribution of the $\ell\nu$ pair from the
$W$-boson decay in the rest frame of the $\ell\nu$ pair,
\begin{displaymath}
  \frac{d^{\,5}\sigma_{\ell\nu}}
       {dy dP_{\rm T}^2 dM^2 d\cos\vartheta d\varphi} =
  \frac{3}{16\pi} 
  \frac{d^3\sigma_W}{dy dP_{\rm T}^2 dM^2}
        {\cal N}(\vartheta,\varphi) \: ,
\end{displaymath}
where the angular-distribution function ${\cal N}(\vartheta,\varphi)$ is
the density of $W$ decays as a function of the polar angle $\vartheta$
and azimuthal angle $\varphi$ of the charged lepton, respectively,
and the charge-specific labels for the $W$-boson cross section and
angular-distribution function are implicit. The decay into the lepton
pair exposes a set of helicity cross sections that characterize the
density matrix of the $W$-boson polarization states that are produced.

\par
\begin{figure}
\includegraphics
   [height=54mm]
   {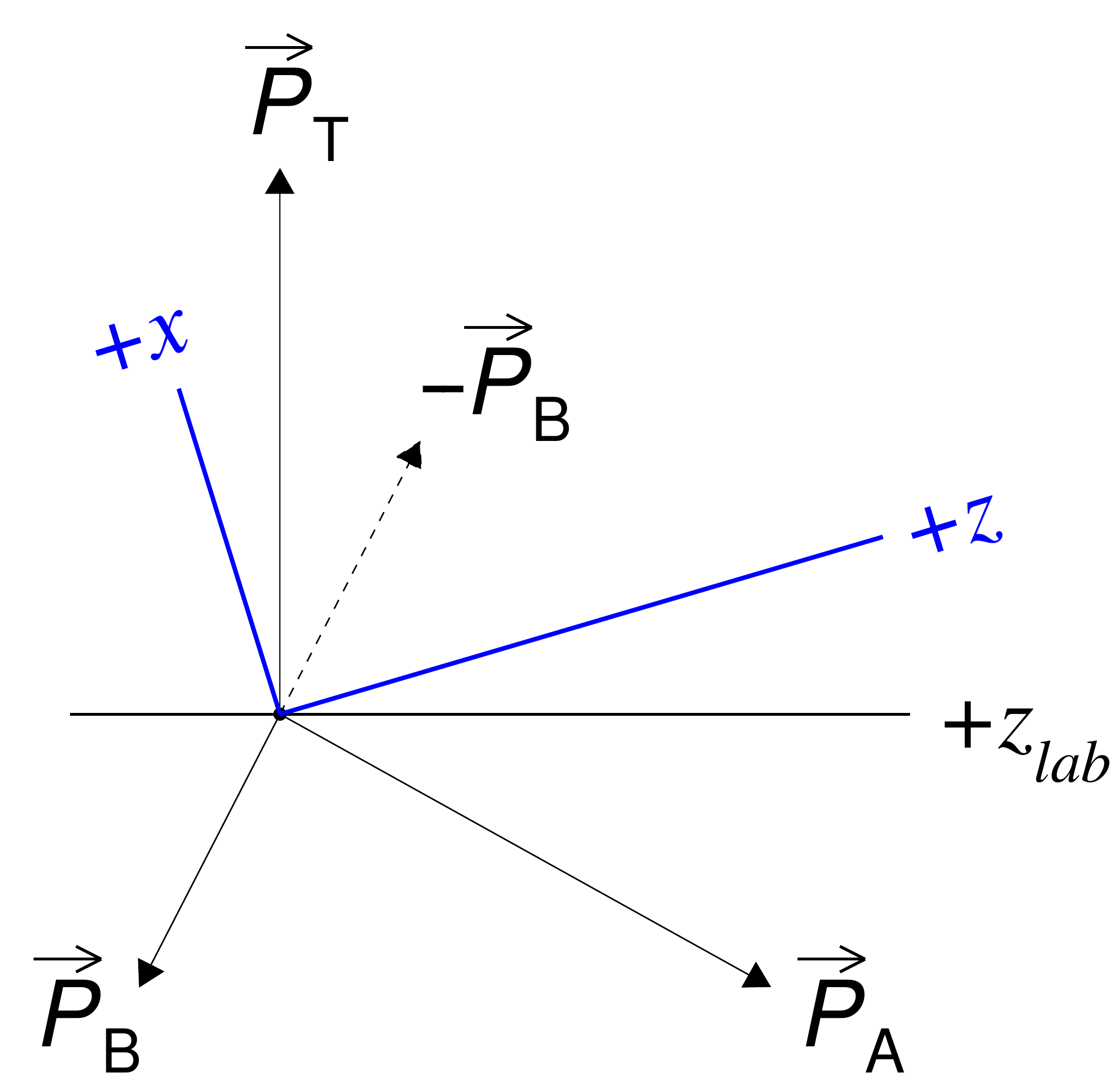}
\caption{\label{fig_CSframe}
Representation of the CS-coordinate axes $(x,z)$ along with the
laboratory $z$ axis ($z_{lab}$). The three axes are in the plane
formed by the proton ($\vec{P}_{\rm A}$) and antiproton
($\vec{P}_{\rm B}$) momentum vectors in the $\ell\nu$-pair rest
frame. The $z$ axis is the angular bisector of $\vec{P}_{\rm A}$
and $-\vec{P}_{\rm B}$, the $y$ axis is along the direction of
$\vec{P}_{\rm B} \times \vec{P}_{\rm A}$, and the $x$ axis is in
the direction away from the transverse component of
$\vec{P}_{\rm A}+\vec{P}_{\rm B}$.
In the limit of vanishing $P_{\rm T}$, the CS and laboratory
axes become equivalent.
}
\end{figure}

In this analysis, the Collins-Soper (CS) rest frame of the $\ell\nu$
pair is used to quantify
${\cal N}(\vartheta,\varphi)$~\cite{CollinsSoperFrame}. 
This frame is reached from the laboratory frame via two Lorentz
boosts, first along the laboratory $z$-axis into the frame where the
$z$ component of the $\ell\nu$-pair momentum vector is zero, followed
by a boost along the transverse component of the $\ell\nu$-pair
momentum vector into its rest frame. A view of the CS frame is shown
in Fig.~\ref{fig_CSframe}.
\par
The angular-distribution function is expressed as
\begin{eqnarray}
{\cal N}(\vartheta,\varphi) & = &
        \: (1 + \cos^2 \vartheta) +  \nonumber \\
  &   & A_0 \:\frac{1}{2} \:
             (1 -3\cos^2 \vartheta) + \nonumber \\
  &   & A_1 \: \sin 2\vartheta
               \cos \varphi +   \nonumber \\
  &   & A_2 \: \frac{1}{2} \:
               \sin^2 \vartheta
               \cos 2\varphi +  \nonumber \\
  &   & A_3 \: \sin \vartheta
               \cos \varphi +   \nonumber \\
  &   & A_4 \: \cos \vartheta + \nonumber \\
  &   & A_5 \: \sin^2 \vartheta
               \sin 2\varphi +  \nonumber \\
  &   & A_6 \: \sin 2\vartheta
               \sin \varphi +   \nonumber \\
  &   & A_7 \: \sin \vartheta
               \sin \varphi \: ,
\label{eqnAngDistr}
\end{eqnarray}
where $A_{0-7}$ are coefficient functions that describe the
nonangular parts of the helicity cross sections relative to
the unpolarized cross section integrated over the
polar angles~\cite{MirkesA0to7a,*MirkesA0to7b}.
In amplitudes at higher order than the tree level, initial-state
interactions of the colliding partons impart transverse
momentum to the boson, affecting the helicity cross sections.
Consequently, $A_{0-7}$ are functions of the $W$ boson
$y$, $P_{\rm T}$, and $M$. They vanish when the boson transverse
momentum is zero, except for $A_4$ whose value is $\pm2$ for
$W^\mp$ decays in QCD calculations at leading order (LO). In
electroweak interactions, the interference between the vector and
axial currents produces the $A_4\,\cos\vartheta$ term.
The $A_{5-7}$ coefficients appear at second
order in the QCD strong-coupling constant, $\alpha_s$, and are
small in the CS frame.
\par
For the $W$-boson cross sections used in comparisons of measurements
to theoretical predictions,
next-to-leading-order (NLO) QCD calculations (of order $\alpha_s$)
and recent PDFs are used. The \textsc{powheg-box}\footnote
{
The \textsc{powheg-box} code is version V2 (svn 3319).
}
implementation of $W$-boson production~\cite{Powheg-Box} and
decay to lepton pairs~\cite{PowhegBoxVBP} provides the
NLO QCD calculation.
It is used as an unweighted partonic event generator.
The NLO-production framework implements a Sudakov form factor that
controls the infrared divergence at low-boson
$P_{\rm T}$~\cite{Sudakov-FFeng, *Sudakov-FFrus}, and an interface
to parton-showering algorithms that avoids double counting. The
\textsc{pythia}~6.41 parton-showering algorithm is used to produce
the hadron-level event~\cite{pythia64}. The combined 
implementation has next-to-leading log resummation accuracy.
Parton fluxes of the incoming proton and antiproton are provided
by the recent NNPDF~3.0 set of NLO PDFs derived with the value of
$\alpha_s = 0.118$ at the $Z$-pole
mass~\cite{ nnpdf301, *nnpdf302, *nnpdf303, *nnpdf304,
      *nnpdf305, *nnpdf306, *nnpdf306e,*nnpdf307}.
The \textsc{powheg-box} calculation with
the NNPDF~3.0 NLO PDFs is the default calculation.
\par
In addition, the \textsc{resbos} NLO calculation
\cite{ResBos1, *ResBos2, *ResBos3, *ResBosc221} with
CTEQ6.6 NLO PDFs~\cite{Cteq66pdf} is used for the ancillary
tuning of $W$-boson production within the \textsc{powheg-box}
calculation. The \textsc{resbos} calculation combines
an NLO fixed-order calculation at high boson $P_{\rm T}$ with
the Collins-Soper-Sterman resummation
formalism~\cite{methodCSS, *wfactorCSS1, *wfactorCSS2, *wfactorCSS3}
at low boson $P_{\rm T}$, which is an all-orders summation
of large terms from gluon emission calculated to
next-to-next-to-leading log accuracy.
The intrinsic $P_{\rm T}$ parameters of \textsc{pythia}~6.41 used
for the default calculation are adjusted\footnote
{
The adjusted parameters and values are:
MSTP(91) = 1, PARP(91) = 1.50~GeV/$c$,
PARP(93) = 12~GeV/$c$, and PARP(64) = 0.4.
}
so that the boson $P_{\rm T}$ distribution of the region below
30~GeV/$c$ is in good agreement with that from the \textsc{resbos}
calculation. The \textsc{resbos} calculation of the $\gamma^*/Z$
$P_{\rm T}$ distribution in $p\bar{p}$ collisions, which is
kinematically similar to $W$-boson production, agrees with the CDF
measurement based on an integrated luminosity corresponding to
2.1~fb$^{-1}$~\cite{zpt21}. 

\par
The current known values for the $W$-boson pole mass $M_W$
and resonant width $\Gamma_W$, 80.385~GeV/$c^2$ and
2.085~GeV~\cite{pdg2016}, respectively, are used in the
\textsc{powheg-box} and \textsc{resbos} calculations.
Both calculations employ resonant line shapes for the
boson-mass distribution with mass-dependent widths.
CDF has modified the \textsc{powheg-box} calculation to
use the recent values of the 
Cabibbo-Kobayashi-Maskawa matrix~\cite{C-ckm, KM-ckm}
elements associated with the
weak-interaction charged current~\cite{pdg2016}.
\par
\begin{figure}
\includegraphics
   [width=85mm]
   {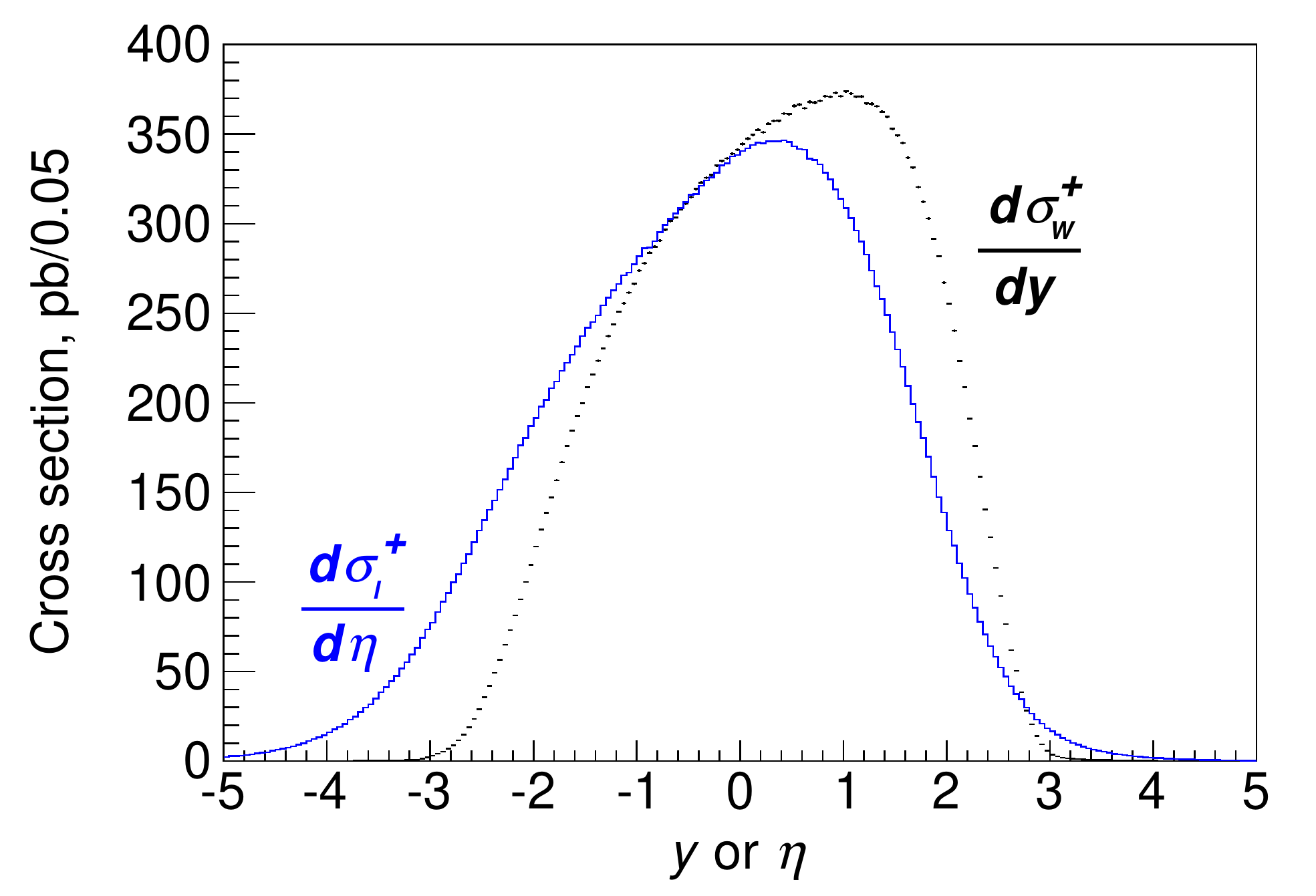}
\caption{\label{fig_pbxSigma}
$d\sigma_W^+/dy$ and $d\sigma_\ell^+/d\eta$ cross sections. The
cross section for $W^+$-boson ($W$-decay charged-lepton)
production in 1.96~TeV proton-antiproton collisions as a
function of the rapidity (pseudorapidity). The cross sections
are from the default calculation.
}
\end{figure}
\begin{figure}
\includegraphics
   [width=85mm]
   {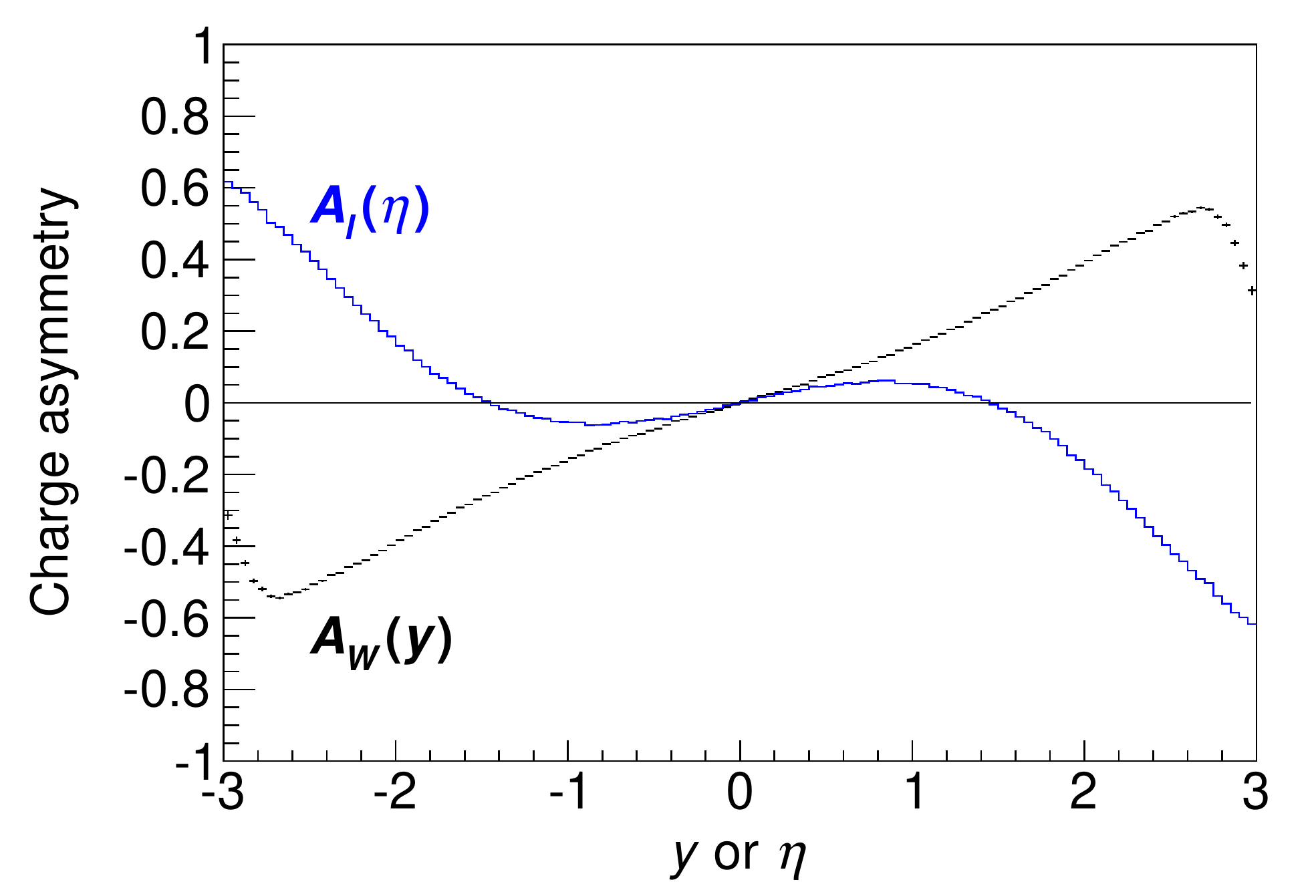}
\caption{\label{fig_pbxAsymm}
$A_W(y)$ and $A_\ell(\eta)$ charge asymmetries.
The charge-dependent $W$-boson ($W$-decay charged-lepton)
yield asymmetries as functions of the rapidity (pseudorapidity).
The asymmetry distributions are antisymmetric and from the default
calculation.
}
\end{figure}

Figure~\ref{fig_pbxSigma} illustrates the $d\sigma_W^+/dy$
and $d\sigma_\ell^+/d\eta$ cross sections from the
\textsc{powheg-box} calculation.
Since the geometry of the colliding $p$ and $\bar{p}$ system is
asymmetric under the reversal of charge and parity (CP), the
$d\sigma_W^-/dy$
and $d\sigma_\ell^-/d\eta$ cross sections in a coordinate frame
whose positive-$z$ axis is oriented along the antiproton direction
are identical to those of the positive-charge cross sections shown
in Fig.~\ref{fig_pbxSigma}, where the positive-$z$ axis is oriented
along the proton direction. In the laboratory frame of
Fig.~\ref{fig_pbxSigma}, $d\sigma_W^-(y)/dy = d\sigma_W^+(-y)/dy$
and $d\sigma_\ell^-(\eta)/d\eta = d\sigma_\ell^+(-\eta)/d\eta$.
Figure~\ref{fig_pbxAsymm} illustrates the boson-level charge
asymmetry $A_W(y)$
and the lepton-level charge asymmetry $A_\ell(\eta)$ based
on the \textsc{powheg-box} calculation.

\section{\label{asymMeasurement}
Measured asymmetries}

Measurements of the charge asymmetry $A_\ell$ within a reconstructed
pseudorapidity interval (bin) can be formally expressed using the observed
cross section $\Delta\sigma = N / ({\cal L}\,\epsilon A)$, where $N$ is
the number of observed signal events after background subtraction,
${\cal L}$ the integrated luminosity,
$\epsilon$ the event reconstruction and selection efficiency, and $A$
the acceptance of events within the kinematic and fiducial restrictions.
When the expressions for the bin-level cross sections of the electrons
and positrons are substituted into Eq.~(\ref{eqnAlepDef}), the common
integrated luminosity terms cancel out yielding
\begin{equation}
  A_\ell = \frac{ N^+/(\epsilon A)^+ - N^-/(\epsilon A)^- }
                { N^+/(\epsilon A)^+ + N^-/(\epsilon A)^- } \: ,
\label{eqnAellpDef}
\end{equation}
where $N^{+(-)}$ and $(\epsilon A)^{+(-)}$, respectively, represent
the $N$ and $\epsilon A$ of positrons (electrons).

\par
In this paper the asymmetry is measured using Eq.~(\ref{eqnAellpDef})
over the positron- and electron-pseudorapidity range of $-3$ to 3 in
uniform bins of width 0.2. As the positrons and electrons in each bin
are within the same region of $\eta$, their reconstruction and
selection efficiencies are expected to be similar except for those
that are not charge symmetric. Relative to a cross-section measurement,
the precision of a ratio of cross-section measurements such as the
asymmetry is far less dependent on accurate measurements of all of the
experimental efficiencies.

\par
Details of the experimental apparatus, data set, simulation of the data,
and corrections to the data used in the measurement follow in
Secs.~\ref{CDFdetector} to \ref{CorrDatSim}.

\section{\label{CDFdetector}
Experimental Apparatus}

The CDF II apparatus is a general-purpose detector~\cite{refCDFII}
at the Fermilab Tevatron, a $p\bar{p}$ collider with
$\sqrt{s}= 1.96$~TeV. The Cartesian coordinates of the detector are
denoted by $x$, $y$, and $z$. The coordinate system is right-handed
with the positive-$z$ axis directed along the proton direction and
the positive-$y$ axis directed vertically upwards.
For particle trajectories, the polar angle $\theta$ is relative to
the positive-$z$ axis and the azimuthal angle $\phi$ is relative to
the positive-$x$ axis.
Detector coordinates, denoted by $(\eta_{\rm det}, \phi)$, are defined
relative to the center of the detector ($z=0$).
\par
The curvature and momentum $\vec{P}$ of a charged particle are
measured in the magnetic
spectrometer, which consists of charged-particle position detectors
(trackers) immersed in a magnetic field. The energy $E$ of photons,
electrons, and hadrons is measured by the calorimeters surrounding the
magnetic spectrometer. The measured energy in the calorimeters
(energy flow) transverse to the beam line is $E_{\rm T} = E \sin \theta$.
\par
The tracking detectors consist of an outer central tracker and an inner
tracker. The central tracker is a 3.1~m long, open-cell drift
chamber~\cite{refCOT} that extends radially from 0.4 to 1.4~m.
Between the Tevatron beam pipe and the central tracker is a 2~m long
silicon-microstrip inner tracker~\cite{refSVXII}. The central
drift-chamber tracker has
96 tracking layers, and the silicon tracker has seven or eight tracking
layers depending on $\eta_{\rm det}$. Both trackers are immersed
in a 1.4~T axial magnetic field produced by a superconducting solenoid
beyond the outer radius of the drift chamber.
Combined, these two trackers provide efficient, high-resolution
tracking and momentum measurement over $|\eta_{\rm det}|<1.3$.
\par
Outside the solenoid is the central-barrel calorimeter that covers
the region $|\eta_{\rm det}|<1.1$~\cite{refCEM,refChad}.
The forward regions,
$1.1<|\eta_{\rm det}|<3.5$, are covered by disk-shaped
end-plug calorimeters~\cite{refPEM,refPES,refPHA}.
The electromagnetic (EM) and hadronic (HAD) sections
of the calorimeters are scintillator-based sampling calorimeters,
transversely segmented into projective towers that point back to the
center of the detector.
The EM-calorimeter energy resolutions measured in test beams with
electrons are $\sigma/E = 13.5\%/\sqrt{E_{\rm T}}$ for the central
calorimeter and $\sigma/E = (16\%/\sqrt{E}) \oplus 1\%$ for the plug
calorimeter, where the symbol $\oplus$ is a quadrature sum,
and $E_{\rm T}$ and $E$ are in units of GeV. Both the central- and
plug-EM calorimeters have preshower and shower-maximum detectors for
electromagnetic-shower identification and centroid measurements.
The shower-maximum detectors within the central- and plug-EM
calorimeters are strip detectors, and are denoted by CES and PES,
respectively.

\par
The combination of the PES detector and silicon tracker provides
enhanced electron-tracking coverage to $|\eta_{\rm det}| = 2.8$.
A PES detector
consists of eight $45^\circ$ wedge-shaped subdetectors assembled
into a disk. Subdetector wedges consist of ``strips'' made of
$5 \times 5$~mm$^2$ scintillator bars organized into two parallel
planes, denoted by $u$ and $v$, that span the length of the fiducial
region of the measurement. The $u$ strips are parallel to one radial
edge, and the $v$ are parallel to the other. 

\par
The presence of neutrinos in $W \rightarrow \ell\nu$ events is
inferred from the energy-flow measurements on all reconstructed
particles in the event.
The transverse momentum of the neutrino balances the vector
sum of the transverse-energy flows. Thus, the negative of this
vector sum, called the missing $\vec{E}_{\rm T}$ and denoted by
$\vec{E}\!\!\!/_{\rm T}$, is an estimator of the neutrino
transverse momentum.

\section{\label{DataSelection}
Data Selection}

The data set, collected over 2002--2011, is the full CDF Run~II sample
corresponding to an integrated $p\bar{p}$ luminosity of 9.1~fb$^{-1}$.
After event selection, the sample consists of $5.8 \times 10^6$ events. 
Section~\ref{ElectronTriggers} reports on the online selection of
events (triggers) for the charge-asymmetry measurement.
Section~\ref{OfflineEleSelection} describes the offline selection
of electron candidates, and Sec.~\ref{EleNeuSelection} describes
the selection of electron-neutrino pairs.

\subsection{\label{ElectronTriggers}
Online event selection}

Event samples enriched in signal candidates are selected by means
of two online triggers, \textsc{central-18} and
\textsc{pem-20\_met-15}. The \textsc{central-18} selection
accepts events containing at least one electron candidate
in the central calorimeter with $E_{\rm T} > 18$~GeV.
Candidates are required to have electromagnetic-shower clusters
in the central calorimeters that are geometrically matched to
tracks from the central tracker. The \textsc{pem-20\_met-15}
selection accepts events with an electron candidate in the
plug calorimeter with $E_{\rm T} > 20$~GeV and with
$E\!\!\!/_{\rm T} > 15$~GeV. Electron candidates in the plug region
are not required to geometrically match a track extrapolation.
Values of the $E_{\rm T}$ and $E\!\!\!/_{\rm T}$ quantities
differ from the corresponding values of the offline quantites
of Sec.~\ref{OfflineEleSelection} due to more refined offline
calibrations and calculation techniques.

\subsection{\label{OfflineEleSelection}
Offline electron selection}

The offline event reconstruction, which includes the application of
standard electron identification and quality requirements,
improves the purity of the sample~\cite{refCDFII}.
Fiducial constraints are applied to ensure that the electron candidates
are reconstructed in instrumented detector regions.
Each electron candidate is required to be associated with a track
whose origin along the beam line $(z_{\rm vtx})$ is restricted to be
within 97\% of the luminous region, $|z_{\rm vtx}| < 60$~cm.

\par
Electron identification in the central region is optimized for
electrons of
$P_{\rm T}>10$~GeV/$c$~\cite{refCDFII}. It uses information from
the central and silicon trackers, the longitudinal and lateral
(tower) segmentation of the electromagnetic and hadronic calorimeter
compartments, and the CES detector within the
electromagnetic calorimeter. The highest quality of signal selection
and background rejection is provided by the trackers in combination
with the CES. An electron candidate must have an associated shower
cluster within
the electromagnetic-calorimeter towers and a CES signal compatible
with the lateral profile of an electromagnetic shower. A candidate
must also be associated with a track that extrapolates to the
three-dimensional position of the CES shower centroid. The
transverse momentum associated with the track must
be consistent with the corresponding electron-shower $E_{\rm T}$ via
an $E/P$ selection if $P_{\rm T} < 50$~GeV/$c$~\cite{refCDFII}.
For both the track matching in the CES and the $E/P$ selection,
allowances are included for bremsstrahlung energy loss in the
tracking volume, where material thickness is on average 20\% of a
radiation length. The ratio of the measured shower energy in
the hadronic calorimeter to that in the electromagnetic calorimeter,
$E_{\rm HAD}/E_{\rm EM}$, must be consistent with that for electrons.

\par
Electron identification in the plug calorimeter
also uses information from the tracker, from the
longitudinal and lateral (tower) segmentation of the
electromagnetic and hadronic calorimeter compartments, and from
the PES detector within the electromagnetic
calorimeter. As the plug-calorimeter geometry differs from the
central geometry, the details of the selection requirements differ.

\par
The plug calorimeters, with sampling planes perpendicular to the
beam line, have much smaller projective towers than the central
calorimeter towers and vary in size as a function
of $|\eta_{\rm det}|$~\cite{refPEM}.
The preshower detector is the first layer of the electromagnetic
calorimeter and it is instrumented and read out separately. As there
are approximately 0.7 radiation lengths of material in front of it,
the energy released in this layer is included in the
electromagnetic-cluster shower energy.

\par
Electrons entering the plug calorimeters have reduced geometrical
acceptance in the central tracker for $|\eta_{\rm det}| > 1$, which
vanishes at $|\eta_{\rm det}| \approx 1.5$. However, the
silicon tracker has good coverage in the forward region, which is
exploited with a calorimetry-seeded tracking algorithm
denoted as ``Phoenix.'' The electron acceptance of this algorithm
is roughly 90\% to $|\eta_{\rm det}| = 2.4$ and decreases beyond
that value but does not vanish. With this algorithm, the track helix
in the magnetic field is specified by the position of the $p\bar{p}$
collision vertex, the three-dimensional position of the
electron in the PES, and the helix curvature. The collision
vertex is reconstructed from other tracks in the event.
The curvature is derived from the $E_{\rm T}$ of the
shower in the electromagnetic calorimeter. Two potential helices
are formed, one for each charge. The algorithm projects each helix into
the silicon tracker to initialize the track reconstruction. If both
projections yield valid tracks, the higher-quality one is selected.
Depending on its vertex location along the beam line, a track
traverses up to eight layers of silicon, of which the track
reconstruction uses the outer seven layers. The innermost layer
has significant electronic noise and is not used. Phoenix tracks
selected for the asymmetry measurement are required to traverse
at least three layers and have at least three silicon signals.
For the high-$E_{\rm T}$ electrons from $\gamma^*/Z$-boson decays,
85\% of the electrons traverse four or more layers for an
average tracking efficiency of about 80\%.

\par
An electron candidate in a plug calorimeter must have a shower
cluster of towers within the electromagnetic calorimeter and an
associated signal in the PES detector. The transverse and lateral
profiles of the shower cluster are required to be consistent with
those obtained from test-beam electrons. 
The transverse profile is evaluated in
a $3 \times 3$ detector-tower grid centered on
the highest-energy tower. The goodness-of-fit $\chi^2$ between this
profile and the expectation based on the shower-centroid location
in the PES detector is denoted by the quantity $\chi^2_{3\times3}$.
The longitudinal profile is measured using $E_{\rm HAD}/E_{\rm EM}$.
Neither a track $P_{\rm T}$ nor an $E/P$ selection requirement is
applied because the reconstruction method correlates the track
momentum to the calorimeter energy.
Charge-misidentification rates of Phoenix tracks increase
significantly with increasing $|\eta_{\rm det}|$ values because the
path lengths of the charged particles within the transverse plane
of the magnetic field decrease from 129 to 23~cm. The
transverse displacements of particles at the track-exit radii of
the PES detector relative to the trajectories of particles in the
absence of a magnetic field vary
quadratically with the path length. The position resolutions
are approximately 1.2~mm.

\par
As electrons from $W \rightarrow e\nu$ decays originate from the
$p\bar{p}$ collision vertex, tracks are required to have
impact parameters $(d_0)$, defined as transverse distances of closest
approach to the beam line, consistent with zero.
Tracks in the central and plug regions are required to have
one or more silicon-detector hits and $|d_0| < 175$~$\mu$m.
These mild requirements are effective
for removing unwanted events from the electron sample.

\par
The high-$E_{\rm T}$ leptons from the production and decay of
$W$ bosons are expected
to be produced in isolation from other particles in the event. 
Consequently, electron candidates are required to be isolated
from other calorimetric activity. The isolation energy,
$E_{\rm iso}$, is defined as the sum of $E_{\rm T}$ over towers
within a cone of radius 0.4 in $(\eta,\phi)$ surrounding the
electron cluster. The towers of the electron cluster are not
included in the sum. For central-electron candidates, the
isolation requirement is $E_{\rm iso}/E_{\rm T} < 0.1$; and
for plug-electron candidates, it is $E_{\rm iso} < 4$~GeV.

\par
As the offline-electron sample contains central electrons from
the $\gamma^*/Z$-production, the following criteria are
applied to reduce the fraction of such electrons. These criteria
improve the efficiency of event processing over the large number
of events in the sample but do not affect the asymmetry
measurement. Events with two or more electrons with
$E_{\rm T} > 18$~GeV and $E\!\!\!/_{\rm T} < 12.5$~GeV are
identified. However, the isolation and $E_{\rm HAD}/E_{\rm EM}$
requirements are not applied. Electron pairs from the production
of $\gamma^*/Z$ bosons are identified following
Ref.~\cite{cdfAfb9eeprd,*cdfAfb9eeprdErr}, and pairs with
invariant masses larger than 40~GeV/$c^2$ are removed.

\par
For central electrons, the selection criteria have an overall
efficiency of about 85\% and result in a high-quality sample of
high-$E_{\rm T}$ electrons. However, the criteria for the plug
region result in a sample with significant background, whose level
varies significantly with the topology of the reconstructed track
in the silicon detector. Track- and electron-quality are combined
and made more stringent depending on the background fraction of
the track topology. Details are presented in
Appendix~\ref{PlugEleSel}. These more stringent critera result in
a sample whose size is 18\% smaller but whose quality is vastly
improved. Overall, the selection efficiency for plug electrons is
about 60\%. After the application of the event selection criteria
of Sec.~\ref{EleNeuSelection}, the purities of the central- and
plug-electron samples are similar.

\subsection{\label{EleNeuSelection}
Event selection}

For the asymmetry measurement, events are required to have high
missing $E_{\rm T}$ and a single high-$E_{\rm T}$ electron.
Electrons are accepted if detected in either the central or
plug calorimeters with the following conditions:
\begin{enumerate}
 \item Central electrons
    \begin{enumerate}
      \item $E_{\rm T} > 25$~GeV;
      \item $E\!\!\!/_{\rm T} > 25$~GeV;
      \item $0.05 < |\eta_{\rm det}| < 1.00$.
    \end{enumerate}
 \item Plug electrons
    \begin{enumerate}
      \item $E_{\rm T} > 25$~GeV;
      \item $E\!\!\!/_{\rm T} > 25$~GeV;
      \item $1.2 < |\eta_{\rm det}| < 2.8$.
    \end{enumerate}
\end{enumerate}
The kinematic variables are based on the energy measured in the
calorimeter and on the track direction.
Detector pseudorapidity $\eta_{\rm det}$ is
defined with the detector coordinates of its shower-centroid
location within the CES or PES detectors.
The missing-$E_{\rm T}$ vector of an event, $\vec{E}\!\!\!/_{\rm T}$,
is derived using energy-flow measurements from the calorimeters.
It is defined as $-\sum_i E_{\rm T}^i \hat{n}_i$,
where the sum is over calorimeter towers, $\hat{n}_i$ is the unit
vector in the azimuthal plane that points from the $p\bar{p}$
collision vertex to the center of the calorimeter tower $i$, and
$E_{\rm T}^i$ is the corresponding transverse energy in that
tower. 

\par
The electron transverse momentum and the missing $E_{\rm T}$
of the event are combined to form the transverse mass of
the boson, $M_{\rm T}$, defined as
$\sqrt{ 2 E_{\rm T}^e E_{\rm T}^\nu (1 - \cos\Delta\phi_{e\nu})}$,
where $E_{\rm T}^e$ is the transverse energy of the electron,
$E_{\rm T}^\nu$ the missing $E_{\rm T}$ of the event,
and $\Delta\phi_{e\nu}$ the azimuthal angle
between them.
The small fraction of events with $M_{\rm T} < 45$~GeV/$c^2$,
which are poorly simulated, is removed.

\section{\label{ExpDatSim}
Signal simulation}

Data corrections are obtained using a simulation of the data events.
The \textsc{pythia}~6.2
event generator~\cite{Pythia621} with CTEQ5L~\cite{Cteq5pdf} PDFs
simulates the LO QCD interaction $q\bar{q}^\prime \rightarrow W$,
along with the initial-state QCD radiation of the colliding quarks
via its parton-shower algorithms; decays the boson via the
$W \rightarrow \ell\nu$ channel; and adds quantum-electrodynamics
(QED) final-state radiation (FSR) to the charged lepton.
Final-state particles not produced in the hard scattering,
referred to as the underlying event (UE), are also simulated.
The boson-$P_{\rm T}$ and UE parameters are derived from
the \textsc{pythia} configuration \textsc{pytune} 101,
a tuning to previous CDF data~\cite{Pythia621,run1CDF-Z,PyTuneAW}.

\par
The simulation model for the production and decay of $W$ bosons
is weighted to resemble the more precise NLO QCD calculation
based on \textsc{powheg-box} and NNPDF~3.0 PDFs,
described in Sec.~\ref{AsymmetryDistr}.
Three event-weight tables are used to improve the agreement.
They are a one-dimensional invariant-mass $(M)$ table with
high resolution and a lower-resolution pair of two-dimensional tables
in the variables $(y_{\rm scl},M)$ and $(y_{\rm scl},P_{\rm T})$,
where $y_{\rm scl}$ is a scaled rapidity defined as
$y_{\rm scl} = y/y_{\rm max}$ with
$y_{\rm max} = \ln \sqrt{s/(M^2 + P_{\rm T}^2)}$. As the correction
steps are not independent,
they are determined using an iterative procedure.
Additionally, the
angular distribution of the $\ell\nu$ pairs is adjusted.
For this correction, the coefficient functions $A_{0-4}$ of
Eq.~(\ref{eqnAngDistr}) are implemented as two-dimensional tables in
the variables $(y_{\rm scl},P_{\rm T}/\sqrt{M^2 + P_{\rm T}^2})$
for the \textsc{pythia} and \textsc{powheg-box}
distributions\footnote
{
The extracted values of $A_0$ are slightly negative for values of
$P_{\rm T}$ near zero. Offsets are added to these table values
so that $A_0 \ge 0$.
},
the values of the ${\cal N}(\vartheta,\varphi)$ functions are
calculated for each event with $A_{5-7} = 0$, and
the ratio of the values is the adjustment event weight.

\par
Generated events are first processed by the event simulation,
which uses \textsc{photos} 2.0 to account for QED FSR
from promptly decaying final-state hadrons and their decay
products~\cite{Photos20a, *Photos20b, Photos20c}, which is not
modeled by \textsc{pythia}. In addition,
multiple $p\bar{p}$ interactions are added by \textsc{pythia}.
This is followed by the CDF~II detector simulation based on
\textsc{geant}-3 and \textsc{gflash}~\cite{nimGflash}.
Standard time-dependent beam and detector conditions are
incorporated in the simulation, including
the $p$ and $\bar{p}$ beam line parameters; the luminous-region profile;
the instantaneous and integrated luminosities per data-taking period;
and the calibration of detector elements, which include electronic
gains and malfunctions. The simulated events are reconstructed,
selected, and analyzed in the same way as the experimental data.
\par
The simulation does not describe kinematic distributions
such as the $E_{\rm T}$ of electrons and the missing $E_{\rm T}$ of
events with sufficient accuracy. Modest adjustments are applied to
bring the simulation into agreement with the data.

\section{\label{CorrOverview}
Correction overview}

A form of the asymmetry expression [Eq.~(\ref{eqnAellpDef})] which
shows the net correction of the $(\epsilon A)^+$ and $(\epsilon A)^-$
terms in the measurement is
$(N^+\! - \rho N^-)/(N^-\! + \rho N^-)$,
where $\rho = (\epsilon A)^+\! / (\epsilon A)^-$.
The $(\epsilon A)^\pm$ terms are evaluated with the simulation after
adjustments are applied to obtain agreement with
the data. Efficiencies and energies of electrons,
energies and distributions of hadrons, and
misidentification rates of the electron charge are suitably adjusted.
All corrections, except those for the rates of
charge misidentification, are independent of the electron charge.

\par
\begin{figure}
\includegraphics
   [width=85mm]
   {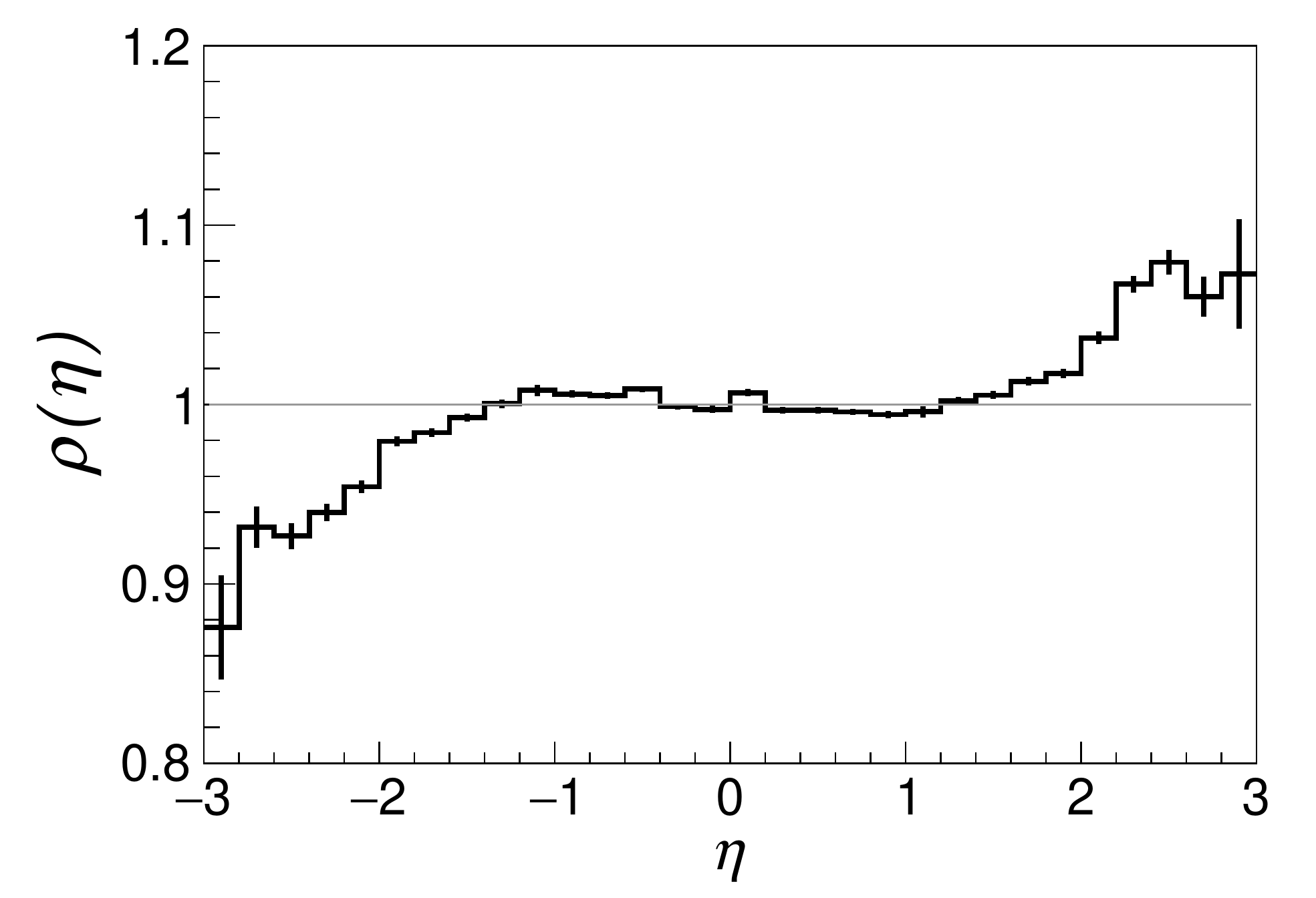}
\caption{\label{fig_reffAccM0}
The ratio $\rho(\eta)$, $(\epsilon A)^+\! / (\epsilon A)^-$, as a
function of $\eta$ with the correct charge assigned to the
reconstructed electrons. The uncertainties shown are statistical.
}
\end{figure}

When the correct charge is assigned to the reconstructed electrons of
the simulation, the values of $(\epsilon A)^+$ and $(\epsilon A)^-$
within an $\eta$ bin are similar, but vary across $\eta$ bins.
Common portions of, and common uncertainties on, $(\epsilon A)^+$
and $(\epsilon A)^-$, including those from the charge-independent
corrections, cancel out to first order in the ratio $\rho$.
Figure~\ref{fig_reffAccM0} shows the ratio as a function of $\eta$.
Changes in the acceptance ratio with $\eta$ are due to the
$E_{\rm T}$ distributions of the leptons, which are similar
in shape when $|\eta| = 0$ but evolve differently as $|\eta|$
increases in value.

\par
An overview of the corrections and the estimation of the backgrounds
in the data is presented in the remainder of this section.
Details are presented in Sec.~\ref{CorrDatSim}.
As many of the charge-symmetric adjustments are influenced by others,
the determination process is iterative. Among all corrections, only
the charge-misidentification rate of plug electrons has a significant
impact on the asymmetry measurement.

\subsection{\label{ovwEleCorr}
Electron corrections}

Corrections for electrons follow Ref.~\cite{cdfAfb9eeprd,*cdfAfb9eeprdErr},
hereafter referred to as the $ee$-pair analysis.
Energy calibrations and efficiency
measurements from the $ee$-pair analysis are used as initial calibrations
in this work because the kinematic properties of the
decay electrons from the production of $\gamma^*/Z$ and $W$ bosons are
similar. Corrections to account for differences in the event
environment and selection criteria are applied.

\par
Energy-scale adjustments are applied to the electron energies of both
the simulation and the data so that observed energies match the generator
level values~\cite{muPcorrMethod}. The adjustments are applied over the
initial corrections. Energy-resolution adjustments are applied to the
simulation so that its electron-$E_{\rm T}$ distributions are in better
agreement with those of the data. Additional adjustments are also needed
in the simulation to account for relative differences in the amounts
of hadronic energy deposited within electron showers.

\subsection{\label{ovwMetCorr}
Hadronic corrections}

The primary source of hadrons in the simulation is the parton shower
associated with the production of the $W$ boson, which approximates
the production of hadrons from initial-state QCD radiation.
A large fraction of the events contains low-energy parton showers,
whose production is nonperturbative.
Additional nonperturbative sources of hadrons that are difficult to
simulate accurately are multiple interactions and the underlying event.
Multiple interactions are independent $p\bar{p}$ interactions within
an event, and vary with the instantaneous luminosity. Their prevalence
and impact vary. The calorimeter response to hadrons
from nonperturbative events is nonlinear and inadequately simulated.
Collectively, the hadrons from all sources are
denoted as the recoil system of hadrons.

\par
All sources affect the missing $E_{\rm T}$ and the reconstructed
$E_{\rm T}$ of the electron. Due to the kinematic restrictions on
the electron and missing $E_{\rm T}$, event acceptances are affected
as well. The calibrations of both electron and recoil-system
quantities are affected by the spatial distribution of the hadrons
relative to the electron.

\par
To orient the spatial distribution of the hadrons, the direction of
the recoil system of hadrons with respect to the electron is specified
with the parameter $\cos(\Delta\phi_{eX})$, where $\Delta\phi_{eX}$ is
the azimuthal angle between the directions of the electron ($e$) and
the recoil system of hadrons ($X$) produced with the $W$ boson.
It is defined in the transverse-momentum frame where the net
transverse-momentum of the electron and neutrino is zero.
The boost to the frame is defined in terms of the electron
$\vec{E}_{\rm T}$ and event missing $\vec{E}_{\rm T}$.

\par
\begin{figure}
\includegraphics
   [width=85mm]
   {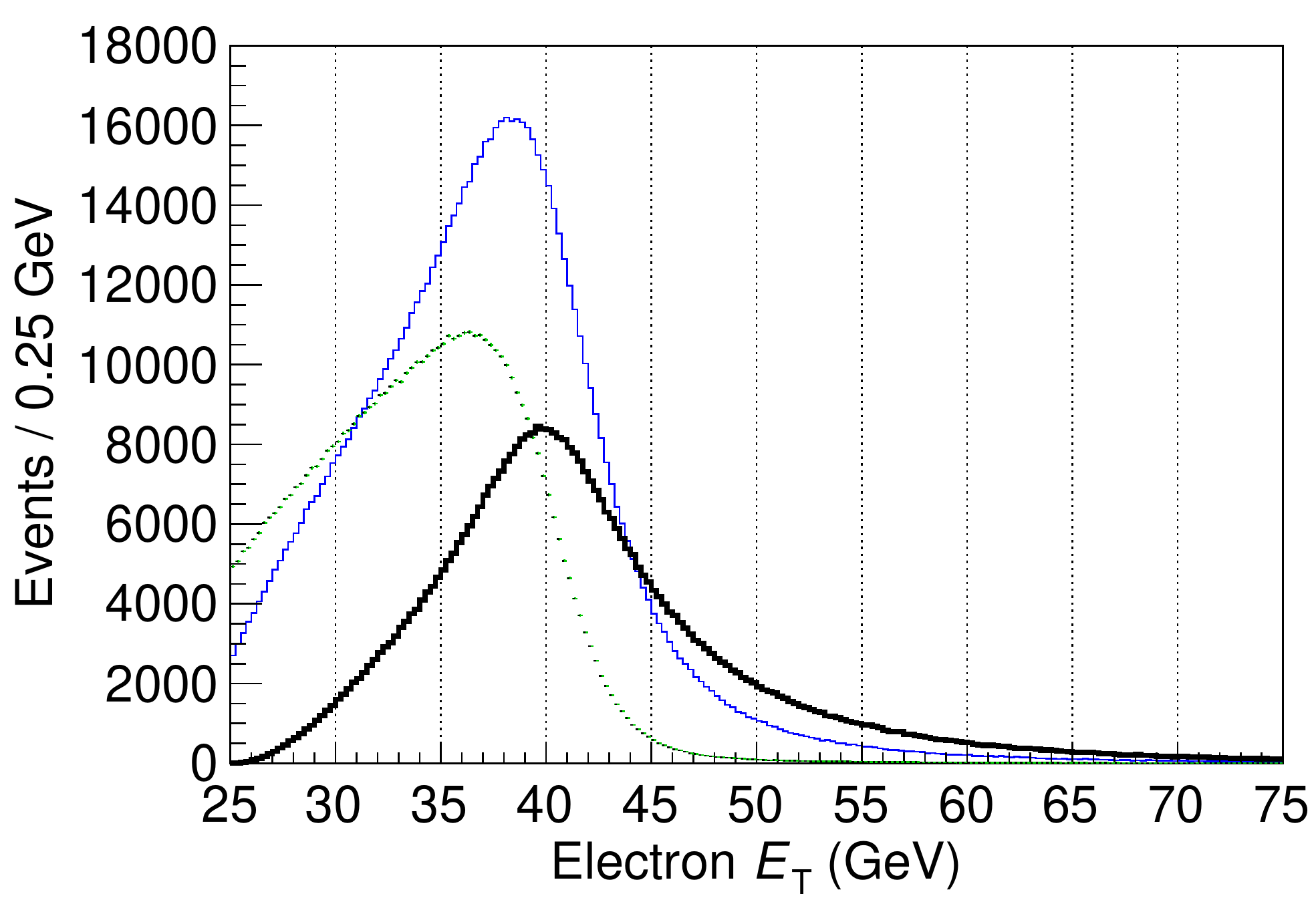}
\caption{\label{fig_selEtMXP}
Simulated $E_{\rm T}$ distributions of the away, transverse,
and toward regions for the electrons in the plug calorimeter.
Distributions for electrons in the central calorimeter are similar.
The bold (black) histogram is for the away region, the light (blue)
histogram is for the transverse region, and the (green) crosses are
for the toward region. Corresponding $E\!\!\!/_{\rm T}$
distributions follow a similar pattern except that away and toward
are reversed because the direction away from the electron
is toward the neutrino and vice versa.
}
\end{figure}

Corrections to quantities of the electron and the recoil system of
hadrons are evaluated in three $\cos(\Delta\phi_{eX})$ ranges,
$-1.0$ to $-0.6$, $-0.6$ to 0.6, and 0.6 to 1.0. As these regions are
relative to the electron, they are denoted as the away, transverse,
and toward regions, respectively. Figure~\ref{fig_selEtMXP}
illustrates the electron-$E_{\rm T}$ distributions
in these regions of $\cos(\Delta\phi_{eX})$. Different kinematic
distributions of the leptons are selected by the acceptances of
each region. These selections have a secondary effect on the
$W$-boson distributions, and the $E_{\rm T}$ distributions of
the recoil system in these regions are only slightly different.
The kinematic separations illustrated in Fig.~\ref{fig_selEtMXP}
expose direction-dependent differences of the simulation relative
to the data.

\par
Energy-scale corrections are first applied to the energies of the
recoil systems of both the data and simulation so that the observed
energies are calibrated to the generator-level
energies~\cite{muPcorrMethod}. Additional corrections are then
applied to the simulation that account for data-to-simulation
differences in the energy and spatial distributions of the
hadrons.

\subsection{\label{ovwBackgrounds}
Backgrounds}

Backgrounds total about 5\%. They are simulated for the following
processes, which produce high-$P_{\rm T}$ electrons:
$\gamma^*/Z \rightarrow ee, \; \tau\tau$, $W \rightarrow \tau\nu$,
dibosons (\textit{WW}, \textit{WZ}, \textit{ZZ}), and $t\bar{t}$ pairs.
The model of the QCD background is extracted from an experimental
data sample independent of the signal sample. Background rates are
normalized to the signal rates, and subtracted from distributions of
the data.

\par
The fraction of the QCD background in the measurement sample depends
on the location of the electron candidate, but overall is under 2\%.
Events from QCD background are due to parton-parton scattering
interactions that result in outgoing partons that fragment into
cascades of hadrons called \textit{jets}. A small fraction of jet
cascades contains electron candidates. QCD events are not expected to
exhibit any $E\!\!\!/_{\rm T}$, but nonzero values are observed
because jets can have reconstructed
energies that differ from their underlying energies, even
significantly, due to the resolution of the detector or the
traversal of hadrons into uninstrumented regions of the detector.
However, the number of events with these instrumental effects
decreases rapidly with increasing $E\!\!\!/_{\rm T}$ values. 

\par
The amount of QCD background in data is determined via a fit of the
QCD model and signal contributions to the $E\!\!\!/_{\rm T}$
distributions.

\subsection{\label{ovwQmisID}
Charge misidentification}

Charge-misidentification rates are significant only in the plug
regions. They are accounted for using measurements on $e^+e^-$
pairs with one electron in the central region and the other in
the plug region. Event selection follows that of the $ee$-pair
analysis. Central-region tracks, whose charges are well
measured, provide the reference charges expected for the
plug-region tracks. Rates are position-dependent and measured
over small regions of $(\eta_{\rm det},\phi)$. As this
division limits the statistical precision of the rates, the
rates are not measured as functions of any other parameters.

\section{\label{CorrDatSim}
Data and simulation corrections}

\subsection{\label{RateNormSim}
Event-rate normalizations}

The default simulation does not model the trigger and reconstruction
efficiencies observed in the data with the desired precision. Event
weights based on the efficiencies derived from the $ee$-pair analysis
are used as the initial correction to the simulation. The event weights
are ratios of the selection efficiencies observed in data to the
simulation versus time, position in the detector volume (denoted
henceforth as detector location), and instantaneous luminosity.
\par
As the electron-selection criteria of the $e\nu$-pair analysis are
more stringent than those of the $ee$-pair analysis, an additional
correction is determined using the $ee$-pair
data. The criteria of the $e\nu$-pair analysis are
applied over those of the $ee$-pair analysis, and efficiencies
calculated for both the data and simulation samples. The efficiency
ratio between the data and simulation provides the additional
correction.

\par
Changes of the Tevatron-luminosity profile over time are measured
using the $e\nu$-pair data and incorporated into the simulation.
The distributions of the location of the $p\bar{p}$-collision vertices
along the beam line $(z_{\rm vtx})$ and the number of multiple
interactions in an event $(n_{\rm vtx})$ changed significantly with
improvements to the beam current and optics of the Tevatron.
Measurements of the $z_{\rm vtx}$ distribution,
which has an rms dispersion of about 30~cm,
are organized into seven time intervals corresponding to the
introductions of major improvements in the Tevatron collider. As
$\eta_{\rm det}$ is a function of $z_{\rm vtx}$, inaccuracies
impact the determination of the acceptance. The $n_{\rm vtx}$
quantity is a measure of the instantaneous luminosity for the event.
Measurements of the $n_{\rm vtx}$ distribution are organized into
calibration periods.

\par
Another luminosity parameter, denoted as the average instantaneous
luminosity, is important for corrections to simulated quantities
over long periods of time. This parameter tracks the effects of the
beam to the event environment over multiyear periods associated with
major changes in the average $\bar{p}$ current circulating within the
Tevatron. Two time intervals need
to be taken into account by the simulation. The first interval covers
calibration periods where the average instantaneous luminosity is
relatively low (low-luminosity period), and the second interval
covers calibration periods where the average instantaneous luminosity
is relatively high (high-luminosity period).

\par
The initial coarse correction is refined toward a better resolution
in time, position, and luminosity using the events of the $e\nu$-pair
analysis. These finer extensions are separate for the events
with central-region and plug-region electrons, and use event-count
ratios between the data and simulation as event weights. All
corrections are functions of $n_{\rm vtx}$. Some are functions of 
the low- or high-luminosity period. Corrections for
detector-location dependencies are functions of $|\eta_{\rm det}|$.
For the central region, the location correction accounts for
data-to-simulation differences of the efficiencies across the CES
detector. For the plug region, the correction accounts
for differences of both the efficiencies across the PES detector and
the position-dependent response of EM-calorimeter towers.

\subsection{\label{ImportantSelVars}
Selection quantities}

The simulated distributions of isolation energy $E_{\rm iso}$
and the plug-electron $\chi^2_{3\times3}$ are adjusted to improve
the agreement with data. Isolation distributions are used in the
analysis of QCD backgrounds.
For electron candidates in the plug region, the adjustments are
important because the criteria for their selection detailed in
Appendix~\ref{PlugEleSel} are more stringent.

\subsection{\label{EnergyCalibrations}
Electron-energy corrections}

The initial corrections to the default calibrations use electron
pairs where one is detected in the central calorimeter and the
other in the plug calorimeter. Corrections to the energy scale
and resolution are functions of $(\eta_{\rm det},\phi)$. The
scale is also dependent on the low- or high-luminosity period.
The initial corrections also account for the extra energy in an
electron shower from the underlying event and multiple interaction
sources as an average correction over all event topologies.
Adjustments to these energy calibrations are derived for the
$e\nu$-pair analysis.

\par
Electron-energy corrections specific to the $e\nu$-pair analysis
are implemented in four steps. The first is the calibration of the
electron energy at the reconstruction level of the simulation to the
event-generator level~\cite{muPcorrMethod}.
Next, the energy scale of the data is aligned with that of the
simulation. The third step accounts for small but location-dependent
differences between the simulation and the data in the amounts of
shower energy from hadronic sources within electron showers.
The $\cos(\Delta\phi_{eX})$-dependent offsets of the simulated
electron-$E_{\rm T}$ distributions are corrected relative to the data.
Finally, the energy
resolution of the simulation is adjusted to improve agreement with
the data.

\par
The calibration of the energy scale of the simulation begins by
associating the reconstruction-level electron with its
generator-level counterpart. Then the electron and its companion
electrons and photons from QED FSR are clustered around the seed
tower as in the electron reconstruction.
The seed tower is based on the reconstructed
electron, and the projection from the $p\bar{p}$ collision vertex
to the tower is achieved by extrapolating the track helix.
The calibration is derived from the distribution of the ratio of
the reconstruction-level energy $E_{\rm rec}$ to the clustered
energy at the generator level $E_{\rm clus}$.
In the vicinity of its peak, which occurs at ratios of about 1.0,
the distribution is approximately Gaussian.
The energy scale is adjusted to make it peak at 1.0.
There are $|\eta_{\rm det}|$-dependent
adjustments of about 0.5\% or less, with the larger shifts being
in the plug region.

\par
The electron-$E_{\rm T}$ distributions are used for the alignment
of the energy scale of the data with that of the simulation. For
most EM-calorimeter towers, the data distributions agree with those
of the simulation without any adjustments. The overall uncertainty
of the energy scale based on the $\chi^2$ between
the data and the simulation is $\pm0.04$\%.

\par
To measure the energy shifts between the simulation and the data
due to the hadrons, events are separated into classes
according to whether the electron is detected in the central or plug
region, instantaneous luminosity, low- or high-luminosity period,
$n_{\rm vtx}$, and the region of $\cos(\Delta\phi_{eX})$.
These groups are denoted as ``standard-calibration groups''.
The shapes of the electron-$E_{\rm T}$ distributions for each
$\cos(\Delta\phi_{eX})$ region are similar to those shown in
Fig.~\ref{fig_selEtMXP}. Along the rising and falling edges about
the peaks of the distributions, small offsets separating the
simulated and experimental data are measured. Observed
offsets are of order 50 (100)~MeV for electrons detected in the
central (plug) calorimeter, and vary in magnitude and sign.

\par
The model for the energy-resolution of the simulation is
$\sigma^2 = c_0^2 E + c_1^2 E^2$, where $\sigma$ is the
resolution, $E$ the electron energy, $c_0$ the sampling term, and
$c_1$ the miscalibration or constant term. The sampling term,
calibrated with test-beam data, is part of the default detector
simulation. Additional adjustments are applied to the constant
terms of both central- and plug-region electrons so that the
electron-$E_{\rm T}$ distributions agree better with the data.

\par
Simulated electrons of the central region have a slightly broader
$E_{\rm T}$ distribution around the peak than in the data.
To reduce mismodeling, the reconstructed energy $E_{\rm rec}$
in the simulation is modified on an event-by-event basis using
$E^\prime_{\rm rec} = E_{\rm clus} - 
                      f_{\rm rms}\;(E_{\rm clus}-E_{\rm rec})$,
where $E^\prime_{\rm rec}$ is the adjusted value and $f_{\rm rms}$ a
parameter. As the $E_{\rm clus}-E_{\rm rec}$ term gives the
fluctuation of the reconstructed energy from its generator-level
value, the $f_{\rm rms}$ parameter rescales the rms of the
fluctuations. The optimization of $f_{\rm rms}$ constrained by the
data  yields $f_{\rm rms} = 0.87 \pm 0.03$.

\par
Simulated electrons of the plug region have a narrower $E_{\rm T}$
distribution around the peak than in the data. This is broadened
on an event-by-event basis by incorporating Gaussian fluctuations
to the energies that effectively increase the constant term $c_1$
beyond its default value of 0.01.
The adjustment is a function of the $|\eta_{\rm det}|$ coordinates
of calorimeter towers and the $p\bar{p}$-interaction count
$n_{\rm vtx}$. Adjusted values for $c_1$ range from 0.021 to 0.056
for increasing values of $|\eta_{\rm det}|$ and $n_{\rm vtx}$.
The uncertainty of the $c_1$
terms is estimated by rescaling all terms with a uniform
factor, propagating the effects to the $E_{\rm T}$ distributions, then
comparing them to the data. This results in a relative uncertainty
of $\pm$4\% on the constant terms.

\par
\begin{figure}
\includegraphics
   [width=85mm]
   {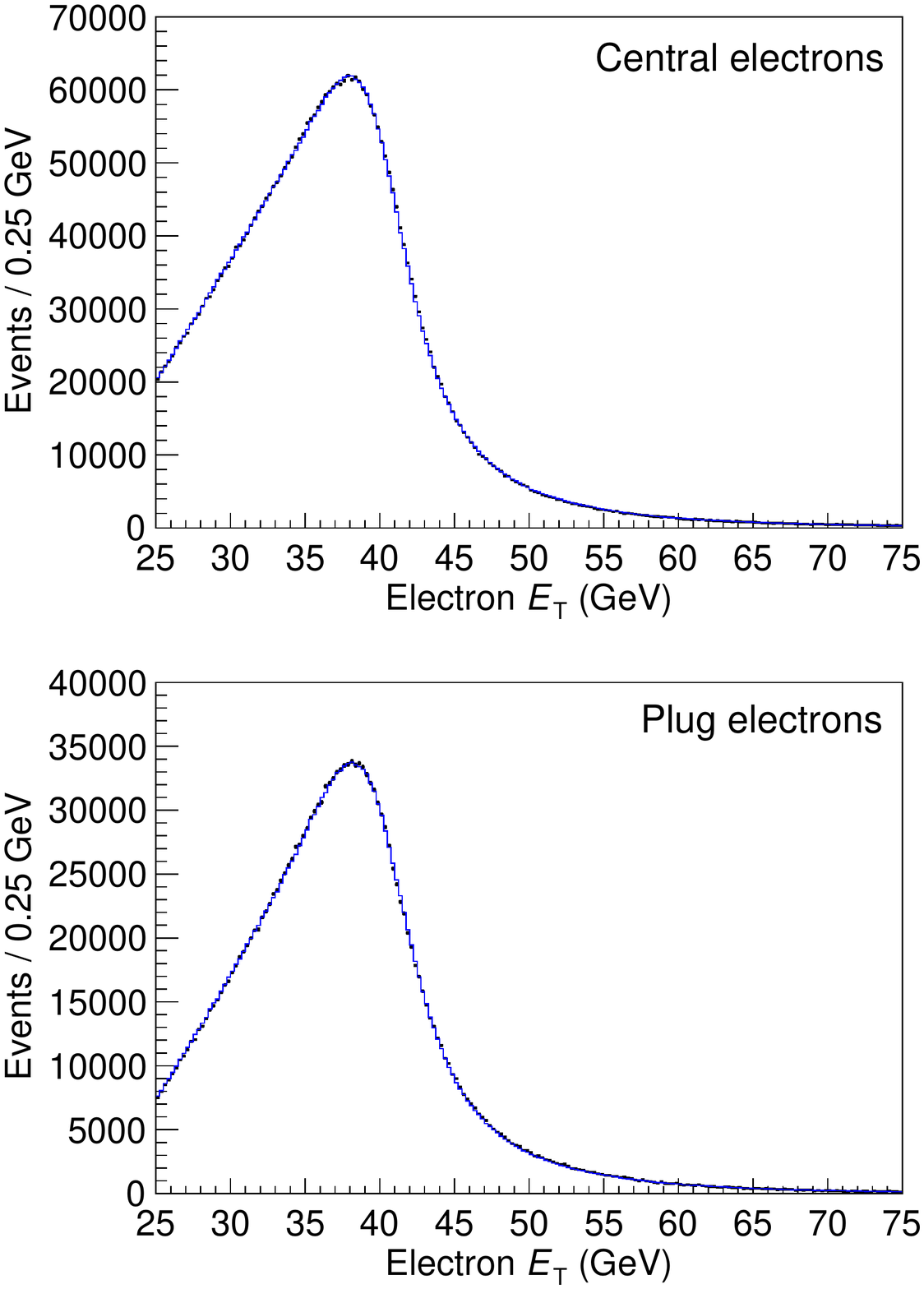}
\caption{\label{fig_cpElcEt2}
Background-subtracted
$E_{\rm T}$ distributions of electron candidates in the central
and plug regions. The crosses are the data and the solid histogram
is the simulation. The comparison between the data and the simulation
for electron candidates of the central (plug) region yields a
$\chi^2$ of 373 (317) per 200 bins.
}
\end{figure}

Figure~\ref{fig_cpElcEt2} shows the $E_{\rm T}$ distributions of
electrons in the central and plug regions after all corrections
are applied, including the hadronic corrections of
Secs.~\ref{HadronCalib} to \ref{METcorr}
and the subtraction of backgrounds described in
Sec.~\ref{ENuBackgrounds}.
All corrections need to be applied because of the correlations
among them.
Adjustments based on the $\cos(\Delta\phi_{eX})$ parameter
significantly reduce the biases affecting different regions of
the simulated-$E_{\rm T}$ distribution relative to the data.

\par
\begin{figure}
\includegraphics
   [width=85mm]
   {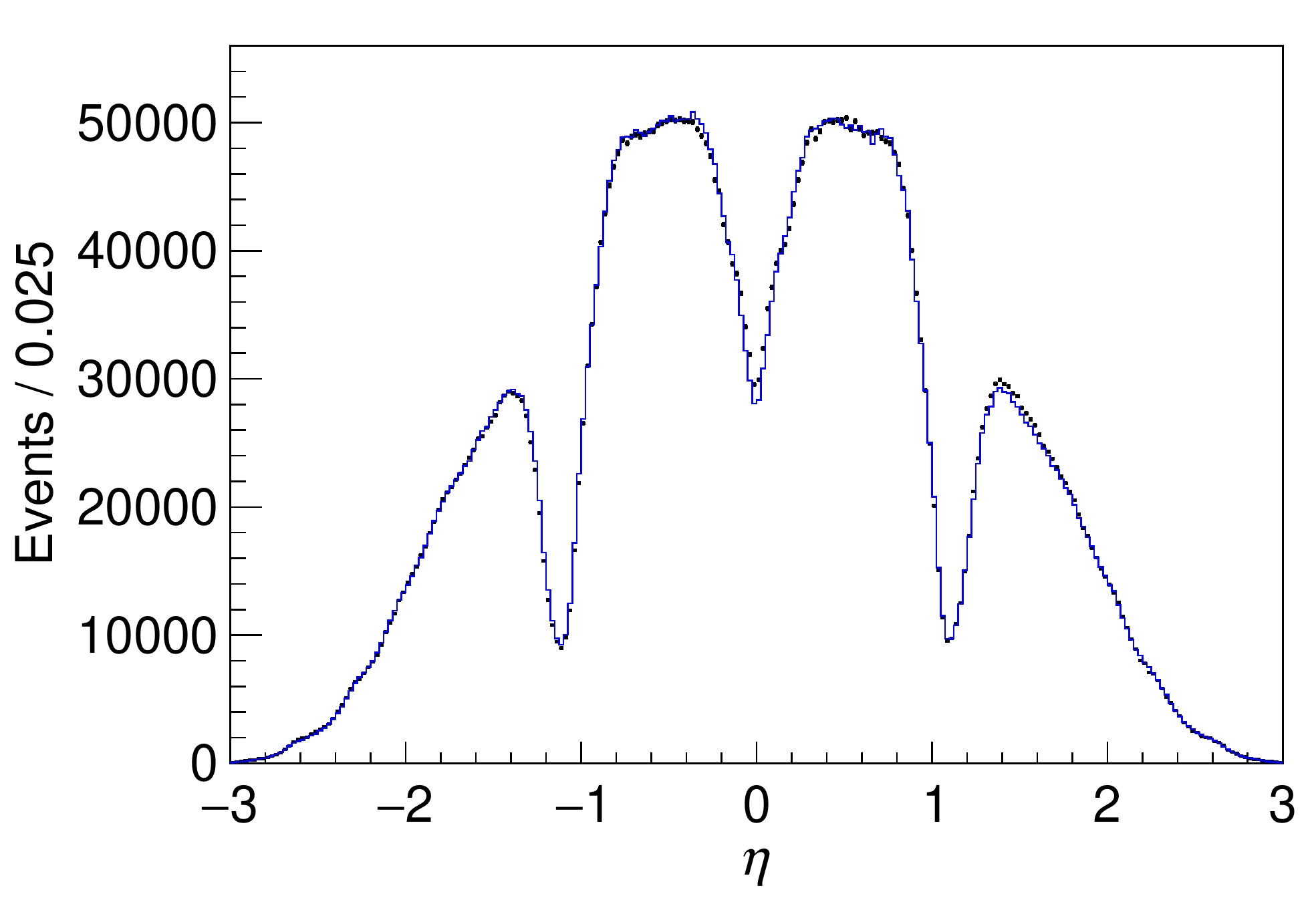}
\caption{\label{fig_elEta2}
Background-subtracted
$\eta$ distribution of the electron candidates used in the measurement
of the asymmetry. The crosses are the data and the solid histogram is
the simulation. All corrections have been applied.
}
\end{figure}

Figure~\ref{fig_elEta2} shows the corresponding $\eta$ distribution
of the electrons.
The valley structures of the distribution reflect the nonfiducial
regions in the central calorimeter ($|\eta| \sim 0$)
and between the central and plug calorimeters ($|\eta| \sim 1.1$).
Electron-$E_{\rm T}$ distributions in $\eta$ subregions covered
by the central and plug calorimeters, $|\eta|$ values above and
below 0.5 for the central calorimeter and $|\eta|$ values above and
below 1.6 for the plug calorimeter, are adequately simulated. The
shapes of the $E_{\rm T}$ distributions of the central (plug)
subregions are similar to those of the central (plug) region in
Fig.~\ref{fig_cpElcEt2}.
Comparisons between the data and the simulation for the electron
candidates of the four subregions yield statistical-$\chi^2$ values
ranging from 360 to 550 for 200 bins.

\subsection{\label{METAna}
Recoil system of hadrons}

The momenta of the $W$ boson ($\vec{P}_{\rm T}^W$) and of the recoil
system of final-state particles directly associated with the boson
production ($\vec{P}_{\rm T}^X$) balance each other in the transverse
plane of an event so that $\vec{P}_{\rm T}^X = -\vec{P}_{\rm T}^W$.
The transverse energy of the recoil system observed in
the detector, denoted by $\vec{E}_{\rm T}^{X{\rm (obs)}}$,
is the quantity $-\vec{E}\!\!\!/_{\rm T}-\vec{E}_{\rm T}^e$,
where $\vec{E}_{\rm T}^e$ represents the uncorrected
contribution of the electron shower to the missing-$E_{\rm T}$
term.

\par
The observed recoil $\vec{E}_{\rm T}^{X{\rm (obs)}}$ is a
combination of the products of the hard collision producing
the $W$ boson and of the other activity in the event such as the
underlying event and multiple interaction sources.
In the QCD parton model, the outgoing parton from the hard
collision recoiling against the $W$ boson fragments into a jet.
At large values of the parton $P_{\rm T}$,
the fragmentation results in a collimated jet of particles in the
detector. Energetic final-state partons with sufficiently low values
of $P_{\rm T}$ fragment into jets where a fraction of the particles
enter the beam hole. Hadrons from softer nonperturbative partons
are distributed more broadly.

\par
The calorimeters are calibrated so that particles that undergo
electromagnetic showering
have $E/P \approx 1$. However, the response
to a hadronic cascade is intrinsically different
from an electromagnetic cascade. In the simulation, the response of
the calorimeters to jet particles is based on the observed
responses from single particles in test beams and collider
data~\cite{refCdfJES}. The measured responses of particles with momenta
down to 0.5~GeV/$c$ are incorporated into the \textsc{gflash} model
of the calorimeter response. For a 2~GeV/$c$ hadron, $E/P$ equals
about 0.65 and increases with the particle momentum. As jets typically
consist of many low-momentum particles, the observed energy of the jet
in the calorimeters is lower than the momentum of the underlying
particles.

\par
For clustered jets, corrections to transform the jet response of the
calorimeters to the momentum of the underlying
jet of particles have been determined~\cite{refCdfJES}. These jet
corrections are validated using events with
$\gamma^*/Z$-bosons produced in association with jets,
where the $\gamma^*/Z$ bosons are reconstructed from electron and
muon pairs. The transverse momentum of the lepton pair serves as the
reference value for that of the jet to be corrected. The resulting
distributions of the difference between the transverse momenta of
the boson and of the jet are peaked close to zero. The shapes of
the distributions are also similar.

\par
The energy scale for the recoil system, which is unclustered, is
investigated using the simulation and the transverse momentum of
the system from the event generator. With the default simulation
of the detector, the bias $P_{\rm T}^X - E_{\rm T}^{X{\rm (obs)}}$
increases approximately linearly with $P_{\rm T}^X$ in
the region above 30~GeV/$c$. Below 30~GeV/$c$, the detector
response to the particles of the recoil system is nonlinear.
Increasing the energy scale by a factor of 1.175 yields a
$\mathcal{O}(1)$ GeV bias, which is approximately constant to about
$\pm 15$\% for the region above 30~GeV/$c$.

\par
The energy-scale result of the recoil-system analysis is
similar in characteristics and values to the jet-energy-scale
calibration result of Ref.~\cite{refCdfJES} for clustered jets.
The recoil system in data with electron pairs from
$\gamma^*/Z$-boson decays provides a test of the scale factor.
The recoil-system bias of electron-pair events is defined
as $P_{\rm T}^{ee} - E_{\rm T}^{X{\rm (obs)}}$,
where $P_{\rm T}^{ee}$ is the transverse momentum of the pair.
As in the simulation, applying a recoil-energy scale
of 1.175 also yields a bias that is approximately
constant for the region above 30~GeV/$c$.

\subsection{\label{HadronCalib}
Hadronic calibrations}

Calibrations associated with the recoil system of hadrons in the
detector are complex because the hadrons are spread across
a large region of the calorimeter, the calorimeter response is
nonlinear, and there are sizeable regions with cracks in the
calorimeter coverage where the response of the simulation is
inadequate~\cite{refCdfJES}. Position-dependent jet corrections
are not applied by default in the missing-$E_{\rm T}$ calculation.

\par
The calibration strategy is to first fix the energy scale of the
recoil system in the data to 1.175, then adjust the corresponding
energy scale in the simulation so that the response matches
that of the data for events with large-$E_{\rm T}^{X{\rm (obs)}}$
values. For the remainder of the calibration, events are
partitioned into the standard-calibration groups. The offsets
between the simulation and data for the $x$ and $y$ components
of the recoil-$E_{\rm T}$ vector and the shapes of the
recoil-$E_{\rm T}$ distributions are corrected.

\par
To determine the energy-scale correction factor of the simulation
relative to
the data, events with $E_{\rm T}^{X{\rm (obs)}} > 30$~GeV are
selected for both the simulation and the data. With this selection,
a scale change alters the profiles of all simulated $E_{\rm T}$
distributions. The simulation scale is expressed as the product of
the data scale and a variable relative scale that is
adjusted using events from the away region of
$\cos(\Delta\phi_{eX})$, where the electron and recoil system are
approximately opposite in azimuth. As the energies of the electron
and recoil-system of hadrons are expected to balance, a scale
misalignment appears as an energy offset between the
electron-$E_{\rm T}$ distributions of the data and the simulation.
These distributions are similar to the away-region distribution of
Fig.~\ref{fig_selEtMXP}, but peaked near 55~GeV.
In order to minimize contamination from multiple interactions,
only the events with $n_{\rm vtx} = 1$ are used in the adjustments.
For the low- and high-luminosity periods of the central-region events,
the relative-scale values after the alignments are $0.959 \pm 0.006$
and $0.958 \pm 0.006$, respectively. For the plug-region events, they
are $0.943 \pm 0.007$ and $0.930 \pm 0.007$, respectively.
These values are used on all $n_{\rm vtx}$ categories.

\par
The distributions of the $x$ and $y$ components of the recoil-$E_{\rm T}$
vector are centered near the origin, and their offsets from the origin
are 1~GeV or less in magnitude, with magnitudes typically increasing with
$n_{\rm vtx}$. Differences between the offsets of the data and the
simulation are measured and applied as event-by-event corrections in the
simulation. Next, the recoil-$E_{\rm T}$ distributions of the simulation
are adjusted to match those of the data using event weights that preserve
normalizations. After these adjustments,
data-to-simulation differences in the
$\cos(\Delta \phi_{eX})$ distributions remain. They vary with
$\cos(\Delta \phi_{eX})$, and do not exceed $\pm10$\% for the distribution
with the largest difference.
The shapes of the simulated-$\cos(\Delta \phi_{eX})$ distributions are
adjusted using event weights. These adjustments modestly improve the
agreements between the data and simulation in the recoil-$E_{\rm T}$
distributions, but have a large impact on the $E\!\!\!/_{\rm T}$
distributions in conjunction with their effects on the
electron-$E_{\rm T}$ distributions.

\par
\begin{figure}
\includegraphics
   [width=85mm]
   {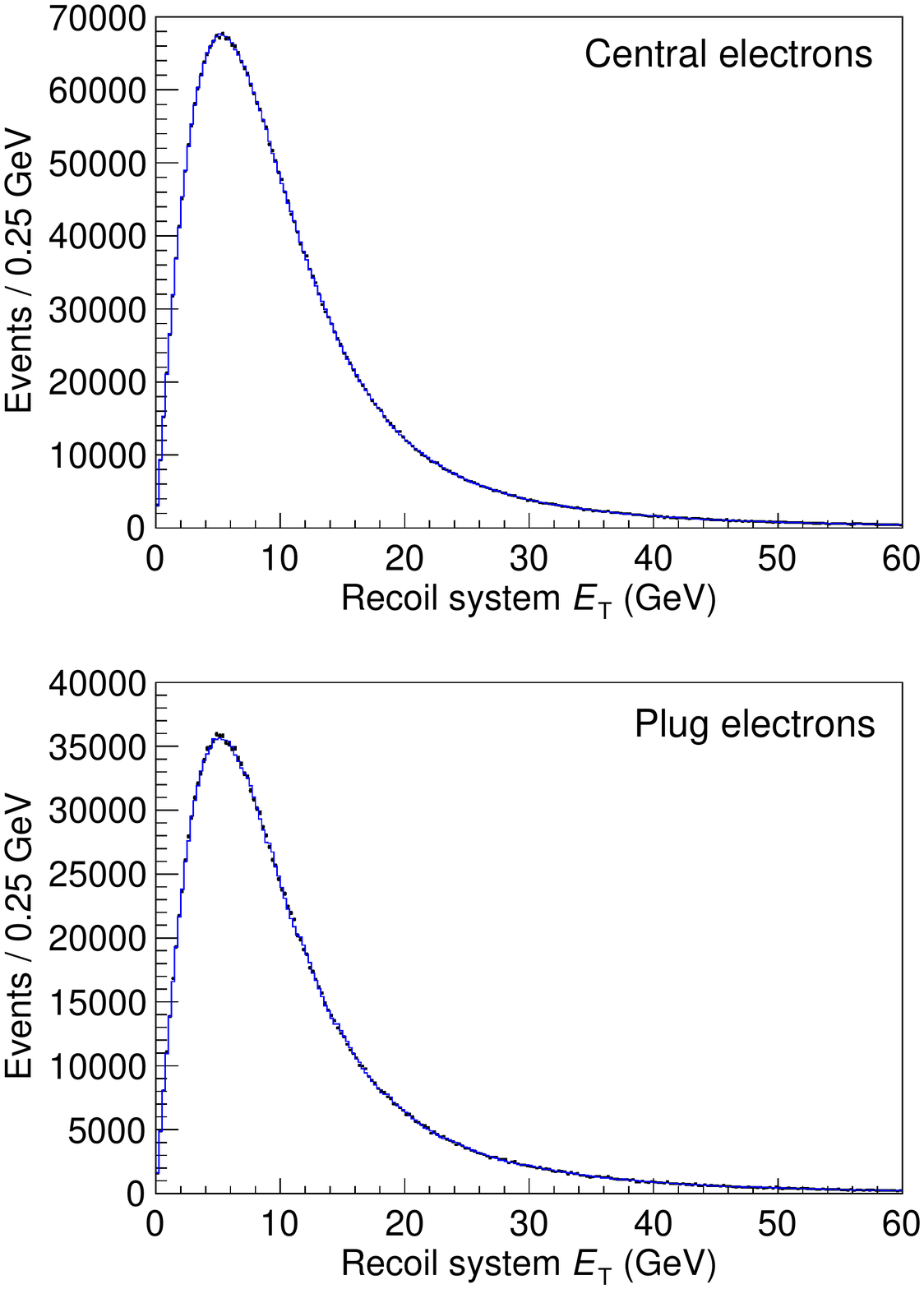}
\caption{\label{fig_cpJtcEt2}
Background-subtracted
$E_{\rm T}^{X{\rm (obs)}}$ distributions for electron candidates
in the central and plug regions. The crosses are the data and the
solid histogram is the simulation. The comparison between the data
and the simulation for electron candidates in the central (plug)
region yields a $\chi^2$ of 256 (402) per 240 bins.
}
\end{figure}

Figure~\ref{fig_cpJtcEt2} shows the
$E_{\rm T}^{X{\rm (obs)}}$ distribution of the recoil system for
electrons in the central and plug regions
after all corrections, including those for the missing-$E_{\rm T}$
of Sec.~\ref{METcorr} and the subtraction of backgrounds described
in Sec.~\ref{ENuBackgrounds}, are applied. The simulated
distributions are expected to similar to those of the data.

\subsection{\label{METcorr}\boldmath 
Missing-$E_{\rm T}$ corrections \unboldmath}

The corrected-$E\!\!\!/_{\rm T}$ vector of an event is a composite
object obtained from the calibrated-$E_{\rm T}^{X{\rm (obs)}}$
vector by incorporating the
contribution of the calibrated-$E_{\rm T}$ vector of the electron.
Corrections to the electron energy and to the distribution of the
electron relative to the recoil system of hadrons tend to have a
significant impact on the missing-$E_{\rm T}$ of events. The
impact is large because the $E_{\rm T}$ of the recoil system of
hadrons is typically smaller than that of the electron and thus
the electron shower is a dominant component of all energy
deposited in the calorimeters.

\par
Plug-region events of the high-luminosity period show small
missing-$E_{\rm T}$ differences between data and simulation at
values greater than 65~GeV. The simulated efficiency of the
underlying electron in this region is slightly
lower than in the data as the default normalization and
efficiency are from optimizations over all events and are not
specific to the high-$E_{\rm T}$ conditions. Adjustments are
applied to the simulated-electron efficiency of the events to
mitigate the differences. The integral of the correction amounts
to under 0.1\% of all events.

\par
\begin{figure}
\includegraphics
   [width=85mm]
   {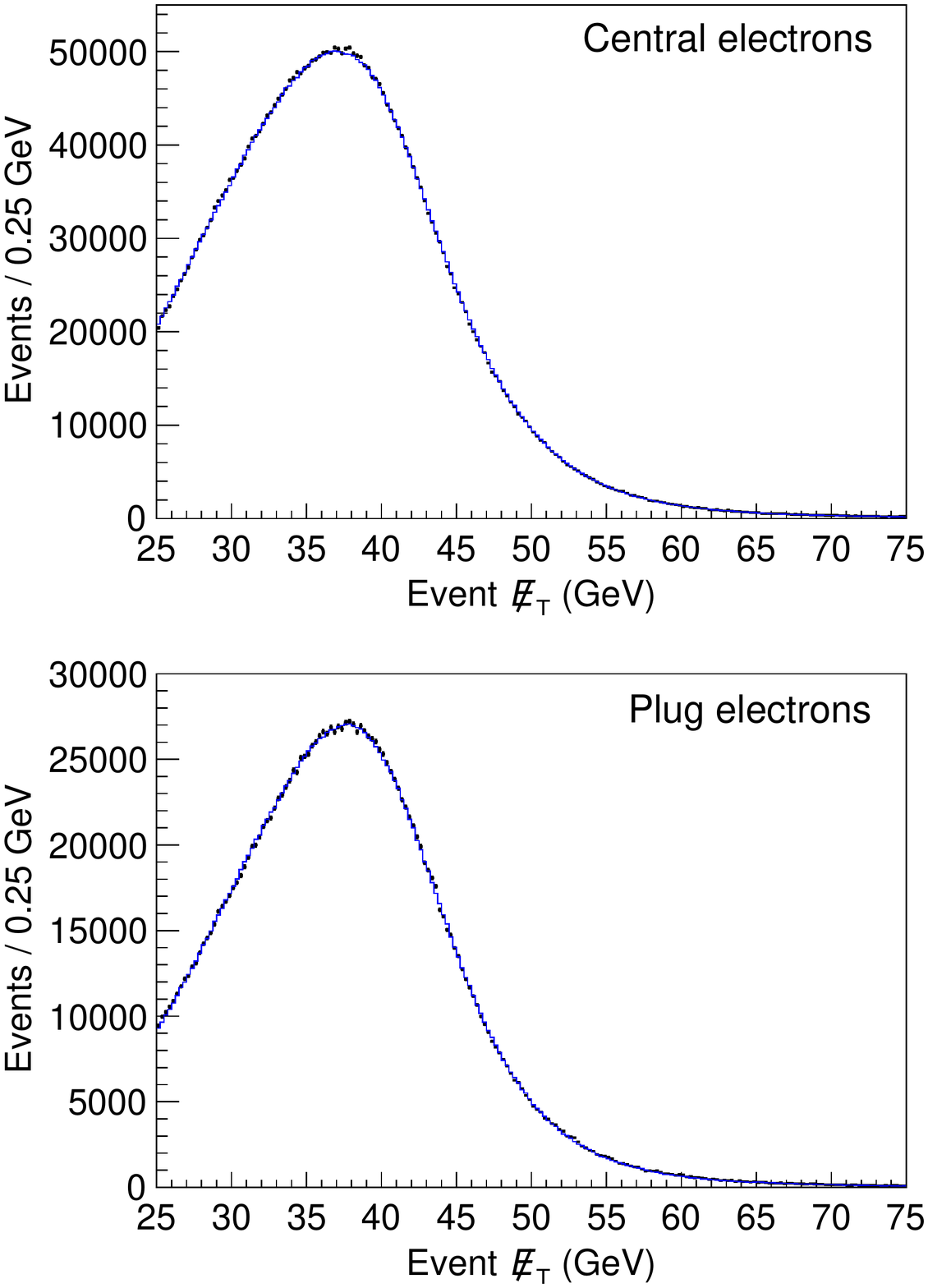}
\caption{\label{fig_cpEvcMET2}
Background-subtracted
$E\!\!\!/_{\rm T}$ distributions for electron candidates in the
central and plug regions. The crosses are the data and the solid
histogram is the simulation. The comparison between the data
and the simulation for electron candidates in the central (plug)
region yields a $\chi^2$ of 260 (310) per 200 bins.
}
\end{figure}

Figure~\ref{fig_cpEvcMET2} shows the
$E\!\!\!/_{\rm T}$ distributions for electrons in the central and
plug regions after all adjustments are applied to the underlying
quantities.
The subtraction of backgrounds discussed in Sec.~\ref{ENuBackgrounds}
is also applied.

\par
\begin{figure}
\includegraphics
   [width=85mm]
   {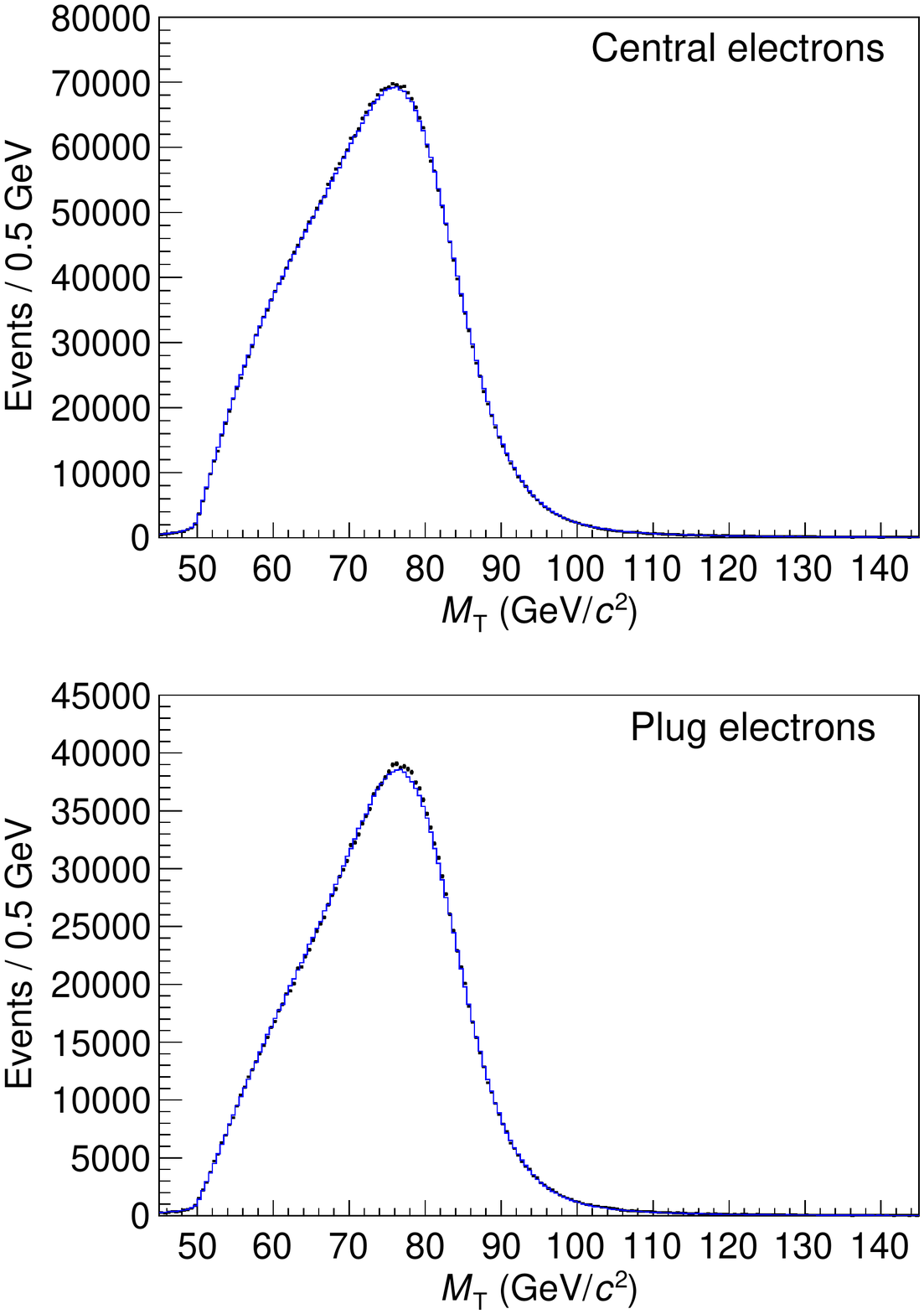}
\caption{\label{fig_cpmenuT2}
Background-subtracted
$M_{\rm T}$ distributions for electron candidates in the
central and plug regions. The crosses are the data and the solid
histogram is the simulation. The comparison between the data
and the simulation for electron candidates in the central (plug)
region yields a $\chi^2$ of 384 (402) per 200 bins.
}
\end{figure}

Figure~\ref{fig_cpmenuT2} shows the transverse mass ($M_{\rm T}$)
distributions for electrons in the central and
plug regions after all adjustments are applied to the underlying
quantities,
including the subtraction of backgrounds discussed in
Sec.~\ref{ENuBackgrounds} . The $M_{\rm T}$ distribution reflects
the azimuthal angular distribution between the electron and
missing-$E_{\rm T}$ quantities of events.

\subsection{\label{ENuBackgrounds}
Backgrounds}

Backgrounds
from the central and plug regions are determined separately.
Backgrounds from the processes
$\gamma^*/Z \rightarrow ee$, $W \rightarrow \tau\nu$, 
$\gamma^*/Z \rightarrow \tau\tau$, dibosons
(\textit{WW}, \textit{WZ}, \textit{ZZ}), $t\bar{t}$ pairs, and
QCD multijets are considered.

\par
Events produced by the $\gamma^*/Z \rightarrow ee$ process can
occasionally have significant amounts of missing-$E_{\rm T}$, similar
to QCD events. Electron showers within uninstrumented
portions of the detector can result in significant amounts of
missing $E_{\rm T}$ being indicated. However, the number of such
events is relatively
small compared to that from the $W \rightarrow e\nu$ process.

\par
The QCD-background sample is derived from the data. Other
backgrounds are derived from \textsc{pythia}~\cite{Pythia621} samples
that are processed with the detector simulation and in which the
integrated luminosity of each sample matches the data.
The diboson and $t\bar{t}$ samples are inclusive and their
normalizations use total cross sections calculated at
NLO~\cite{MCFM345} and
next-to-next-to-leading order~\cite{ttbarNNLO}, respectively.
The $W \rightarrow \tau\nu$ and $Z \rightarrow \tau\tau$ sample
normalizations use the total cross sections from \textsc{pythia}
multiplied by the 1.4 ratio of the NLO-to-LO cross sections.
The $\gamma^*/Z \rightarrow ee$ sample is the signal sample of
the $ee$-pair analysis~\cite{cdfAfb9eeprd,*cdfAfb9eeprdErr}, 
and data-constrained normalizations derived therein are utilized.
Sample normalizations as mentioned above are referred to as the
default normalizations. All normalizations are implemented as
event weights. This allows background events to be subtracted
from (added to) event distributions via the use of negative
(positive) weights.

\par
Candidates of the QCD sample are a subset of the events that
fail the event-selection criteria. Events in this sample fail
the $E_{\rm HAD}/E_{\rm EM}$ criterion, but satisfy all other
electron-identification criteria except the isolation criterion.
This subset definition enhances the fraction of QCD events, and,
limits the events to those whose kinematic distributions are
closer to those of the QCD events within the signal sample due
to the similarity of the selections.
For the background in plug-region
events, the additional requirement on the transverse-shower shape
described in Appendix~\ref{PlugEleSel} is removed because it
severely limits the size of the sample.

\par
The QCD sample includes events from non-QCD processes with
high-$E_{\rm T}$ electrons. These events are modeled
using the same \textsc{pythia} samples of the
$W \rightarrow e\nu$ and background processes. However,
events are required to pass the selection criteria for the QCD
sample. Most of the events are from the $W \rightarrow e\nu$
process.

\par
\begin{figure}
\includegraphics
   [width=85mm]
   {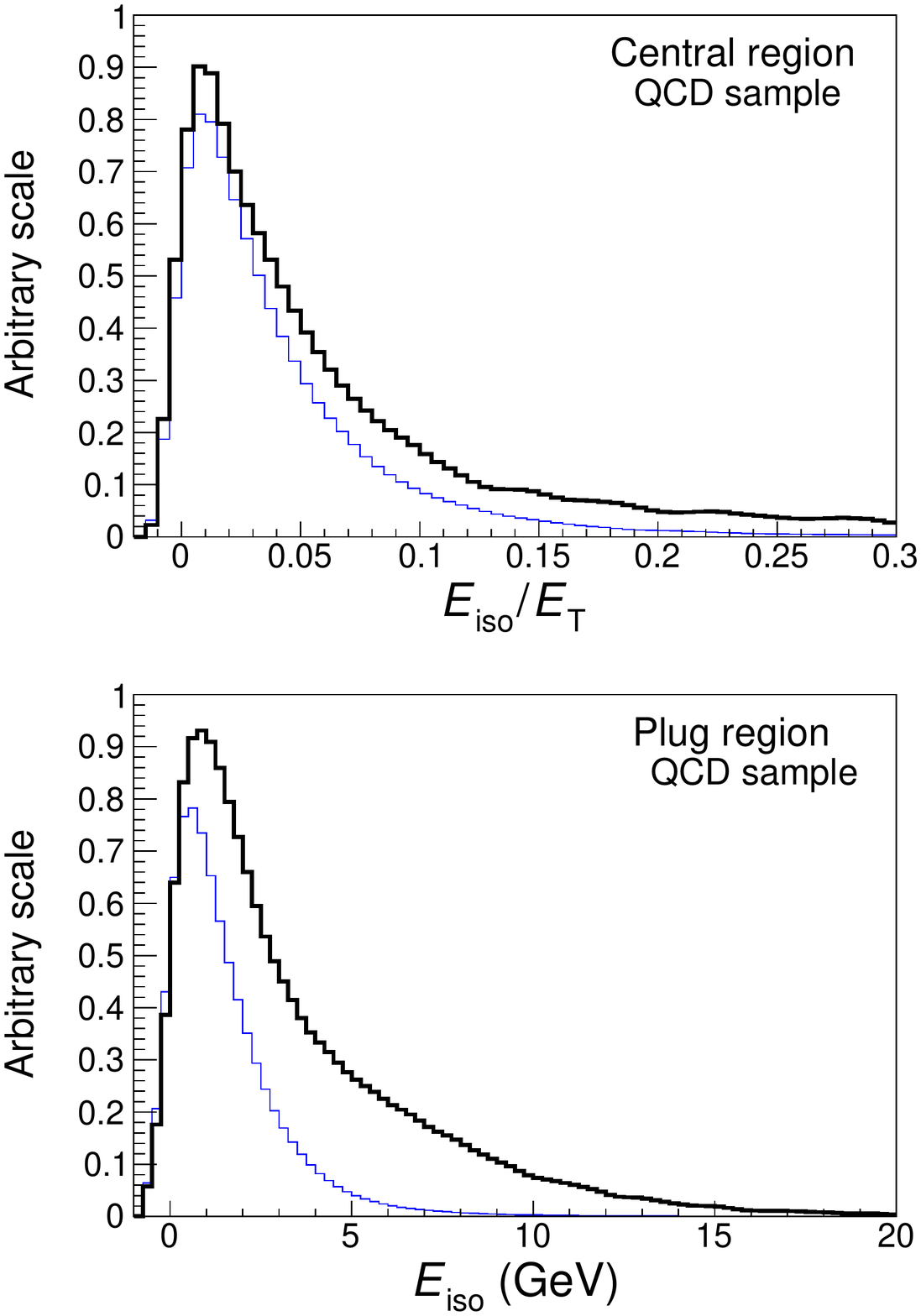}
\caption{\label{fig_cpEisoAH2}
Shape of the isolation distributions of events
in the central and plug regions of the QCD sample. The bold (black)
histogram is the data and the lighter (blue) histogram is
the expected contribution of the non-QCD processes.
}
\end{figure}

Figure \ref{fig_cpEisoAH2} illustrates
the shape of the isolation distributions from the QCD sample
for events of the central and plug regions.
Also shown is the expected contribution of the non-QCD
component, normalized relative to the QCD component.
To normalize the non-QCD component of the sample to
the QCD component, events are split into two disjoint sets based
on the isolation energy of the electron candidate. These
collections are denoted as the tight- and loose-isolation sets.
Events with central-region candidates having
$E_{\rm iso}/E_{\rm T} < 0.05$ and events with plug-region
candidates having $E_{\rm iso} < 2$~GeV are assigned to the
tight-isolation set. Unassigned events form the loose-isolation set.

\par
The sole purpose of the tight-isolation set is to provide the
normalizations of the non-QCD processes within the QCD sample.
The QCD events of the loose-isolation set are used as the model
for the QCD background within the signal sample. Normalizations
of the various components in the QCD samples are determined for
each standard-calibration group.

\par
Normalizations for the non-QCD processes are derived using the
$E\!\!\!/_{\rm T}$ distribution of events. Events from QCD processes
are concentrated in the region with $E\!\!\!/_{\rm T}$ smaller than
approximately 35~GeV. At higher values, $e\nu$ events from the
production of $W$ bosons dominate while the contribution from QCD
production is small. A single scale applied to the default
normalizations of the non-QCD processes is adjusted so that the
simulated distribution in the region above 35~GeV is in better
agreement with that of the data. As the loose-isolation set includes
events from non-QCD processes, their contribution is subtracted to
yield the model of the QCD background within the signal sample.

\par
All events of the model are used for the subtraction of the QCD
background from the signal sample, \mbox{i.e.}, the isolation
requirement of the signal sample is not applied. Normalizations for
the background are derived using the $E\!\!\!/_{\rm T}$ distributions
of signal events for the data and simulation. The event yields of the
simulation and QCD background are adjusted in a two-parameter fit so
that their sum matches that of the data.

\par
Backgrounds from QCD processes are larger in the plug region and
suffer from insufficiently accurate predictions.
To improve the agreement between
the observations and the predictions, events with electron candidates
in the plug region are subdivided further into smaller groups based on
the additional selection criteria for electrons described in the
Appendix, and the levels of the QCD background therein determined.

\par
\begin{figure}
\includegraphics
   [width=80mm]		
   {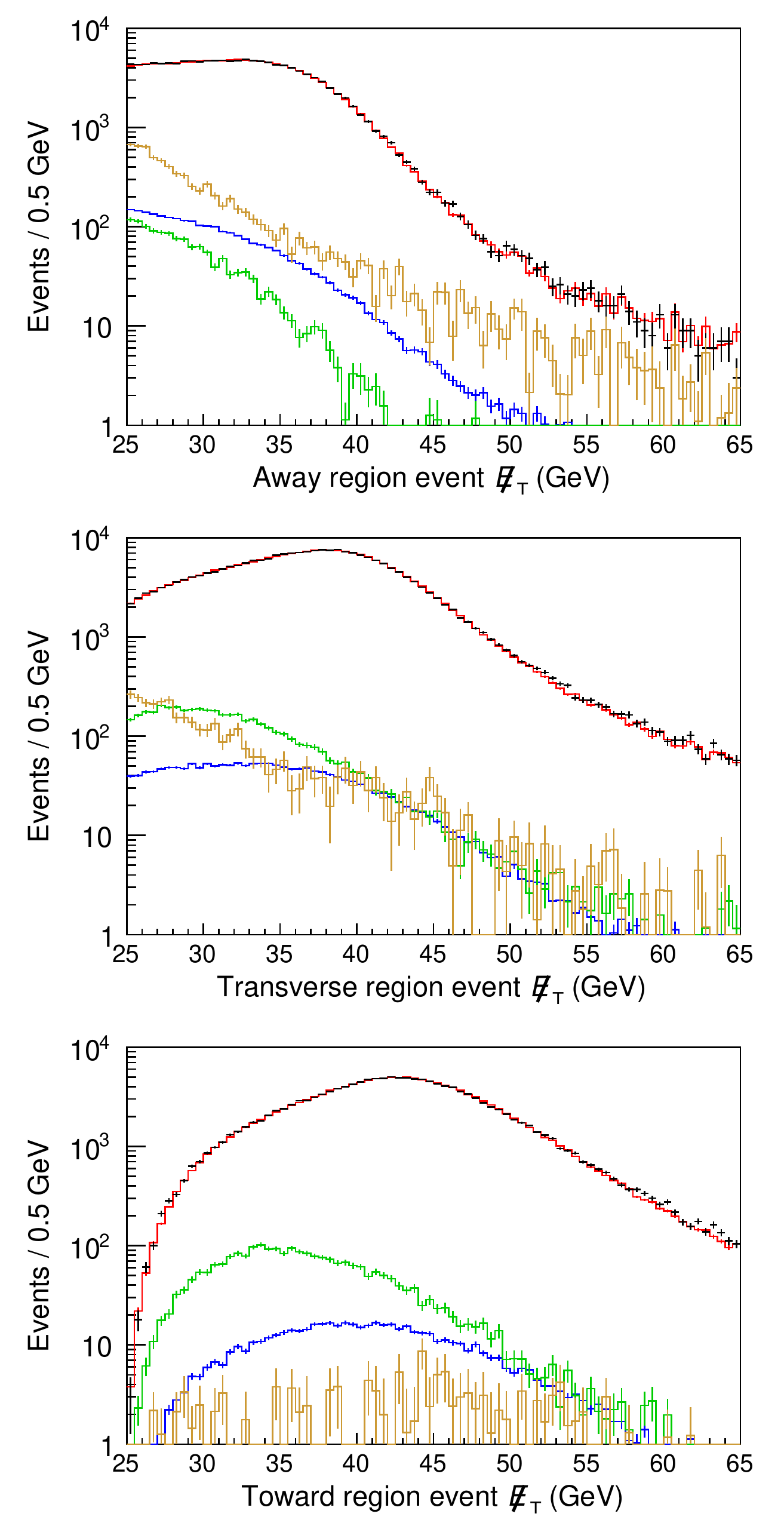}
\caption{\label{fig_plgEvcMET1113}
$E\!\!\!/_{\rm T}$ distributions for electron candidates in the
plug calorimeter from the high-luminosity period with
$n_{\rm vtx}=2$ in the $\cos(\Delta\phi_{eX})$ regions.
The (black) crosses are the data, the (red) histogram
overlapping the data is the simulation with all backgrounds,
and the lower histograms are the backgrounds for QCD (brown),
$\gamma^*/Z \rightarrow ee$ (blue), and
$W \rightarrow \tau\nu$ (green).
For the away region, the backgrounds from top to bottom are
QCD, $\gamma^*/Z \rightarrow ee$, and $W \rightarrow \tau\nu$.
For the toward region, the order is reversed.
For the transverse region, the backgrounds from top to
bottom along the $y$ axis are QCD, $W \rightarrow \tau\nu$,
and $\gamma^*/Z \rightarrow ee$.
}
\end{figure}

Examples of the QCD backgrounds in the missing-$E_{\rm T}$ distributions
of plug-region electrons from the high-luminosity period and $n_{\rm vtx}=2$
for the away, transverse, and toward regions of $\cos(\Delta\phi_{eX})$
is shown in Fig.~\ref{fig_plgEvcMET1113}, where only the
larger backgrounds are shown to reduce the overlap of histograms.
Distributions for electron candidates in the central calorimeter
are similar.

\par
The toward-region distribution is suppressed at low
missing-$E_{\rm T}$ values because most of its events have geometries
where the missing-$E_{\rm T}$ vectors are in opposite directions
relative to those of the electron and recoil systems. As the QCD
background in this region is small, the data inputs to the fit do not
constrain the QCD normalization. Consequently, the normalization is
fixed to a value determined from an extrapolation that uses the
$\cos(\Delta\phi_{eX})$ distribution of the QCD background, which
decreases exponentially as the value of $\cos(\Delta\phi_{eX})$
increases. All toward-region distributions are similar and treated
the same way.

\par
Plug-region events of the high-luminosity period have small
differences between data and simulation in the electron-$E_{\rm T}$
distribution for $E_{\rm T}>65$~GeV. In this
region, the predicted amount of QCD background is large, exceeding
the signal at $E_{\rm T}>80$~GeV. Adjustments
are applied to the QCD-background shape to mitigate the differences.
The integral of the correction amounts to 0.1\% of all events.

\par
\begin{figure}
\includegraphics
   [width=85mm]
   {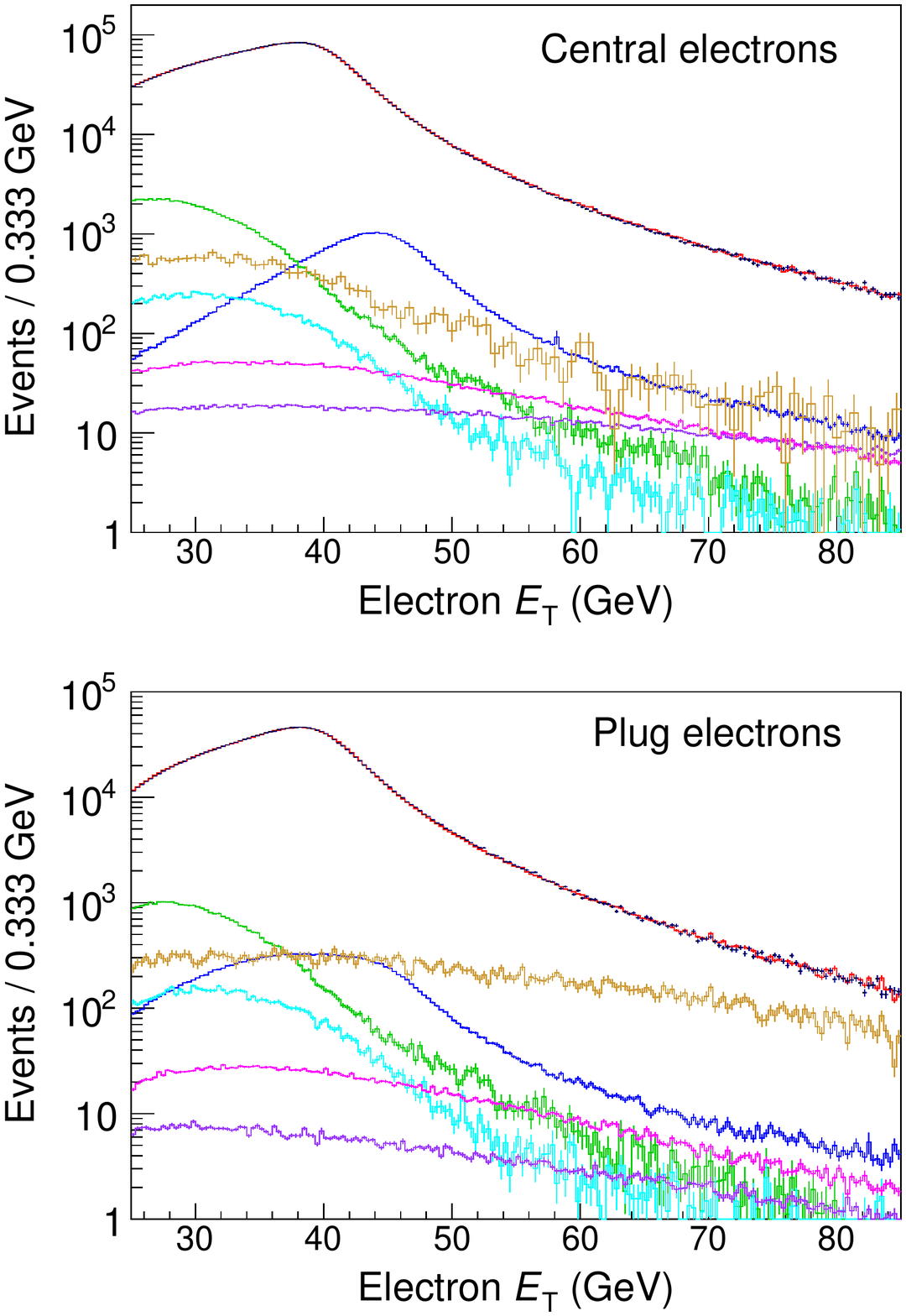}
\caption{\label{fig_cpEleETbkg2}
$E_{\rm T}$ distributions of electrons in the central and plug
regions. The data
are the crosses (black) and the simulation with all backgrounds
is the histogram (red) overlapping the data. The
individual backgrounds are the lower histograms, and from the
top to bottom along the left edge of the plot they are the
$W \rightarrow \tau\nu$ (green), QCD (brown),
$\gamma^*/Z \rightarrow \tau^+\tau^-$ (cyan),
$\gamma^*/Z \rightarrow ee$ (blue), diboson (magenta),
and $t\bar{t}$ (purple) contributions. For plug electrons,
$\gamma^*/Z \rightarrow ee$ (blue) peaks around 40~GeV.
}
\end{figure}

The data samples consist of approximately
\mbox{$3 \: 819 \: 000$} events for the central region and
\mbox{$2 \: 003 \: 000$} events for the plug region.
Table~\ref{tblBkgrFrac} lists the background composition.
\begin{table}
\caption{\label{tblBkgrFrac}
Background composition. Relative uncertainties
of the QCD backgrounds are 12\% for the central-region electrons
and 6\% for the plug-region electrons. All simulated backgrounds
have a 6\% relative uncertainty associated with the integrated
luminosity~\cite{cdfR2CLC} with the exception of the
$\gamma^*/Z \rightarrow ee$ background, which is well constrained by
the electron-pair data~\cite{cdfAfb9eeprd,*cdfAfb9eeprdErr}.
}
\begin{ruledtabular}
\begin{tabular}{cccc}
Component     & \multicolumn{2}{c}{\text{Background fraction (\%)}}  \\
              & \multicolumn{1}{c}{\text{Central region}} &
	        \multicolumn{1}{c}{\text{Plug region}}         \\ \hline
$W\rightarrow\tau\nu$              &   1.78      &    1.62    \\
QCD                                &   0.91      &    1.98    \\
$\gamma^*/Z\rightarrow ee$         &   1.09      &    0.96    \\
$\gamma^*/Z\rightarrow \tau\tau$   &   0.29      &    0.35    \\
Diboson                            &   0.14      &    0.13    \\
$t\bar{t}$                         &   0.08      &    0.04    \\
\end{tabular}
\end{ruledtabular}
\end{table}
The fully corrected electron-$E_{\rm T}$ distributions, including the
individual contributions from the various background processes,
is shown in Fig.~\ref{fig_cpEleETbkg2} for events in the central
and plug regions.

\subsection{\label{ChargeMisID1}
Charge-misidentification rates}

The rates of charge misidentification for central-region tracks are
small and are due to interactions of the electrons with the material in
the tracking volume. For tracks associated with plug electrons,
the misidentification rates are significant, and increase with
$|\eta_{\rm det}|$. No charge bias is detected in the track
reconstruction. The charge bias and rates of charge misidentification
are studied using $e^+e^-$ pairs from $\gamma^*/Z$-boson production.
The event selection follows the $ee$-pair analysis except that the
track requirements of the $e\nu$-pair analysis are applied.

\par
For the study of the charge bias of plug-region electrons, the
dielectron masses are limited to the range 66 to 116~GeV/$c^2$.
Pairs with one electron in the central region and the other in the
plug region are used to determine the charge bias for plug electrons.
The central-region electron provides the reference charge for the
measurement. A positive charge is assigned to the plug electron if
the central-electron charge is negative, and vice versa. The bias
is measured using the asymmetry $A_{\rm b} = (N^+ - N^-)/(N^+ + N^-)$,
where $N^\pm$ is the number of plug electrons with $\pm$ charges.
As a function of $|\eta_{\rm det}|$, the asymmetry of plug electrons
is consistent with zero and integrates to $-0.001 \pm 0.002$.

\par
For the measurement of the rates of charge misidentification, the
dielectron masses are limited to exceed 40~GeV/$c^2$. For pairs where
both electrons are in the central region, the electron with the
largest $E_{\rm T}$ provides the reference charge for the measurement
of the misidentification rate of the opposing electron. For pairs where
one electron is in the central region and the other is in the plug
region, the central-region electron provides the reference charge
for the measurement of the misidentification rate of the plug-region
electron.

\par
Rates of charge misidentification are measured on a $44 \times 8$
$(\eta_{\rm det}, \phi)$ grid which reflects the transverse
segmentation of the calorimeter towers.
Subdivisions along the $\eta_{\rm det}$ direction
correspond to the 44 azimuthal rings of towers, numbered from
0 to 43. The low (high) edge of ring 0 (43), which is adjacent to
the beam line, is located at $\eta_{\rm det} = -3.5$ (3.5);
however, these rings are not in the fiducial region.
Subdivisions along the $\phi$ direction correspond to a $45^\circ$
section of adjacent towers in a ring. Each subdivision in $\phi$
is denoted as a sector.

\par
The sector subdivisions along the $\phi$ direction match the
underlying wedge structure of the PES detector, which provides
the exit point for track finding in the plug region.
Each PES wedge is aligned as a single
unit with the track detectors of the central region.

\par
\begin{figure}
\includegraphics
   [width=85mm]
   {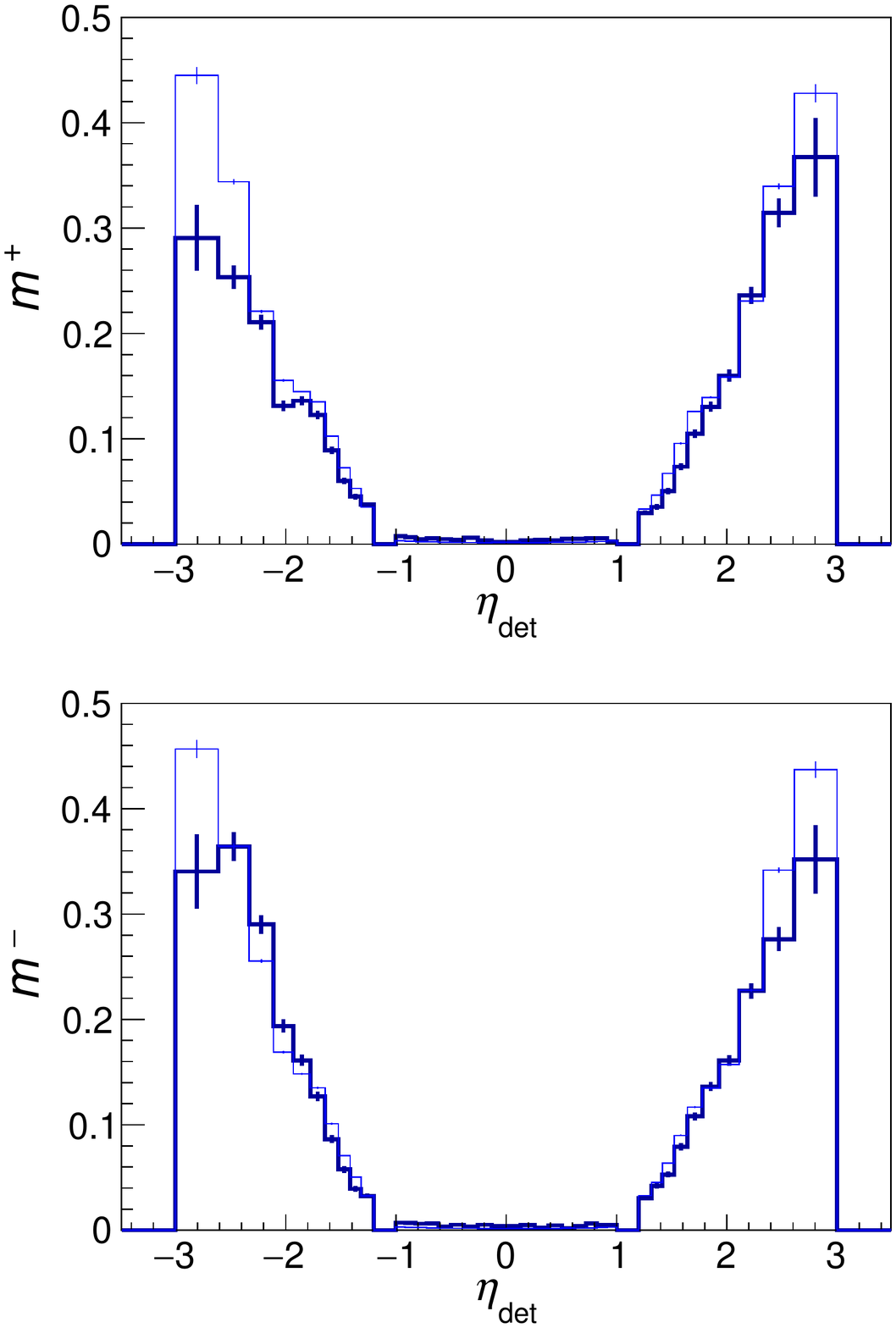}
\caption{\label{fig_zeeMIDe2}
Charge-misidentification rates $m^\pm$ averaged over the sectors
of a ring, versus $\eta_{\rm det}$.
The bold (black) histogram is for the data, and the lighter (blue)
histogram for the simulation.
}
\end{figure}

The rate of misidentification, $m^\pm$, is the fraction of
observed (simulated) particles with expected (known) charge $\pm$
reconstructed with the wrong charge.
Figure~\ref{fig_zeeMIDe2} illustrates the
average misidentification rates of electrons and positrons,
where the tower rings are drawn at their locations in
$\eta_{\rm det}$ space. Zones with null rates are not in the
fiducial region. In the plug region $(|\eta_{\rm det}| > 1.1)$,
rate variations among the $\phi$-sectors of a ring are
significant for both the data and simulation. The data rates
include the effects from discrepancies between the true orientation
of each PES wedge and that specified by the alignments. For the
simulation, the alignments are exact.

\par
For the individual sectors of ring 1 (42) towers near the beam line,
the numbers of events available for the rate measurements range from
tens down to a handful. The quality of the measurements is
inadequate. Consequently, a rate based on the combined events
of rings 1 and 2 (42 and 41) within the same PES wedge is used.
The rates of these rings are correlated because they
are adjacent in the same strip detector. However, there
are $\eta_{\rm det}$ dependencies as shown in Fig.~\ref{fig_zeeMIDe2}.
To account for these dependencies, the
combined rate of rings 1 and 2 (42 and 41) of a PES wedge is scaled
by the ratio of the integrated rate over all sectors of ring 1 (42)
to that of the combined rate integrated over all sectors.
The scaled results are consistent with the original ring 1 (42) rates.

\subsection{\label{ChargeMisID2}
Charge-misidentification corrections}

The numbers of events with correctly and incorrectly reconstructed
charges for the simulation are given by $N_{\rm t}^\pm (1-m^\pm)$ and
$N_{\rm t}^\pm m^\pm$, respectively, where $N_{\rm t}^\pm$ is the
number of events with the truth-level charge specified in the
superscript. Charge-misidentification corrections based on the
measured rates from $ee$ pairs are applied as event weights to the
simulated events.

\par
For $ee$ pairs, the weight for events with incorrectly reconstructed
charges is the data-to-simulation ratio
$m_{ee \text{-} d}/m_{ee \text{-} s}$, where the $\pm$
superscripts for the truth-level charge are suppressed for clarity,
and the $m_{ee \text{-} d}$ and $m_{ee \text{-} s}$ symbols are the
misidentification rates observed in the data and simulation,
respectively. For $e\nu$ pairs, only $m_{e\nu \text{-} s}$ is
measured, and it differs from $m_{ee \text{-} s}$.
In addition to the alignment of the PES wedge, there is
an effect from the differences in the $P_{\rm T}$ distributions of
the electrons (positrons) from $\gamma^*/Z$- and $W$-boson decays.

\par
The corrected misidentification rate for $e\nu$ pairs,
$m_{e\nu \text{-} s}^\prime$, is given by the product
$m_{ee \text{-} d} \: (m_{e\nu \text{-} s}/m_{ee \text{-} s})$.
The leading term is the measured rate from $ee$ pairs,
and the following term in parentheses is a relative correction
that accounts for the additional effect of electron-$P_{\rm T}$
differences between $W$- and $\gamma^*/Z$-boson decays.
The event-weight correction for charge-misidentified events is
$m_{e\nu \text{-} s}^\prime / m_{e\nu \text{-} s}$, and
$(1 - m_{e\nu \text{-} s}^\prime) / (1 - m_{e\nu \text{-} s})$
for events with the correct charge.

\par
To confirm the efficacy of the charge-misidentification
corrections to the simulated rate of positrons and electrons
for the $W \rightarrow e\nu$ process, the following asymmetry
$A_\pm  = (n^+ - n^-) /(n^+ + n^-)$, where $n^{+(-)}$ is the number
of positrons (electrons) as a function of $\eta_{\rm det}$,
is used. All of the charge-independent corrections discussed in this
section are applied. Charge-misidentification corrections affect the
numerator difference, but leave the denominator sum unchanged.
Prior to the application of the charge-misidentification corrections,
the $A_\pm$ asymmetries of the simulation differed from those of
the data for $|\eta_{\rm det}| > 1.5$. Disagreements increased with
increasing values of $|\eta_{\rm det}|$ and the amount of disagreement
varied with the $\phi$-sector locations of the positrons and
electrons. The amounts of disagreement are large, significant, and
different for positive and negative $\eta_{\rm det}$ regions within a
$\phi$-sector. After the application of the charge-misidentification
corrections, the differences of the $\phi$-sector asymmetries of the
simulation relative to the data are significantly decreased.

\section{\label{AsymMeas}\boldmath
The $A_\ell$ Measurement}\unboldmath

Equation~(\ref{eqnAellpDef}) is the basis of the asymmetry measurement.
The corrections discussed
in Sec.~\ref{CorrDatSim} are incorporated into the evaluation of the
$N^+$, $N^-$, $(\epsilon A)^+$, and $(\epsilon A)^-$ quantities.
Using the simulation, the product of the efficiency and acceptance
is derived bin-by-bin with the formula
$(\epsilon A)^\pm = N_{\rm r}^\pm / N_{\rm g}^\pm$, where
$N_{\rm r}^\pm$ is the number of reconstructed and selected events
in a bin of the reconstructed pseudorapidity, and $N_{\rm g}^\pm$
is the number of accepted events at the event-generation level in the
corresponding bin of generated pseudorapidity.
In the determination of the generated-level acceptance, the kinematic
restrictions on the reconstructed quantities
$E_{\rm T}^e > 25$~GeV, $E_{\rm T}^\nu > 25$~GeV,
and $M_{\rm T} > 45$~GeV/$c^2$ are applied to the corresponding
generator-level quantities.

\par
\begin{figure}
\includegraphics
   [width=85mm]
   {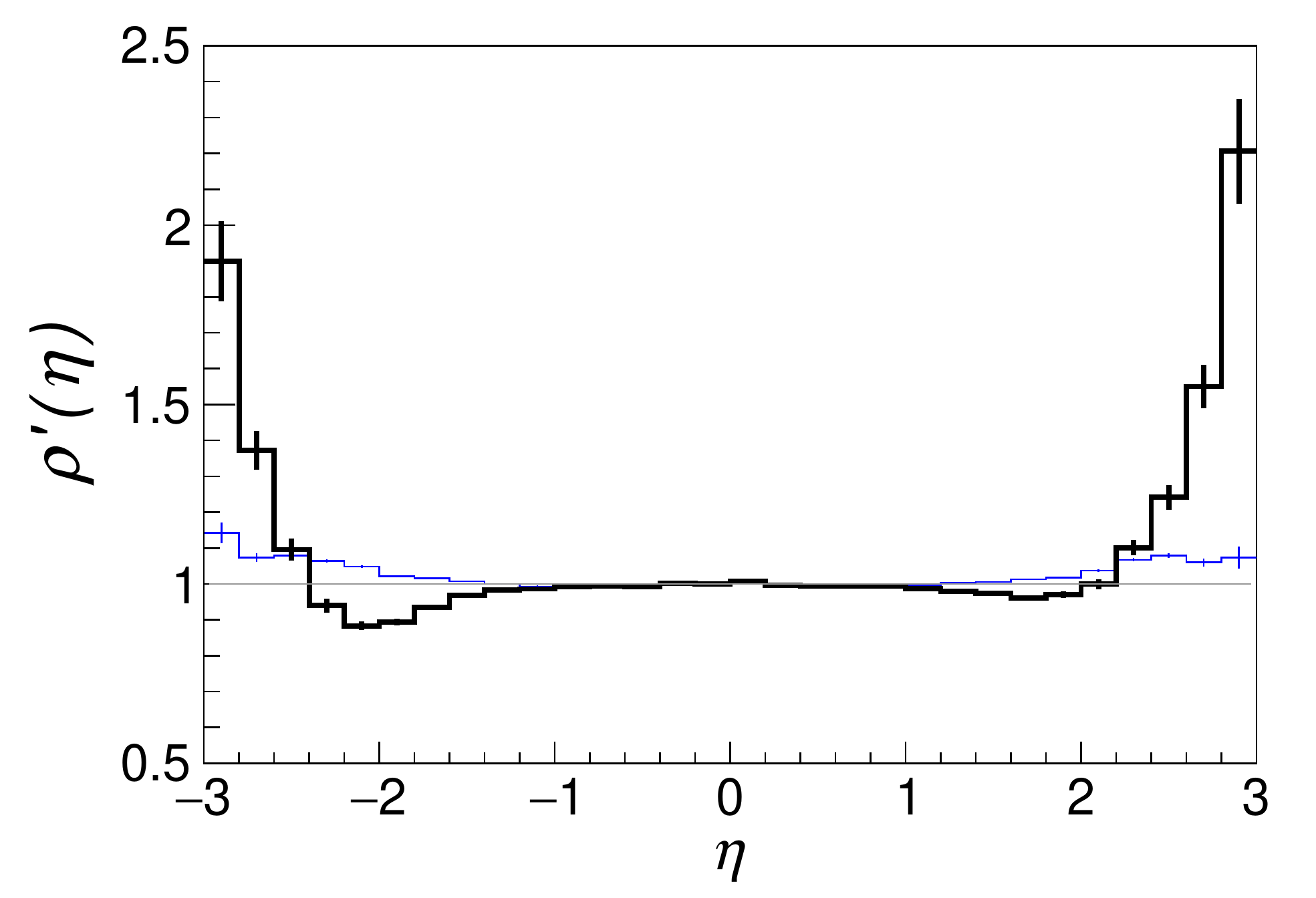}
\caption{\label{fig_reffAcc}
Dependence of $\rho^\prime(\eta)$ on $\eta$.
The bold (black) histogram includes the accounting of charge
misidentification while the lighter (blue) histogram does not.
The uncertainties shown are statistical and are evaluated
bin-by-bin. The horizontal (gray) line is a unit-value reference.
}
\end{figure}

Alternatively, the asymmetry can be measured using
$(N^+\! - \rho N^-)/(N^+\! + \rho N^-)$ or
$(N^+\!/\rho \,- N^-)/(N^+\!/\rho \,+ N^-)$, where
$\rho = (\epsilon A)^+\! / (\epsilon A)^-$.
Figure~\ref{fig_reffAcc} shows the function $\rho^\prime(\eta)$
defined as $1/\rho$ for the $\eta < 0$ region and
$\rho$ for the $\eta \ge 0$ region. 
The effect of the $\rho^\prime(\eta)$ correction on the measurement
can be gauged using $D(\eta) = 1-2m(\eta)$, where $D(\eta)$
approximates the ratio of the uncorrected to corrected
asymmetry, and $m(\eta)$ is the mean charge-misidentification rate
of both charges.

\par
As the $A_\ell$ distribution is an antisymmetric function of the
pseudorapidity, events from the $\eta \ge 0$ and
$\eta < 0$ regions provide independent measurements of the
asymmetry distribution. They are combined to improve the statistical
precision. Prior to the combination, the measurement over the
$\eta<0$ region is transformed via the ``CP-folding'' operation,
$A_\ell(\eta) \rightarrow -A_\ell(-\eta)$. The measurement over the
$\eta<0$ region, the measurement over the $\eta \ge 0$ region, and
the combined measurement are generically denoted by the symbol
$A_\ell(|\eta|)$.

\par
\begin{figure}
\includegraphics
   [width=85mm]
   {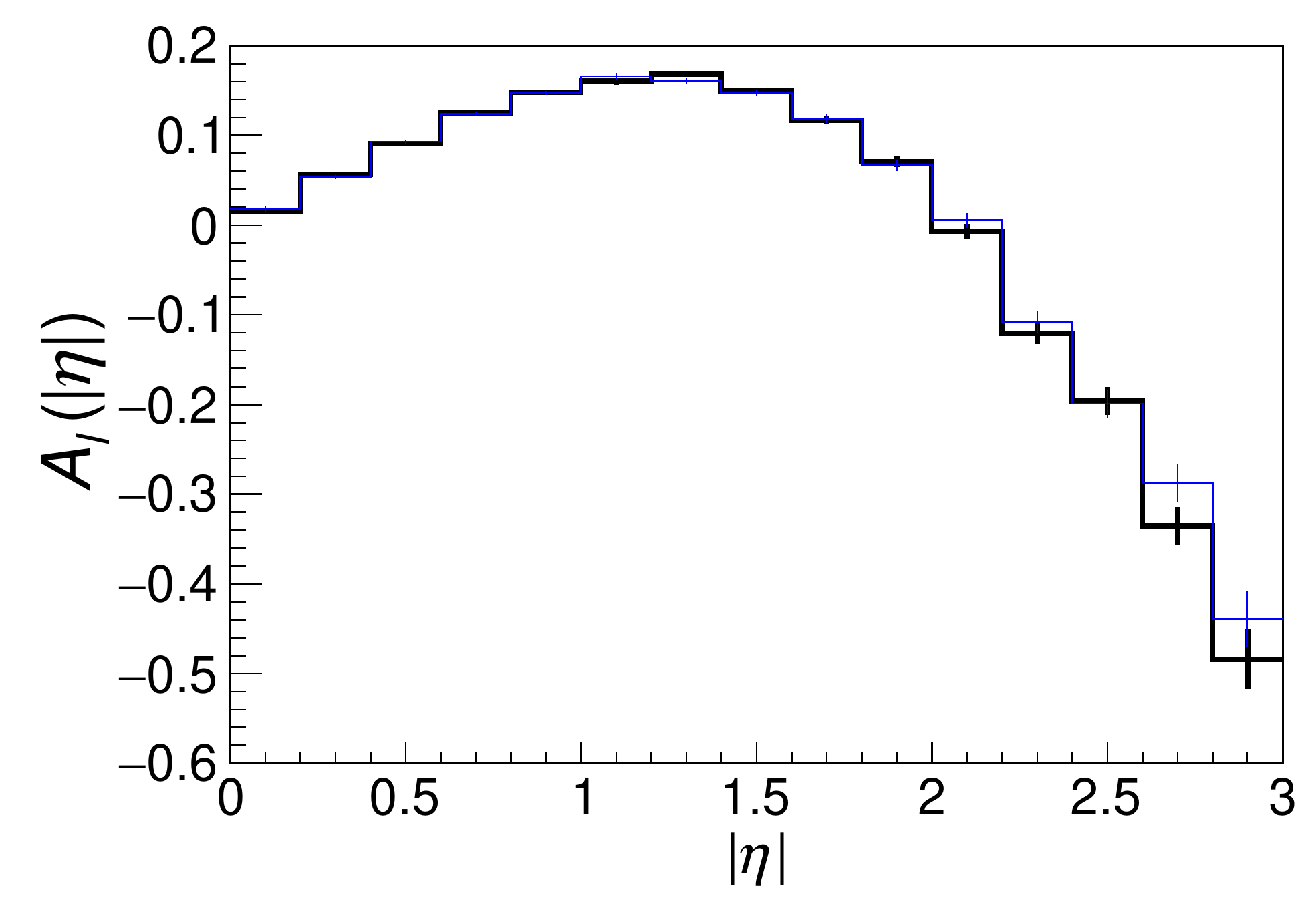}
\caption{\label{fig_epmEllAsym}
Observed lepton asymmetry $A_\ell(|\eta|)$ as a function of $|\eta|$.
The bold (black) histogram is the measurement in the $\eta \ge 0$
region, and the lighter (blue) histogram is in the $\eta < 0$
region. The measurements include all corrections, and the
uncertainties are statistical only, and evaluated bin-by-bin.
Bin-centering corrections are not a part of the measurements.
}
\end{figure}

The fully corrected measurements of the $A_\ell(|\eta|)$
from the $\eta \ge 0$ and $\eta < 0$ regions are shown in
Fig.~\ref{fig_epmEllAsym}.
The uncertainties shown in Fig.~\ref{fig_epmEllAsym} are the
diagonal elements of the covariance matrix for the data, which
is discussed in the next paragraph.
The $\chi^2$ comparison between the $A_\ell(|\eta|)$ measurements
of the $\eta \ge 0$ and $\eta < 0$ regions yields a value of
11 over the 15 bins. As the measurements are consistent, they
are combined into the CP-folded asymmetry.

\par
Measurement uncertainties require a covariance matrix since the
uncertainty of an $\eta$ bin $i$ and of another bin $j$ are
correlated from the measurement uncertainties of the
charge-misidentification rates in conjunction with the 30~cm
spread of the $p\bar{p}$-collision vertex along the beam line.
The charge-misidentification rate for the tracks traversing a
detector region $(\eta_{\rm det})$ affects the asymmetry
uncertainties of multiple $\eta$ bins because those tracks are
from electrons produced
over a wide range of $\eta$. The covariance matrix is
calculated using the simulation, which provides the distributions
for $N_{\rm r}^\pm$, $N_{\rm g}^\pm$, and $N_{\rm t}^\pm$.
Recall that $N_{\rm r}^\pm$ is the number of reconstructed and
selected events in a bin of the reconstructed pseudorapidity,
$N_{\rm g}^\pm$ is the number of accepted events at the
event-generation level in the corresponding bin of the
generated pseudorapidity, and $N_{\rm t}^\pm$ is the analog of
$N_{\rm r}^\pm$ but with the true charges of the generator level.

\par
In Eq.~(\ref{eqnAellpDef}), the $N_{\rm r}^\pm/(\epsilon A)^\pm$
terms are the corrected event counts $N_{\rm c}^\pm$ in a 
data $\eta$ bin. The $N_{\rm r}^\pm$ and
$(\epsilon A)^\pm$ components contain sums of events from
different detector regions.
Expressions for
the first-order fluctuations of $N_{\rm c}^\pm$ due to input
uncertainties are derived in terms of the fluctuations from
its component $N_{\rm r}^\pm$ and $(\epsilon A)^\pm$
sums. Fluctuation distributions for the
$N_{\rm r}^\pm$ terms are based on the statistical precision
of the data, while those for $(\epsilon A)^\pm$ are based on
that of the simulation. For both terms, the uncertainties of
their $N_{\rm r}^\pm$ values are estimated using
$N_{\rm r}^\pm = N_{\rm t}^\pm (1-m^\pm) + N_{\rm t}^\mp m^\mp$.
Charge-misidentification related uncertainties consist of
two components, those from the binomial distribution among the
number of events with correctly and incorrectly reconstructed
charges, and those from the measured values of $m^\pm$.
Uncertainties for the measured values of $m^\pm$ are systematic
uncertainties of $(\epsilon A)^\pm$, but they are accounted for
here with the statistical uncertainties of $(\epsilon A)^\pm$.

\par
To obtain the final expression for the fluctuations of
the asymmetry measurement, the expressions derived for the
fluctuations of $N_{\rm c}^\pm$ are incorporated into the
asymmetry, Eq.~(\ref{eqnAellpDef}).
Then, the covariances of fluctuations between the $\eta$ bins
of the measurement are calculated. 
Two covariance matrices are calculated as there are two asymmetry
measurements, the base measurement with 30 bins covering the
range $-3 < \eta < 3$ and the combination of the
$\eta \ge 0$ and $\eta < 0$ measurements with half the
number of bins.

\par
\begin{figure}
\includegraphics
   [width=85mm]
   {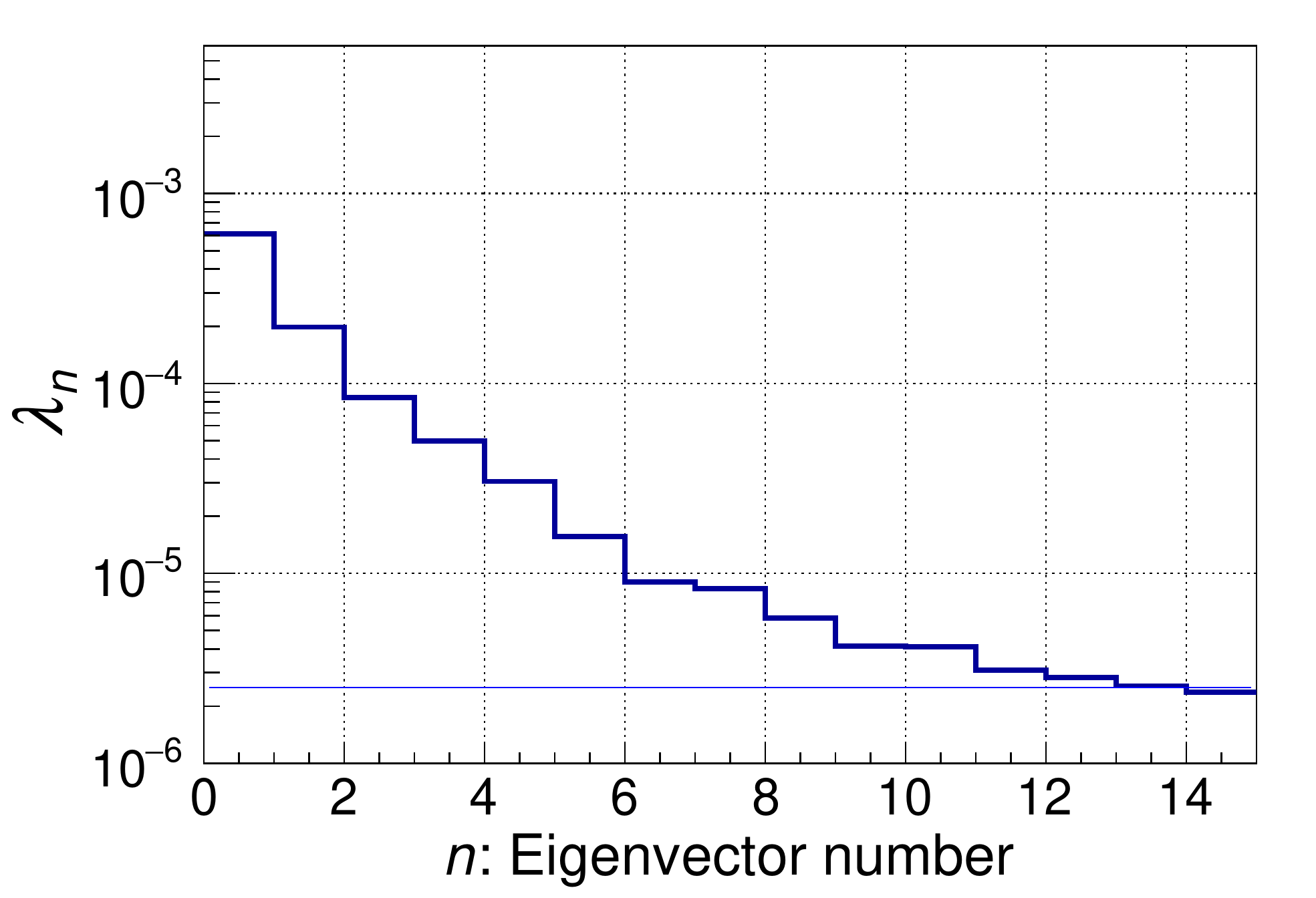}
\caption{\label{fig_covEigenV1}
Eigenvalues $\lambda_n$ as a function of the eigenvector number
$n$ of the covariance matrix for data uncertainties. The eigenvalues,
ordered from the largest to the smallest, are numbered from 0
to 14, respectively.
The bold (black) histogram shows the eigenvalues,
and the lighter (blue) horizontal line corresponds to the smallest
value of the diagonal terms of the covariance matrix.
}
\end{figure}

The covariance matrix of uncertainties for the combined
measurement, denoted by \mbox{\boldmath$V$\unboldmath},
is expanded and inverted to the error matrix using singular-value
decomposition methods. As this is a real-valued
symmetric $15 \times 15$ matrix, its 15 eigenvalues and
eigenvectors are the rank-1 matrix components in the
decomposition of the covariance matrix and of the error matrix
\begin{eqnarray}
\mbox{\boldmath$V$\unboldmath}      & = & \sum_n \lambda_n  \:
                      |v_n\rangle \langle v_n| 
				\; {\rm and} 
		      \nonumber \\ 
\mbox{\boldmath$V$\unboldmath}^{-1} & = & \sum_n \lambda_n^{-1} \:
                      |v_n\rangle \langle v_n| \,
		      \label{svdErrExp} , 
\end{eqnarray}
where $\lambda_n$ and $|v_n\rangle$ are the eigenvalues and
eigenvectors of \mbox{\boldmath$V$\unboldmath}, respectively,
and $|v_n\rangle \langle v_n|$ represents a vector projection
operator in the notation of Dirac bra-kets. In the basis space
of the eigenvectors where the error matrix is diagonal, the
$\chi^2$ comparison of a calculation to the data is
$\sum_n \Delta_n^2 / \lambda_n$, where $\Delta_n$ is the
difference between a calculation and the data along the
$n{\rm th}$ eigenvector, and $\lambda_n$ represents the
squared uncertainty of the difference.
Figure~\ref{fig_covEigenV1} shows the eigenvalues.
Also shown is the smallest value of the diagonal terms of the
covariance matrix. As this value is on par with the smallest
eigenvalue, the error matrix does not have any anomalously small
eigenvalues that need regulation.

\section{\label{systUncerts}
Systematic uncertainties}

Systematic uncertainties are evaluated for the electron-energy scales
and resolutions, the recoil-system energy scale of the simulation
relative to the data, the $|\eta_{\rm det}|$-dependent scale factors,
the backgrounds, and the PDFs. Of these, the PDF uncertainties are
the largest.

\subsection{\label{systUncerts1}
Non-PDF uncertainties}

For plug-region events, additional
uncertainties are evaluated for the correction of the QCD-background
shape at large $E_{\rm T}$ values of the electron, the correction of
the simulation efficiency relative to that for data at large
$E\!\!\!/_{\rm T}$ values, and the charge-dependent bias of the track
reconstruction. Except for the backgrounds, the uncertainties affect
the $(\epsilon A)^\pm$ components of the asymmetry measurement.
All systematic uncertainties, except those from the PDFs, are small
in relation to the statistical uncertainties on the data. The
various categories of systematic uncertainties are treated as
uncorrelated.

\par
To obtain most of the systematic uncertainties, the corresponding
measurement uncertainties are propagated to the
asymmetry. The systematic uncertainties of the
$|\eta_{\rm det}|$-dependent scale factors are derived from
their statistical uncertainties. One standard-deviation shifts
of all $|\eta_{\rm det}|$ bins are coherently propagated to the
asymmetry measurement to obtain upper-limit estimates of the
systematic uncertainties. 
Systematic uncertainties of plug region events due to the
QCD-background shape and simulation-efficiency corrections at large
$E\!\!\!/_{\rm T}$ values are taken to be half the difference of the
asymmetries observed with and without the correction. For the
uncertainty from the plug track-finding bias, the uncertainty of
the integrated value of the bias $A_{\rm b}$ is propagated to the
asymmetry measurement.

\subsection{\label{systUncerts2}
PDF uncertainties}

\par
Systematic uncertainties due to the PDFs used in the simulation enter
through the $(\epsilon A)^\pm$ corrections, which depend on the asymmetry.
For $(\epsilon A)^+$, the explicit expression is
\begin{equation*}
\frac{N_{\rm r}^+}{N_{\rm g}^+} = (1-m^+) \frac{N_{\rm t}^+}{N_{\rm g}^+}
   + m^- \frac{N_{\rm t}^-}{N_{\rm g}^-} \frac{N_{\rm g}^-}{N_{\rm g}^+}
   \;,
\end{equation*}
where $N_{\rm g}^-/N_{\rm g}^+$ equals $(1-A_\ell)/(1+A_\ell)$.
The expression for $(\epsilon A)^-$ is obtained by interchanging the $+$
and $-$ charge superscripts. The $N_{\rm t}^\pm / N_{\rm g}^\pm$ ratios
are the combined efficiencies and acceptances when $m^\pm = 0$.

\par
The implementation of PDFs from the NNPDF collaboration that is used in
this paper is the ensemble set of 100 equally probable PDFs based on the
fit to the input data, along with a default or best-fit PDF. For such
probabilistic PDFs, the prediction is the average value of
$A_\ell(|\eta|)$ calculated over the ensemble, and the dispersion rms
about the average is the PDF uncertainty of the prediction.
These uncertainties are correlated across $|\eta|$ bins.

\par
The simulation is used to calculate the covariance matrix of
uncertainties due to PDF effects. For the calculation of the
corrected number of events $N_{\rm r}^\pm/(\epsilon A)^\pm$ in
the expression for the asymmetry $A_\ell$, the data term
$N_{\rm r}^\pm$ of the numerator is fixed to its default value
from the simulation. The denominator term $(\epsilon A)^\pm$
is modified for each ensemble PDF to include its effect on its
asymmetry relative to the default.
Covariance sums are evaluated using the differences of asymmetries
calculated with the modified values of $(\epsilon A)^\pm$ relative
to the default asymmetries.
The eigenvalues and eigenvectors of the covariance
matrix are determined with the method used for the covariance matrix
of data uncertainties described at the end of Sec.~\ref{AsymMeas}.
The three largest eigenvalues are comparable to or larger in value than
those of the covariance matrix for data uncertainties, but the others
are smaller.

\subsection{\label{systUncerts3}
Total systematic uncertainties}

\begin{figure}
\includegraphics
   [width=85mm]
   {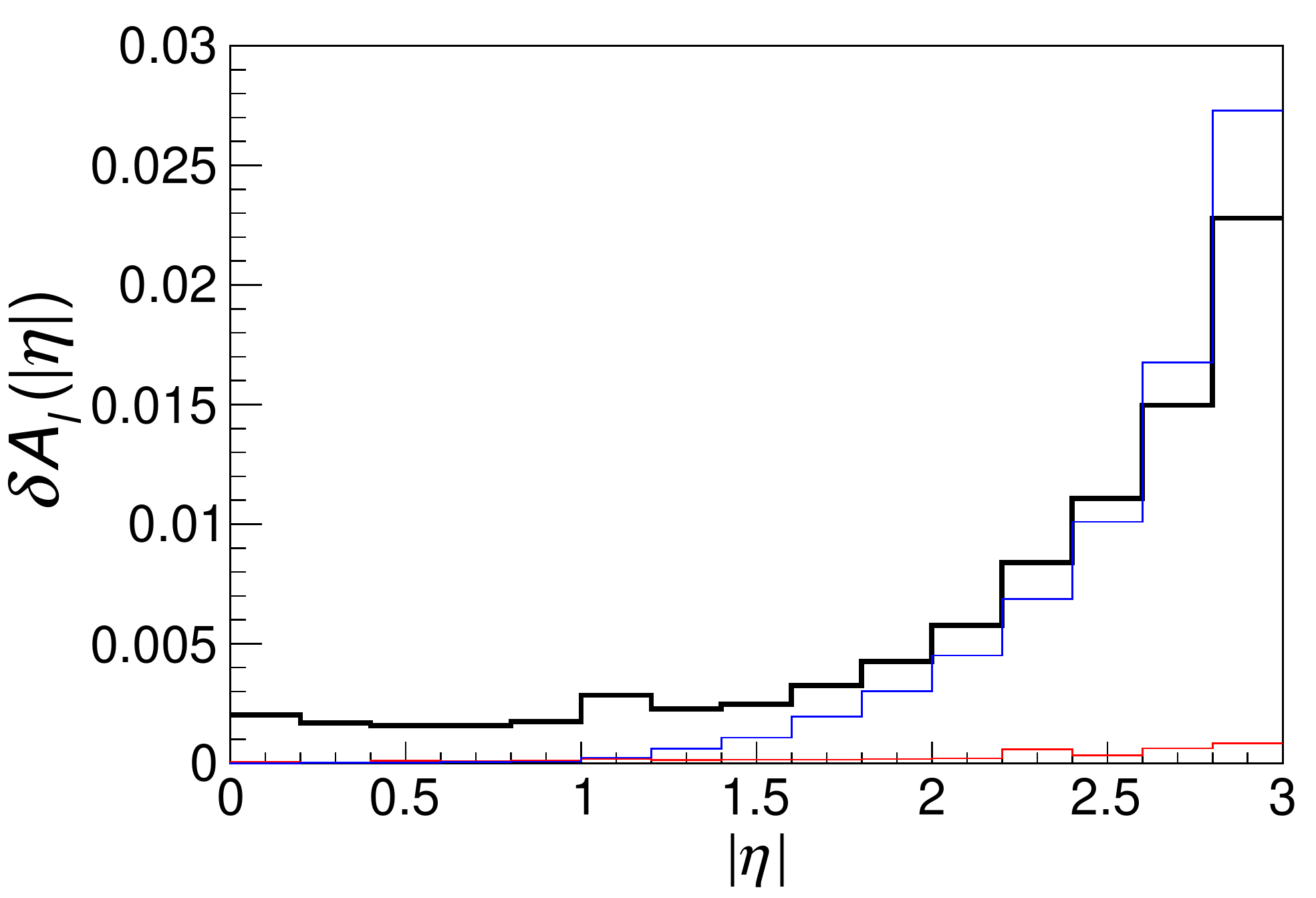}
\caption{\label{fig_systErr}
$\delta A_\ell(|\eta|)$ as a function of $|\eta|$, where $\delta A_\ell$
denotes the data, PDF, or non-PDF uncertainty. The bold (black) histogram
shows the statistical uncertainty of the data. The lowest (red)
histogram is the non-PDF systematic uncertainty.
The remaining lighter-shaded (blue) histogram is the uncertainty due to
PDFs.
}
\end{figure}

A summary of the minimum and maximum values of the systematic
uncertainties from each non-PDF source across the $|\eta|$ bins
of the measurement is shown in Table~\ref{tbl_systUncert}.
\begin{table}
\caption{\label{tbl_systUncert}
Minimum and maximum values of the systematic uncertainties from
each non-PDF source for $A_\ell$ over the $|\eta|$ bins of the
measurement. Except for the specific sources of the plug region,
the minimum values correspond to $|\eta| \leq 1$ electrons and
the maximum values to $|\eta| > 1$. In general, uncertainty
values increase with increasing $|\eta|$ values.
}
\begin{ruledtabular}
\begin{tabular}{lcc}
Source                   &  Minimum value      &  Maximum value     \\ \hline
Electron-energy scale    &  $3.1 \times 10^{-6}$   &  $2.4 \times 10^{-4}$  \\
Electron-energy resolution
			 &  $5.2 \times 10^{-6}$   &  $5.9 \times 10^{-5}$  \\
Recoil-energy scale      &  $6.5 \times 10^{-6}$   &  $3.6 \times 10^{-4}$  \\
Efficiency-scale factor  &  $1.2 \times 10^{-6}$   &  $6.8 \times 10^{-4}$  \\
Backgrounds		 &  $7.6 \times 10^{-6}$   &  $5.4 \times 10^{-4}$  \\
Plug-QCD shape           &  $8.1 \times 10^{-6}$   &  $1.5 \times 10^{-4}$  \\
Plug high-$E\!\!\!/_{\rm T}$ efficiency
			 &  $1.7 \times 10^{-6}$   &  $3.2 \times 10^{-5}$  \\
Plug track-finding bias  &  $6.0 \times 10^{-7}$   &  $6.1 \times 10^{-5}$  \\
\end{tabular}
\end{ruledtabular}
\end{table}
Table~\ref{tbl_systUncertEta} shows the data, non-PDF, and PDF
uncertainties across the $|\eta|$ bins of the measurement,
\begin{table}
\caption{\label{tbl_systUncertEta}
Data, non-PDF, and PDF uncertainties of each $|\eta|$ bin. The data and
PDF uncertainties are taken from the diagonal terms of their respective
covariance matrices. The non-PDF entry is the quadrature sum of the non-PDF
uncertainties specified in Table~\ref{tbl_systUncert}. All uncertainties
are bin-by-bin.
}
\begin{ruledtabular}
\begin{tabular}{cccc}
$|\eta|$ bin  & Data & Non-PDF & PDF \\ \hline
0.0--0.2 & $2.0\times 10^{-3}$  & $6.8\times 10^{-5}$  & $4.0\times 10^{-6}$ \\
0.2--0.4 & $1.7\times 10^{-3}$  & $3.2\times 10^{-5}$  & $2.0\times 10^{-5}$ \\
0.4--0.6 & $1.6\times 10^{-3}$  & $1.1\times 10^{-4}$  & $3.3\times 10^{-5}$ \\
0.6--0.8 & $1.6\times 10^{-3}$  & $8.0\times 10^{-5}$  & $4.9\times 10^{-5}$ \\
0.8--1.0 & $1.7\times 10^{-3}$  & $1.1\times 10^{-4}$  & $6.7\times 10^{-5}$ \\
1.0--1.2 & $2.8\times 10^{-3}$  & $1.8\times 10^{-4}$  & $2.3\times 10^{-4}$ \\
1.2--1.4 & $2.3\times 10^{-3}$  & $1.3\times 10^{-4}$  & $6.2\times 10^{-4}$ \\
1.4--1.6 & $2.5\times 10^{-3}$  & $1.5\times 10^{-4}$  & $1.1\times 10^{-3}$ \\
1.6--1.8 & $3.3\times 10^{-3}$  & $1.5\times 10^{-4}$  & $2.0\times 10^{-3}$ \\
1.8--2.0 & $4.2\times 10^{-3}$  & $1.8\times 10^{-4}$  & $3.0\times 10^{-3}$ \\
2.0--2.2 & $5.8\times 10^{-3}$  & $2.1\times 10^{-4}$  & $4.5\times 10^{-3}$ \\
2.2--2.4 & $8.4\times 10^{-3}$  & $5.9\times 10^{-4}$  & $6.9\times 10^{-3}$ \\
2.4--2.6 & $1.1\times 10^{-2}$  & $3.4\times 10^{-4}$  & $1.0\times 10^{-2}$ \\
2.6--2.8 & $1.5\times 10^{-2}$  & $6.3\times 10^{-4}$  & $1.7\times 10^{-2}$ \\
2.8--3.0 & $2.3\times 10^{-2}$  & $8.3\times 10^{-4}$  & $2.7\times 10^{-2}$ \\
\end{tabular}
\end{ruledtabular}
\end{table}
and Fig.~\ref{fig_systErr} shows the corresponding plot of the
data, PDF, and non-PDF uncertainties presented in
Table~\ref{tbl_systUncertEta}.

\par
For the total systematic uncertainty, the non-PDF and PDF components
are combined. The non-PDF components are negligible, but are added in
quadrature with the diagonal elements of the covariance matrix of
PDF uncertainties.

\section{\label{finalResults}
Results}

The final CDF measurement of the charge asymmetry $A_\ell$ using the
electrons from the production and decay of $W$ bosons is presented in
Table~\ref{tblAsymMeas},
\begin{table}
\caption{\label{tblAsymMeas}
Electron-asymmetry results compared with predictions of the
NLO calculations.
The NNPDF~3.0 and NNPDF~3.1 columns respectively show the predictions
of the NLO calculations with the NNPDF~3.0 and NNPDF~3.1 PDFs, and
the PDF uncertainties shown are evaluated bin-by-bin.
The measurement uncertainties shown are from the diagonal elements of
the covariance matrix for the measurement.
}
\begin{ruledtabular}
\begin{tabular}{cccc}
$|\eta|$ bin  & Measurement      & NNPDF~3.0        & NNPDF~3.1 \\
              & $(\times 10)$ & $(\times 10)$      & $(\times 10)$ \\ \hline
0.0--0.2 & $0.164 \pm 0.020$  & $0.185 \pm 0.011$  & $0.184 \pm 0.004$ \\
0.2--0.4 & $0.549 \pm 0.017$  & $0.549 \pm 0.032$  & $0.545 \pm 0.011$ \\
0.4--0.6 & $0.921 \pm 0.016$  & $0.893 \pm 0.049$  & $0.887 \pm 0.016$ \\
0.6--0.8 & $1.246 \pm 0.016$  & $1.198 \pm 0.063$  & $1.190 \pm 0.019$ \\
0.8--1.0 & $1.479 \pm 0.018$  & $1.442 \pm 0.073$  & $1.433 \pm 0.023$ \\
1.0--1.2 & $1.634 \pm 0.029$  & $1.600 \pm 0.081$  & $1.588 \pm 0.025$ \\
1.2--1.4 & $1.647 \pm 0.024$  & $1.640 \pm 0.086$  & $1.621 \pm 0.027$ \\
1.4--1.6 & $1.487 \pm 0.027$  & $1.525 \pm 0.090$  & $1.496 \pm 0.030$ \\
1.6--1.8 & $1.178 \pm 0.038$  & $1.214 \pm 0.094$  & $1.182 \pm 0.035$ \\
1.8--2.0 & $0.688 \pm 0.052$  & $0.679 \pm 0.100$  & $0.656 \pm 0.039$ \\
2.0--2.2 & $-0.009 \pm 0.073$  & $-0.082 \pm 0.109$  & $-0.072 \pm 0.046$ \\
2.2--2.4 & $-1.149 \pm 0.109$  & $-1.036 \pm 0.124$  & $-0.958 \pm 0.058$ \\
2.4--2.6 & $-1.976 \pm 0.150$  & $-2.134 \pm 0.153$  & $-1.930 \pm 0.083$ \\
2.6--2.8 & $-3.115 \pm 0.225$  & $-3.302 \pm 0.210$  & $-2.889 \pm 0.137$ \\
2.8--3.0 & $-4.605 \pm 0.356$  & $-4.428 \pm 0.324$  & $-3.680 \pm 0.248$ \\
\end{tabular}
\end{ruledtabular}
\end{table}
along with the default NLO prediction using the
NNPDF~3.0~\cite{ nnpdf301, *nnpdf302, *nnpdf303, *nnpdf304,
		*nnpdf305, *nnpdf306, *nnpdf306e,*nnpdf307}.
Uncertainties of the measurement are represented by the sum
of the covariance matrices for the statistical uncertainties
and the systematic uncertainties. Table~\ref{tblAsymMeas}
also includes a calculation using the NNPDF~3.1 PDFs derived
with the value of $\alpha_s = 0.118$ at the $Z$-pole
mass~\cite{nnpdf301A}.
For consistency with the measurement, the predictions using the
NNPDF~3.0 and 3.1 PDFs are restricted to the kinematic region of
$P_{\rm T}^e > 25$~GeV/$c$, $P_{\rm T}^\nu > 25$~GeV/$c$,
and $M_{\rm T} > 45$~GeV/$c^2$.

\par
The input data used in the global fits for the NNPDF~3.0 PDFs
do not include any Tevatron measurements of the charge asymmetry in
the production of $W$ bosons but do include the lepton-charge asymmetry
measurements in the electron and muon channels from CMS using $pp$
collisions at $\sqrt{s} = 7$~TeV~\cite{CMS7Ae1,CMS7Am1}.
The input data for the NNPDF~3.1 PDFs include the
final Tevatron measurements of the lepton-charge asymmetry in the
electron and muon channels from D0~\cite{D0196Am2,D0196Ae2,*D0196Ae2E}.
This inclusion significantly reduces the uncertainty relative to
NNPDF~3.0. In addition, the ensemble methodology for NNPDF~3.1 is more
robust in that the ensemble represents a better sampling of the
probability distribution of the fit to input data.

\par
\begin{figure}
\includegraphics
   [width=85mm]
   {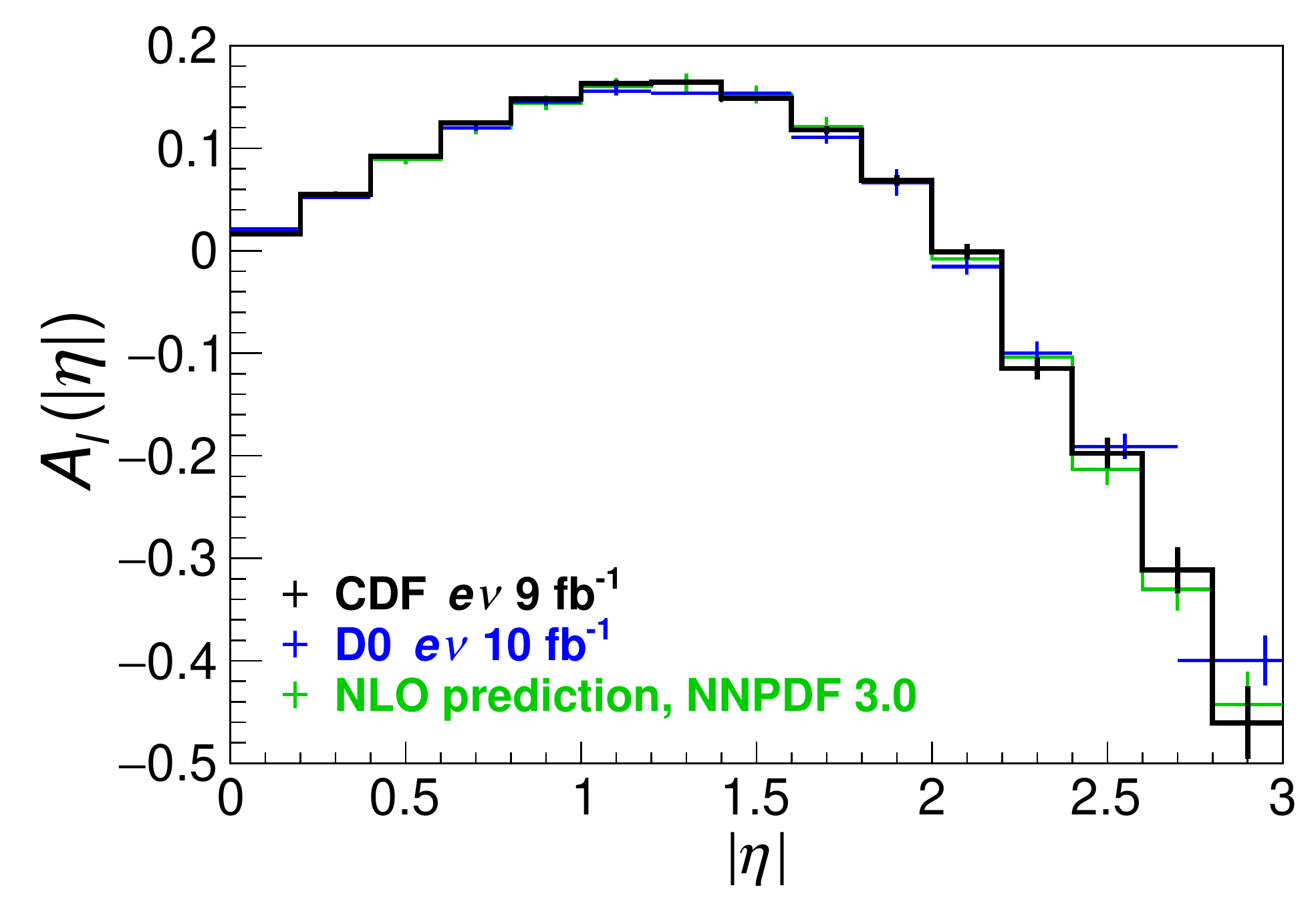}
\caption{\label{fig_ecpEllAsym30}
CP-folded distributions of $A_\ell(|\eta|)$ as a function of $|\eta|$.
The bold
(black) histogram is the result of this measurement, and the uncertainties
shown are evaluated bin-by-bin and include both statistical and
systematic contributions. The (blue) crosses represent the measurement
from D0~\cite{D0196Ae2,*D0196Ae2E}. For D0, the bin size is also 0.2
$|\eta|$-units wide, except for these regions:
\mbox{1.2--1.6}, \mbox{2.4--2.7}, and \mbox{2.7--3.2}. Both the CDF and
D0 measurements use all the data from Run~II of the Tevatron Collider.
The thin (green) histogram is the NLO prediction using the NNPDF~3.0 PDFs.
}
\end{figure}
\begin{figure}
\includegraphics
   [width=85mm]
   {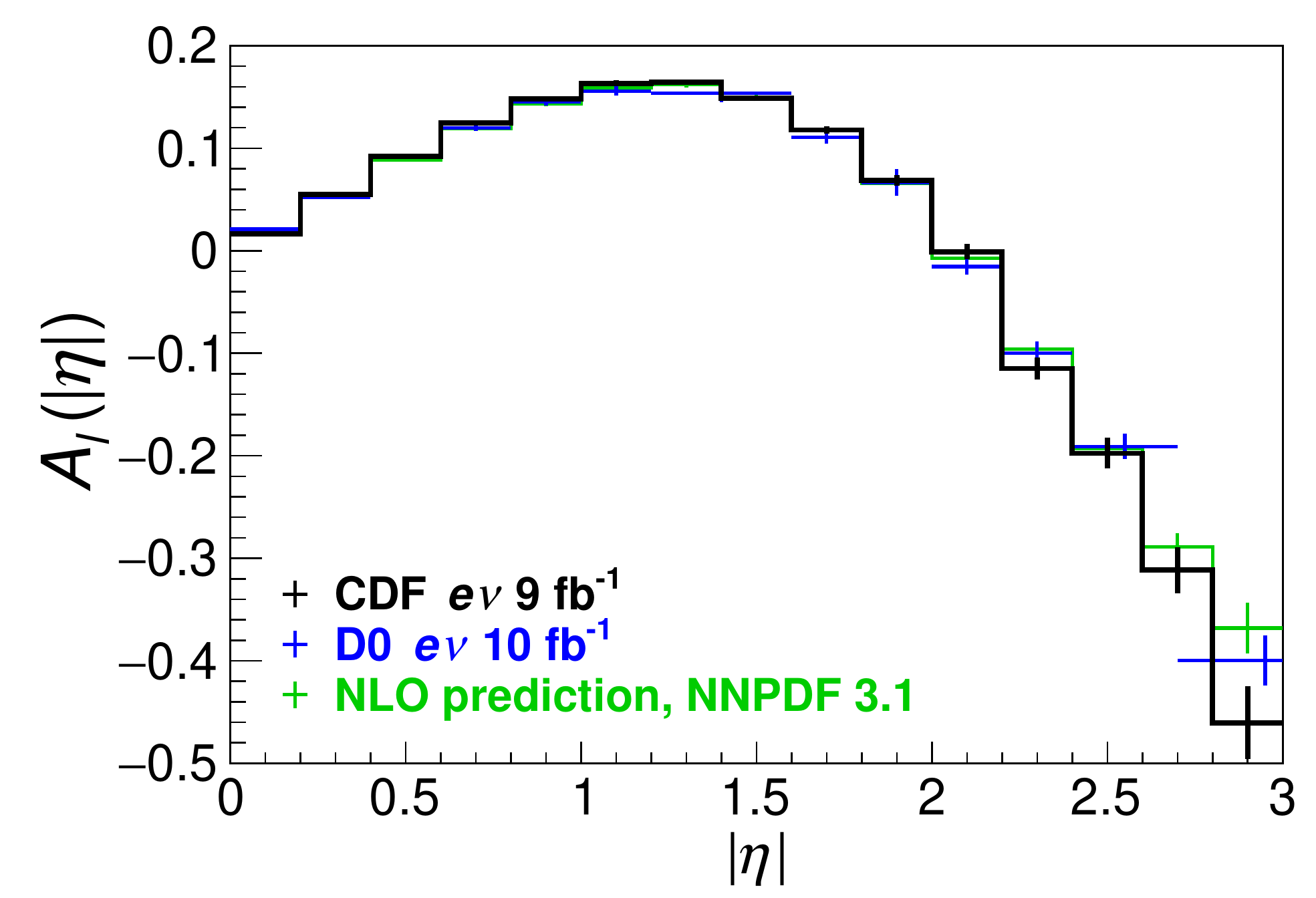}
\caption{\label{fig_ecpEllAsym31}
Distributions of
$A_\ell(|\eta|)$ as a function of $|\eta|$ for the Tevatron measurements
previously shown in Fig.~\ref{fig_ecpEllAsym30}, along with the NLO prediction
with NNPDF~3.1 PDFs. The bold (black) histogram is the result of this
measurement and the (blue) crosses the
D0 measurement~\cite{D0196Ae2,*D0196Ae2E}. The thin (green) histogram is
the NLO prediction using the NNPDF~3.1 PDFs.
}
\end{figure}

Figures \ref{fig_ecpEllAsym30} and \ref{fig_ecpEllAsym31} show the
final results for the charge asymmetry using the electrons at
the Tevatron from this measurement and D0~\cite{D0196Ae2,*D0196Ae2E}.
All uncertainties presented in Table~\ref{tblAsymMeas} and shown in
Figs.~\ref{fig_ecpEllAsym30} and \ref{fig_ecpEllAsym31} are bin-by-bin,
and do not reflect their correlations with the uncertainties of
neighboring bins. For the data, interbin correlations increase from
about 0.03 to about 0.80 as $|\eta|$ increases from 0 to 3.0.

\par
\begin{figure}
\includegraphics
   [width=85mm]
   {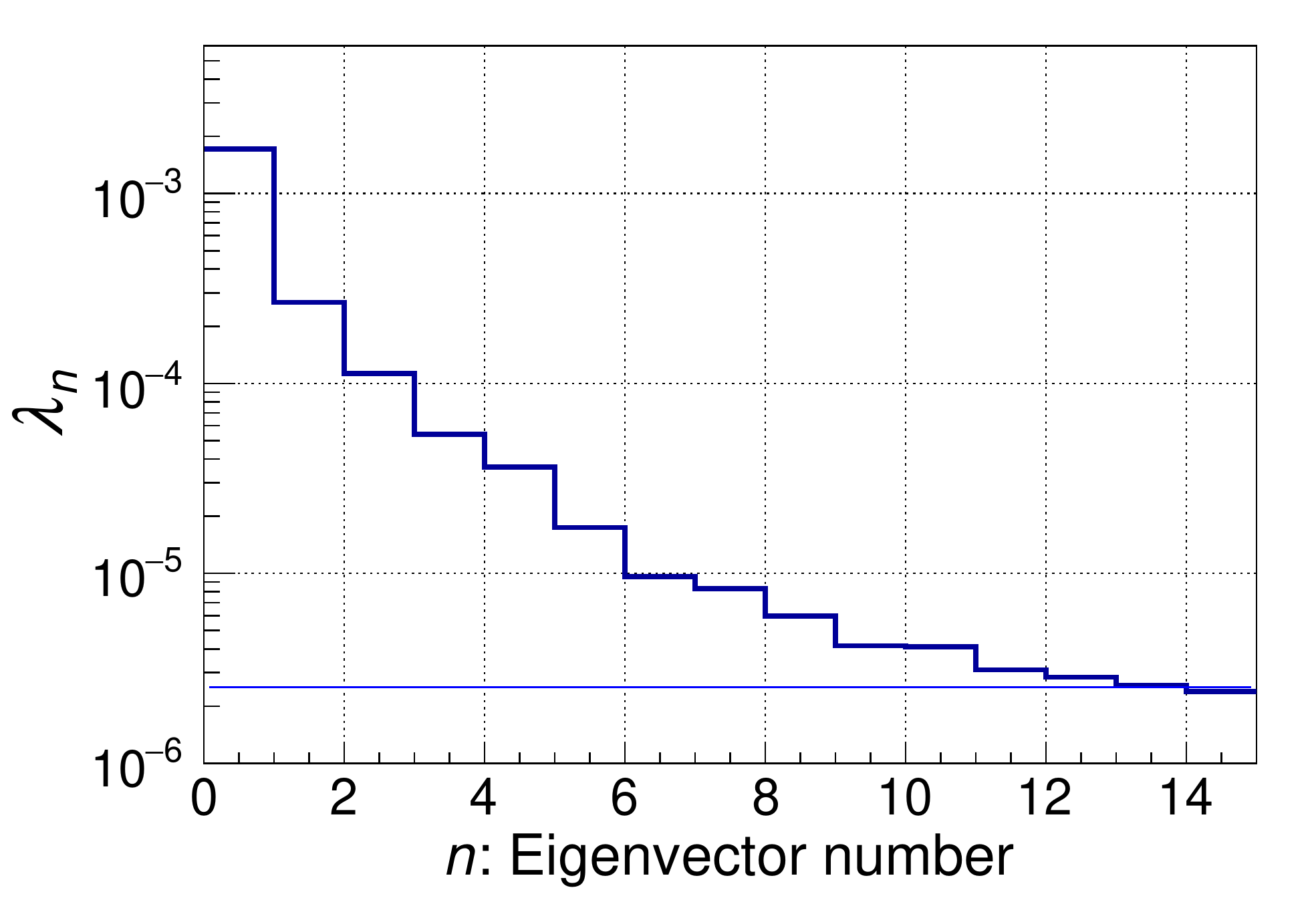}
\caption{\label{fig_covEigenV2}
Eigenvalues $\lambda_n$ as a function of the eigenvector number $n$
from the covariance matrix of the asymmetry measurement.
The bold (black) histogram shows the eigenvalues,
and the lighter (blue) horizontal line corresponds to the smallest
value of the diagonal terms of the covariance matrix.
}
\end{figure}

To compare the CDF measurement with predictions, the $\chi^2$ statistic
is evaluated over all bins using the error matrix of the measurement.
The eigenvalues of the corresponding covariance matrix are shown in
Fig.~\ref{fig_covEigenV2}.
The comparison of the asymmetry measurement with the prediction
derived from the NNPDF~3.0 (3.1) ensemble yields the $\chi^2$ value
32.6 (44.9) for the 15 bins of the measurement. 
Calculation of the corresponding $\chi^2$ value with the bin-by-bin
uncertainties of the measurement shown in Table~\ref{tblAsymMeas}
instead of the error matrix yields 26.4 (41.2).

\par
\begin{figure}
\includegraphics
   [width=85mm]
   {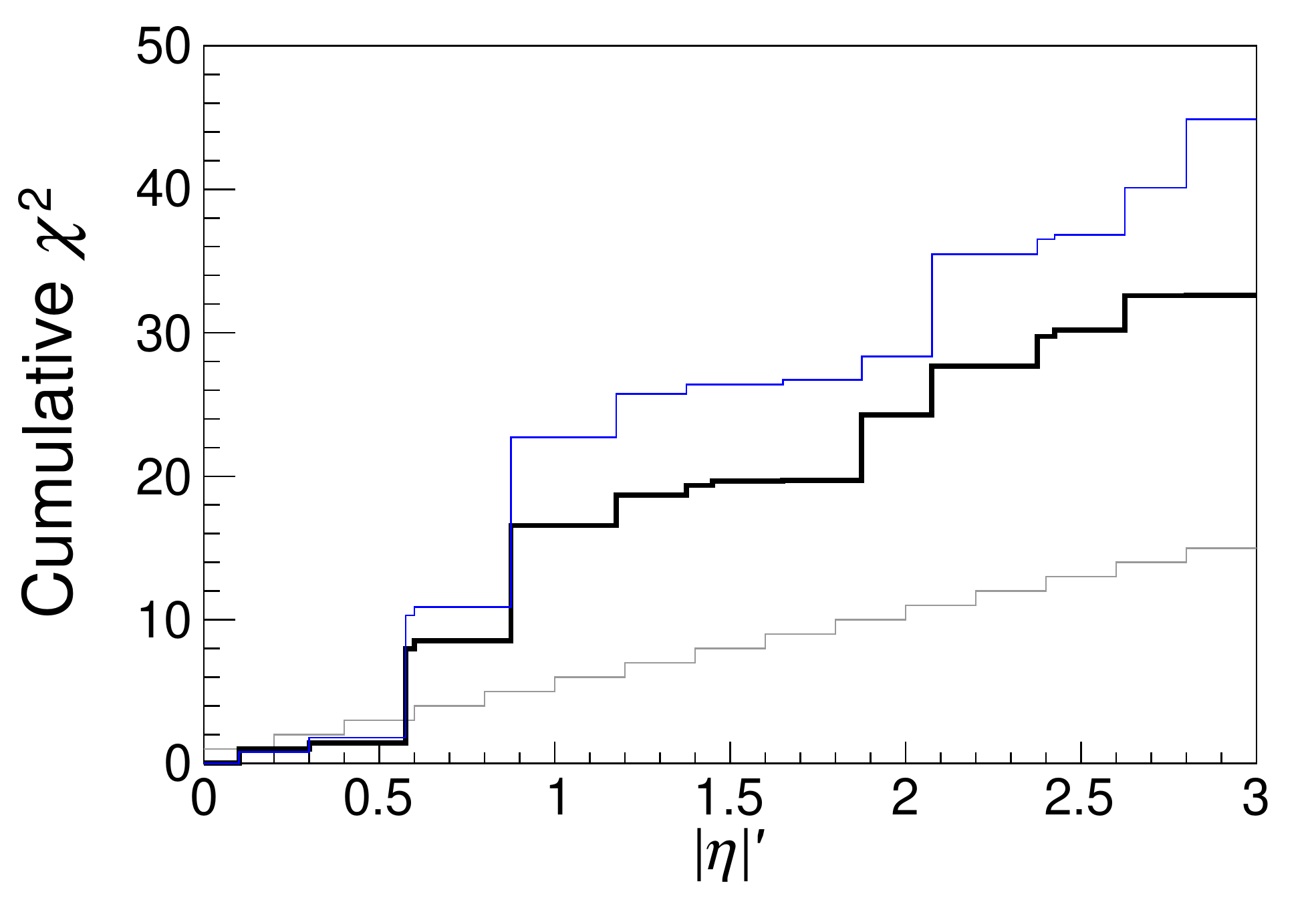}
\caption{\label{fig_cumChisqNN3}
Cumulative $\chi^2$ as a function of $|\eta|^\prime$.
The bold (black) [light (blue)] histogram is for NNPDF~3.0 [3.1].
The lowermost (gray) histogram is for an ideal 15-bin measurement
and a prediction whose underlying physics matches that of the
measurement.
Differences between the underlying PDFs of the data and that of
the calculation result in differences between the ideal and
observed $\chi^2$ distributions.
}
\end{figure}

The cumulative-$\chi^2$ distribution versus $|\eta|$ as a function
of $|\eta|$ is used to
assess how the goodness-of-fit varies across the $|\eta|$ bins.
For the $\chi^2$ evaluated with the error matrix, the $\chi^2$
increment per eigenvector covers several $|\eta|$ bins. Consequently,
the increment for each eigenvector is associated with its expectation
value of the $|\eta|$-bin centers, $0.1+0.2j$, where $j$ is the number
of the $|\eta|$ bin, which ranges from 0 to 14. The expectation value is
denoted by $|\eta|^\prime$. Figure~\ref{fig_cumChisqNN3} shows the
cumulative-$\chi^2$ distributions
versus $|\eta|^\prime$ for the NNPDF~3.0 and 3.1 predictions. These
distributions show that the measurement can tighten the constraints
to the PDFs over a broad region, $|\eta| > 0.5$.

\par
Since each of the ensemble PDFs is equally probable, the distribution of
$\chi^2$ values from the comparisons between the measurement and the
individual predictions from each of the ensemble PDFs quantifies
the consistency between the ensemble and the underlying PDFs of the
measurement.
\begin{figure}
\includegraphics
   [width=85mm]
   {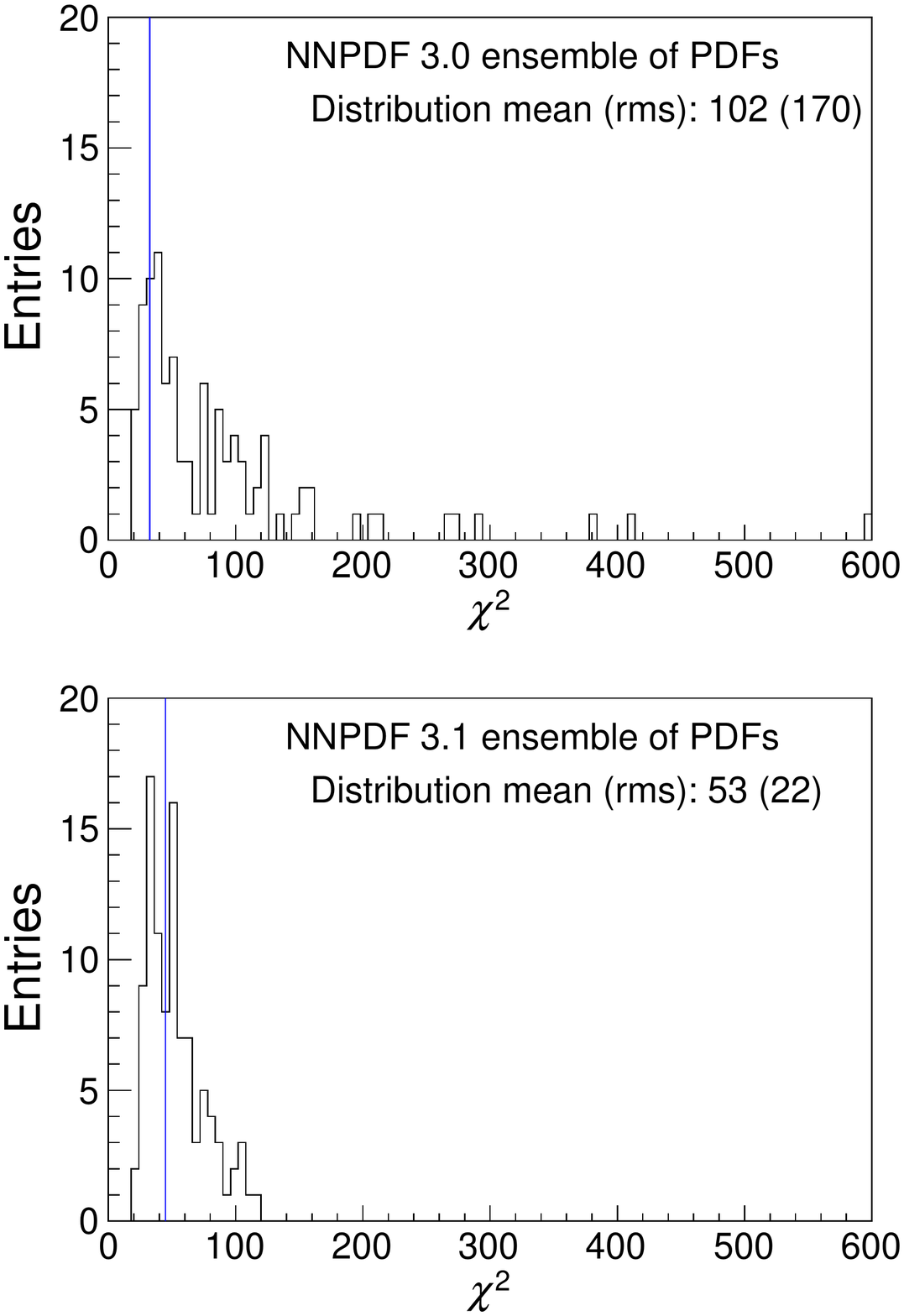}
\caption{\label{fig_chi2NNPDF3X2}
Distribution of $\chi^2$ values from the comparisons between the
measurement and the individual predictions over 15 bins for the NNPDF~3.0
and NNPDF~3.1 ensembles of PDFs. The light (blue) vertical line of each
panel show the $\chi^2$ value of the ensemble-averaged prediction
for NNPDF~3.0 (32.6) and NNPDF~3.1 (44.9).
For NNPDF~3.0, one of the ensemble PDFs gives
a $\chi^2$ value of 1548, and is not shown.
For NNPDF~3.1, all values are contained within the panel.
}
\end{figure}
Figure~\ref{fig_chi2NNPDF3X2} shows the $\chi^2$
distributions for the NNPDF~3.0 and NNPDF~3.1
ensemble of PDFs.
As the mean and rms of the NNPDF~3.1 ensemble distribution are much smaller
than those of NNPDF~3.0, the NNPDF~3.1 ensemble is thus found to be
a more robust representation of the Tevatron PDFs and their uncertainties.

\par
The inclusion of this measurement in global PDF fits will improve the
precision of the PDFs over the kinematic region for the Tevatron.
Numerical tables for the measurement and its covariance matrix of
uncertainties are provided as supplemental materials to this paper.
Also included are numerical tables of the $\chi^2$ values for each PDF
of the NNPDF~3.0 and NNPDF~3.1 ensembles.

\section{\label{theEndSummary}
Summary}

The yield asymmetry between positrons and electrons from
the decays of $W^\pm$ bosons produced in $p\bar{p}$ collisions at the
center-of-momentum energy of 1.96~TeV is measured as a function of
the electron pseudorapidity using the full Run~II data set of CDF,
corresponding to 9.1~fb$^{-1}$ of integrated luminosity. Results are in
Table~\ref{tblAsymMeas}, and Figs.~\ref{fig_ecpEllAsym30} and
\ref{fig_ecpEllAsym31}. The uncertainties in the results are dominated
approximately equally by the statistical precision of the data and the
effect of PDF uncertainties on the modeling of acceptance and
efficiencies.

\par
At the Tevatron collider, the asymmetry is sensitive to the slope
of the ratio of $d$- to $u$-quark parton-distribution functions of
the proton versus the Bjorken-$x$ parameter.
Inclusion of this asymmetry measurement in global fits to PDFs
will reduce the overall uncertainties of the PDFs within
the kinematic region of Tevatron collisions.

\begin{acknowledgments}

This document was prepared by the CDF collaboration using the resources of the
Fermi National Accelerator Laboratory (Fermilab), a U.S.  Department of Energy,
Office of Science, HEP User Facility. Fermilab is managed by Fermi Research
Alliance, LLC (FRA), acting under Contract No.  DE-AC02-07CH11359.  We thank
the Fermilab staff and the technical staffs of the participating institutions
for their vital contributions. This work was supported by the U.S. Department
of Energy and National Science Foundation; the Italian Istituto Nazionale di
Fisica Nucleare;
the U.S. Department of Energy and the Italian Istituto Nazionale di Fisica
Nucleare summer-student exchange program for the years 2014 and 2015;
the Ministry of Education, Culture, Sports, Science and
Technology of Japan; the Natural Sciences and Engineering Research Council of
Canada; the National Science Council of the Republic of China; the Swiss
National Science Foundation; the A.P. Sloan Foundation; the Bundesministerium
f\"ur Bildung und Forschung, Germany; the National Research Foundation of Korea
(Grants No. 2018R1A6A1A06024970); the Science and Technology Facilities Council
and the Royal Society, United Kingdom; the Russian Foundation for Basic
Research; the Ministerio de Ciencia e Innovaci\'{o}n, and Programa
Consolider-Ingenio 2010, Spain; the Slovak R\&D Agency; the Academy of Finland;
and the Australian Research Council (ARC).

\end{acknowledgments}

\appendix
\section{\label{PlugEleSel}
Plug electron selection}

The background in the plug-electron sample
varies significantly with the topology of the reconstructed track
in the silicon detector. The purity is adjusted as a function of
the quality parameters of the electron candidate, which are
the goodness-of-fit
between the measured and expected transverse-shower shapes
$\chi^2_{3\times3}$, and the goodness-of-fit between the
track helix and the hits attached to the helix $\chi^2_{\rm trk}$.

\par
Tracks in the plug region are reconstructed with the calorimetry-seeded
tracking algorithm (``Phoenix''), which searches for hits in seven
layers of the silicon detector. The tracks are characterized by
two parameters, $(n_l,\: n_h)$, where $n_l$ is the number
of fiducial layers of the silicon tracker traversed by the particle,
and $n_h$ is the number of hits detected in those layers that are
associated with the track by the algorithm. Multiple hits per layer can
be attached to the track by the algorithm. The number of layers is
a prediction based on the track-helix parameters and a simplified
model of the silicon-detector geometry. It restricts the electron
candidate to a region of $|\eta_{\rm det}|$.

\par
The maximum allowed $\chi^2_{\rm trk}/n_h$ and $\chi^2_{3\times3}$
values are both 10 in the default selection. For electron candidates
with lower quality tracks, the maximum $\chi^2$ values are
reduced to improve the signal purity.
For events with high-$E_{\rm T}$ electrons, the distributions
of these quantities are peaked at values of about 1.0 and decrease
exponentially beyond the peak. For background events, the
distributions are broad and relatively uniform across the $\chi^2$
values in relation to those for the electrons.
The events used to adjust the
maximum values must pass the asymmetry-measurement criteria of
Sec.~\ref{EleNeuSelection}, except for the $E\!\!\!/_{\rm T}$
criterion. After an adjustment, the $E\!\!\!/_{\rm T}$
distribution of the event is used to evaluate independently
the purity of the electron sample from $W$-boson decays.
The results are shown in Table~\ref{tblPlugSelMatrix}.
\begin{table}
\caption{\label{tblPlugSelMatrix}
Selection criteria matrix for plug-region electrons. The
index $n_l$ is the number of fiducial layers in
the silicon tracker traversed by the particle, and $n_h$
is the number of hits detected in those layers that are
associated with the track. In each
table-entry pair, the first value is the maximum value for the
$\chi^2_{\rm trk}/n_h$ quantity, and the second is the maximum
value for the $\chi^2_{3\times3}$ quantity.
}
\begin{ruledtabular}
\begin{tabular}{cccccc}
        & $n_h=3$   & $n_h=4$   & $n_h=5$  & $n_h=6$  & $n_h=7$  \\ \hline
$n_l=7$ & $2.5/3.0$ & $2.5/4.0$ & $10/10$  & $10/10$  & $10/10$  \\
$n_l=6$ & $2.5/4.0$ & $2.5/5.0$ & $10/10$  & $10/10$  & $10/10$  \\
$n_l=5$ & $3.0/3.0$ & $10/3.0$  & $10/3.0$ & $10/3.0$ & $10/3.0$  \\
$n_l=4$ & $10/3.0$  & $10/3.0$  & $10/3.0$ & $10/3.0$ & $10/3.0$  \\
$n_l=3$ & $10/3.0$  & $10/3.0$  & $10/3.0$ & $10/3.0$ & $10/3.0$
\end{tabular}
\end{ruledtabular}
\end{table}

\par
In addition, the lateral-shower profile measured in the PES detector,
which consists of 5~mm wide scintillator strips, is required to be
consistent with that of an electron. The profile is measured with the
ratio of the shower energy observed in five strips relative to nine
strips, $R_{5/9}$. For EM showers, the $R_{5/9}$ distribution is peaked
near the value of 0.9. The consistency criterion is $R_{5/9}>0.75$.

\par
\begin{figure}
\includegraphics
   [width=85mm]
   {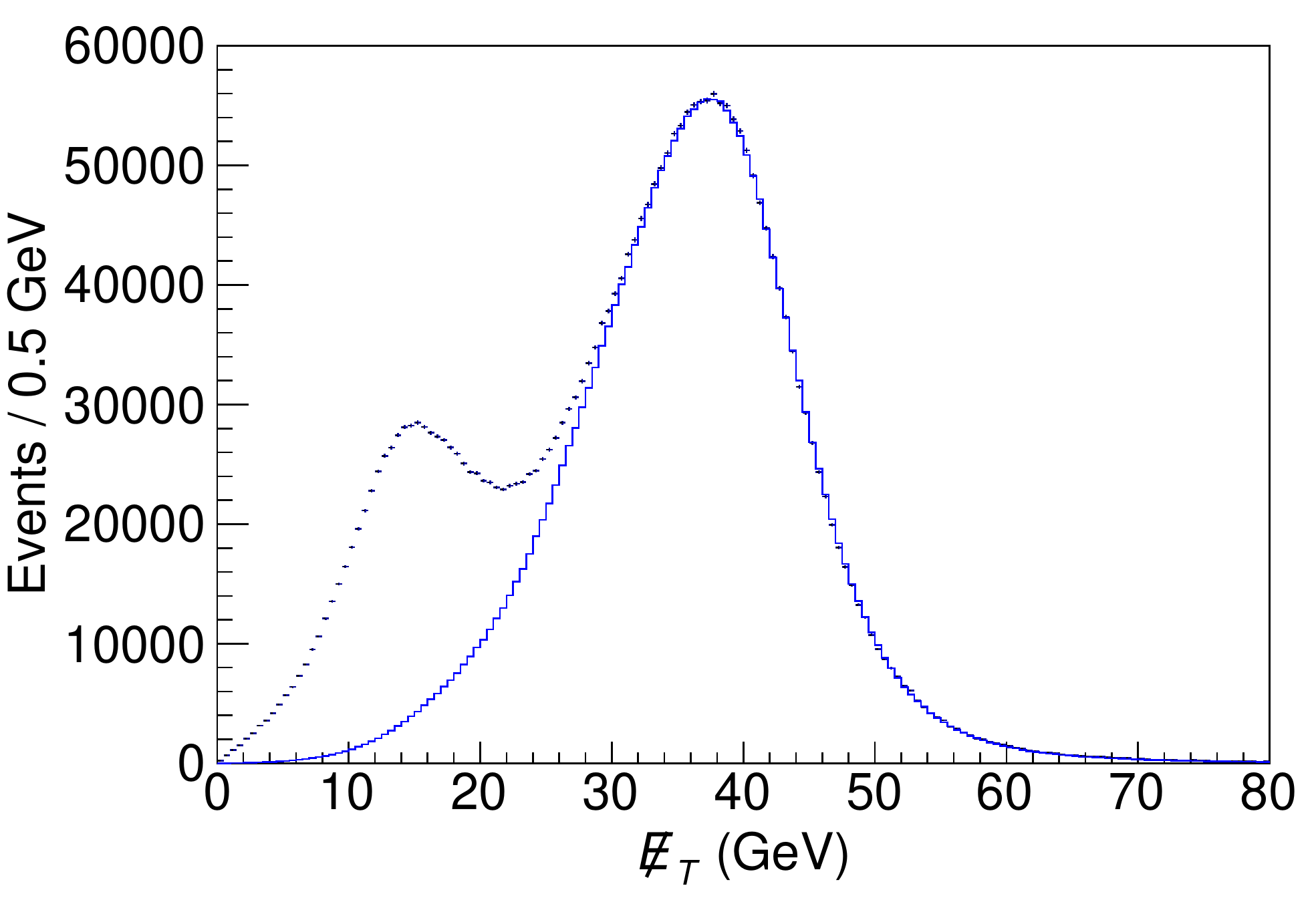}
\caption{\label{figA1_plugMETsel}
$E\!\!\!/_{\rm T}$ distribution for plug-region electrons accepted
by the additional selection criteria. The (black) crosses are the
data, and the (blue) histogram is the simulation.
The online-trigger selection allows the peak at low values of
$E\!\!\!/_{\rm T}$.
}
\end{figure}
\begin{figure}
\includegraphics
   [width=85mm]
   {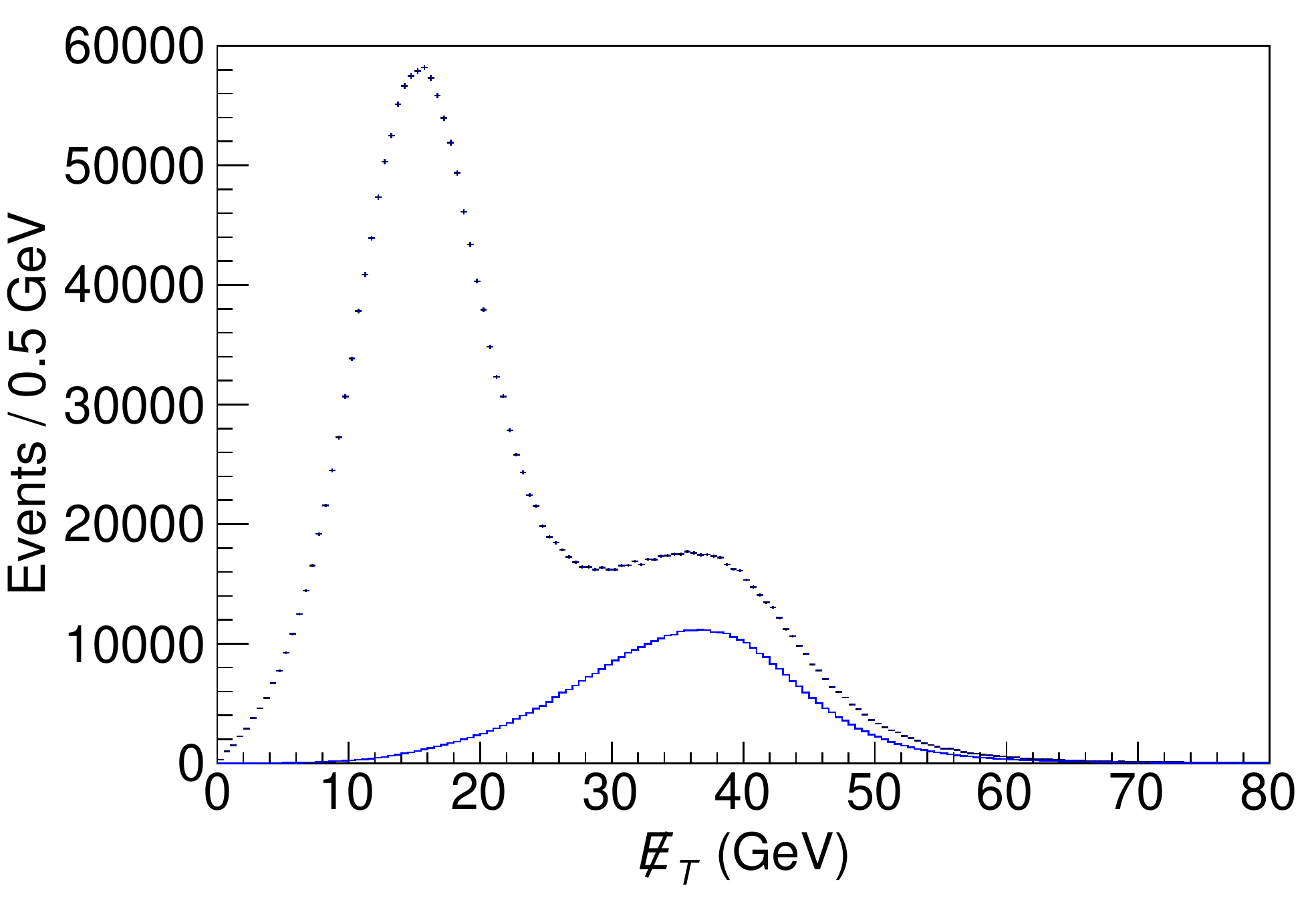}
\caption{\label{figA1_plugMETcut}
$E\!\!\!/_{\rm T}$ distribution for plug-region electrons rejected by
the additional selection criteria. The (black) crosses are the data,
and the (blue) histogram is the simulation, which contains about
18\% of the events without the additional requirements. The
online-trigger selection allows the peak at low values of
$E\!\!\!/_{\rm T}$.
}
\end{figure}

The $E\!\!\!/_{\rm T}$ distribution after the application
of the additional selection criteria is shown in
Fig.~\ref{figA1_plugMETsel}.
The $E\!\!\!/_{\rm T}$ distribution for events passing the default
selection criteria but failing the additional criteria is shown
in Fig.~\ref{figA1_plugMETcut}.
The simulation is described in Secs.~\ref{ExpDatSim} and
\ref{CorrDatSim}.

\bibliography{cdfWAsym9enu}

\providecommand{\noopsort}[1]{}\providecommand{\singleletter}[1]{#1}%
\begin{thebibliography}{76}%
\makeatletter
\providecommand \@ifxundefined [1]{%
 \@ifx{#1\undefined}
}%
\providecommand \@ifnum [1]{%
 \ifnum #1\expandafter \@firstoftwo
 \else \expandafter \@secondoftwo
 \fi
}%
\providecommand \@ifx [1]{%
 \ifx #1\expandafter \@firstoftwo
 \else \expandafter \@secondoftwo
 \fi
}%
\providecommand \natexlab [1]{#1}%
\providecommand \enquote  [1]{``#1''}%
\providecommand \bibnamefont  [1]{#1}%
\providecommand \bibfnamefont [1]{#1}%
\providecommand \citenamefont [1]{#1}%
\providecommand \href@noop [0]{\@secondoftwo}%
\providecommand \href [0]{\begingroup \@sanitize@url \@href}%
\providecommand \@href[1]{\@@startlink{#1}\@@href}%
\providecommand \@@href[1]{\endgroup#1\@@endlink}%
\providecommand \@sanitize@url [0]{\catcode `\\12\catcode `\$12\catcode
  `\&12\catcode `\#12\catcode `\^12\catcode `\_12\catcode `\%12\relax}%
\providecommand \@@startlink[1]{}%
\providecommand \@@endlink[0]{}%
\providecommand \url  [0]{\begingroup\@sanitize@url \@url }%
\providecommand \@url [1]{\endgroup\@href {#1}{\urlprefix }}%
\providecommand \urlprefix  [0]{URL }%
\providecommand \Eprint [0]{\href }%
\providecommand \doibase [0]{http://dx.doi.org/}%
\providecommand \selectlanguage [0]{\@gobble}%
\providecommand \bibinfo  [0]{\@secondoftwo}%
\providecommand \bibfield  [0]{\@secondoftwo}%
\providecommand \translation [1]{[#1]}%
\providecommand \BibitemOpen [0]{}%
\providecommand \bibitemStop [0]{}%
\providecommand \bibitemNoStop [0]{.\EOS\space}%
\providecommand \EOS [0]{\spacefactor3000\relax}%
\providecommand \BibitemShut  [1]{\csname bibitem#1\endcsname}%
\let\auto@bib@innerbib\@empty
\bibitem [{\citenamefont {Drell}\ and\ \citenamefont
  {Yan}(1970{\natexlab{a}})}]{DrellYan}%
  \BibitemOpen
  \bibfield  {author} {\bibinfo {author} {\bibfnamefont {S.~D.}\ \bibnamefont
  {Drell}}\ and\ \bibinfo {author} {\bibfnamefont {T.-M.}\ \bibnamefont
  {Yan}},\ }\href {\doibase 10.1103/PhysRevLett.25.316} {\bibfield  {journal}
  {\bibinfo  {journal} {Phys. Rev. Lett.}\ }\textbf {\bibinfo {volume} {25}},\
  \bibinfo {pages} {316} (\bibinfo {year} {1970}{\natexlab{a}})}\BibitemShut
  {NoStop}%
\bibitem [{\citenamefont {Drell}\ and\ \citenamefont
  {Yan}(1970{\natexlab{b}})}]{DrellYanE}%
  \BibitemOpen
  \bibfield  {author} {\bibinfo {author} {\bibfnamefont {S.~D.}\ \bibnamefont
  {Drell}}\ and\ \bibinfo {author} {\bibfnamefont {T.-M.}\ \bibnamefont
  {Yan}},\ }\href {\doibase 10.1103/PhysRevLett.25.902.2} {\bibfield  {journal}
  {\bibinfo  {journal} {Phys. Rev. Lett.}\ }\textbf {\bibinfo {volume} {25}},\
  \bibinfo {pages} {902} (\bibinfo {year} {1970}{\natexlab{b}})}\BibitemShut
  {NoStop}%
\bibitem [{\citenamefont {Bjorken}\ and\ \citenamefont
  {Paschos}(1969)}]{BjorkenX}%
  \BibitemOpen
  \bibfield  {author} {\bibinfo {author} {\bibfnamefont {J.~D.}\ \bibnamefont
  {Bjorken}}\ and\ \bibinfo {author} {\bibfnamefont {E.~A.}\ \bibnamefont
  {Paschos}},\ }\href {\doibase 10.1103/PhysRev.185.1975} {\bibfield  {journal}
  {\bibinfo  {journal} {Phys. Rev.}\ }\textbf {\bibinfo {volume} {185}},\
  \bibinfo {pages} {1975} (\bibinfo {year} {1969})}\BibitemShut {NoStop}%
\bibitem [{\citenamefont {{F. Abe \emph{et al.}}}(1998)}]{CDF180Aem}%
  \BibitemOpen
  \bibfield  {author} {\bibinfo {author} {\bibnamefont {{F. Abe \emph{et
  al.}}}} (\bibinfo {collaboration} {CDF Collaboration}),\ }\href {\doibase
  doi.org/10.1103/PhysRevLett.81.5754} {\bibfield  {journal} {\bibinfo
  {journal} {Phys. Rev. Lett.}\ }\textbf {\bibinfo {volume} {81}},\ \bibinfo
  {pages} {5754} (\bibinfo {year} {1998})}\BibitemShut {NoStop}%
\bibitem [{\citenamefont {{D. Acosta \emph{et al.}}}(2005)}]{CDF196Ae1}%
  \BibitemOpen
  \bibfield  {author} {\bibinfo {author} {\bibnamefont {{D. Acosta \emph{et
  al.}}}} (\bibinfo {collaboration} {CDF Collaboration}),\ }\href {\doibase
  10.1103/PhysRevD.71.051104} {\bibfield  {journal} {\bibinfo  {journal} {Phys.
  Rev. D}\ }\textbf {\bibinfo {volume} {71}},\ \bibinfo {pages} {051104(R)}
  (\bibinfo {year} {2005})}\BibitemShut {NoStop}%
\bibitem [{\citenamefont {{V. M. Abazov \emph{et
  al.}}}(2008{\natexlab{a}})}]{D0196Am1}%
  \BibitemOpen
  \bibfield  {author} {\bibinfo {author} {\bibnamefont {{V. M. Abazov \emph{et
  al.}}}} (\bibinfo {collaboration} {D0 Collaboration}),\ }\href {\doibase
  10.1103/PhysRevD.77.011106} {\bibfield  {journal} {\bibinfo  {journal} {Phys.
  Rev. D}\ }\textbf {\bibinfo {volume} {77}},\ \bibinfo {pages} {011106(R)}
  (\bibinfo {year} {2008}{\natexlab{a}})}\BibitemShut {NoStop}%
\bibitem [{\citenamefont {{V. M. Abazov \emph{et
  al.}}}(2008{\natexlab{b}})}]{D0196Ae1}%
  \BibitemOpen
  \bibfield  {author} {\bibinfo {author} {\bibnamefont {{V. M. Abazov \emph{et
  al.}}}} (\bibinfo {collaboration} {D0 Collaboration}),\ }\href {\doibase
  10.1103/PhysRevLett.101.211801} {\bibfield  {journal} {\bibinfo  {journal}
  {Phys. Rev. Lett.}\ }\textbf {\bibinfo {volume} {101}},\ \bibinfo {pages}
  {211801} (\bibinfo {year} {2008}{\natexlab{b}})}\BibitemShut {NoStop}%
\bibitem [{\citenamefont {{V. M. Abazov \emph{et al.}}}(2013)}]{D0196Am2}%
  \BibitemOpen
  \bibfield  {author} {\bibinfo {author} {\bibnamefont {{V. M. Abazov \emph{et
  al.}}}} (\bibinfo {collaboration} {D0 Collaboration}),\ }\href {\doibase
  10.1103/PhysRevD.88.091102} {\bibfield  {journal} {\bibinfo  {journal} {Phys.
  Rev. D}\ }\textbf {\bibinfo {volume} {88}},\ \bibinfo {pages} {091102(R)}
  (\bibinfo {year} {2013})}\BibitemShut {NoStop}%
\bibitem [{\citenamefont {{V. M. Abazov \emph{et
  al.}}}(2015{\natexlab{a}})}]{D0196Ae2}%
  \BibitemOpen
  \bibfield  {author} {\bibinfo {author} {\bibnamefont {{V. M. Abazov \emph{et
  al.}}}} (\bibinfo {collaboration} {D0 Collaboration}),\ }\href {\doibase
  10.1103/PhysRevD.91.032007} {\bibfield  {journal} {\bibinfo  {journal} {Phys.
  Rev. D}\ }\textbf {\bibinfo {volume} {91}},\ \bibinfo {pages} {032007}
  (\bibinfo {year} {2015}{\natexlab{a}})}\BibitemShut {NoStop}%
\bibitem [{\citenamefont {{V. M. Abazov \emph{et
  al.}}}(2015{\natexlab{b}})}]{D0196Ae2E}%
  \BibitemOpen
  \bibfield  {author} {\bibinfo {author} {\bibnamefont {{V. M. Abazov \emph{et
  al.}}}} (\bibinfo {collaboration} {D0 Collaboration}),\ }\href {\doibase
  10.1103/PhysRevD.91.079901} {\bibfield  {journal} {\bibinfo  {journal} {Phys.
  Rev. D}\ }\textbf {\bibinfo {volume} {91}},\ \bibinfo {pages} {079901}
  (\bibinfo {year} {2015}{\natexlab{b}})}\BibitemShut {NoStop}%
\bibitem [{\citenamefont {{T. Aaltonen \emph{et al.}}}(2009)}]{CDF196Awe1}%
  \BibitemOpen
  \bibfield  {author} {\bibinfo {author} {\bibnamefont {{T. Aaltonen \emph{et
  al.}}}} (\bibinfo {collaboration} {CDF Collaboration}),\ }\href {\doibase
  10.1103/PhysRevLett.102.181801} {\bibfield  {journal} {\bibinfo  {journal}
  {Phys. Rev. Lett.}\ }\textbf {\bibinfo {volume} {102}},\ \bibinfo {pages}
  {181801} (\bibinfo {year} {2009})}\BibitemShut {NoStop}%
\bibitem [{\citenamefont {{V. M. Abazov \emph{et al.}}}(2014)}]{D0196Awe1}%
  \BibitemOpen
  \bibfield  {author} {\bibinfo {author} {\bibnamefont {{V. M. Abazov \emph{et
  al.}}}} (\bibinfo {collaboration} {D0 Collaboration}),\ }\href {\doibase
  10.1103/PhysRevLett.112.151803} {\bibfield  {journal} {\bibinfo  {journal}
  {Phys. Rev. Lett.}\ }\textbf {\bibinfo {volume} {112}},\ \bibinfo {pages}
  {151803} (\bibinfo {year} {2014})}\BibitemShut {NoStop}%
\bibitem [{\citenamefont {{V. M. Abazov \emph{et
  al.}}}(2015{\natexlab{c}})}]{D0196Awe1E}%
  \BibitemOpen
  \bibfield  {author} {\bibinfo {author} {\bibnamefont {{V. M. Abazov \emph{et
  al.}}}} (\bibinfo {collaboration} {D0 Collaboration}),\ }\href {\doibase
  10.1103/PhysRevLett.114.049901} {\bibfield  {journal} {\bibinfo  {journal}
  {Phys. Rev. Lett.}\ }\textbf {\bibinfo {volume} {114}},\ \bibinfo {pages}
  {049901} (\bibinfo {year} {2015}{\natexlab{c}})}\BibitemShut {NoStop}%
\bibitem [{\citenamefont {Bodek}\ \emph {et~al.}(2008)\citenamefont {Bodek},
  \citenamefont {Chung}, \citenamefont {Han}, \citenamefont {McFarland},\ and\
  \citenamefont {Halkiadakis}}]{WasyNuWeights}%
  \BibitemOpen
  \bibfield  {author} {\bibinfo {author} {\bibfnamefont {A.}~\bibnamefont
  {Bodek}}, \bibinfo {author} {\bibfnamefont {Y.}~\bibnamefont {Chung}},
  \bibinfo {author} {\bibfnamefont {B.-Y.}\ \bibnamefont {Han}}, \bibinfo
  {author} {\bibfnamefont {K.}~\bibnamefont {McFarland}}, and\ \bibinfo
  {author} {\bibfnamefont {E.}~\bibnamefont {Halkiadakis}},\ }\href {\doibase
  10.1103/PhysRevD.77.111301} {\bibfield  {journal} {\bibinfo  {journal} {Phys.
  Rev. D}\ }\textbf {\bibinfo {volume} {77}},\ \bibinfo {pages} {111301(R)}
  (\bibinfo {year} {2008})}\BibitemShut {NoStop}%
\bibitem [{\citenamefont {{G. Aad \emph{et al.}}}(2011)}]{ATLAS7Am1}%
  \BibitemOpen
  \bibfield  {author} {\bibinfo {author} {\bibnamefont {{G. Aad \emph{et
  al.}}}} (\bibinfo {collaboration} {ATLAS Collaboration}),\ }\href {\doibase
  10.1016/j.physletb.2011.05.024} {\bibfield  {journal} {\bibinfo  {journal}
  {Phys. Lett. B}\ }\textbf {\bibinfo {volume} {701}},\ \bibinfo {pages} {31}
  (\bibinfo {year} {2011})}\BibitemShut {NoStop}%
\bibitem [{\citenamefont {{M. Aaboud \emph{et al.}}}(2017)}]{ATLAS7Aem1}%
  \BibitemOpen
  \bibfield  {author} {\bibinfo {author} {\bibnamefont {{M. Aaboud \emph{et
  al.}}}} (\bibinfo {collaboration} {ATLAS Collaboration}),\ }\href {\doibase
  10.1140/epjc/s10052-017-4911-9} {\bibfield  {journal} {\bibinfo  {journal}
  {Eur. Phys. J. C}\ }\textbf {\bibinfo {volume} {77}},\ \bibinfo {pages} {367}
  (\bibinfo {year} {2017})}\BibitemShut {NoStop}%
\bibitem [{\citenamefont {{S. Chatrchyan \emph{et al.}}}(2011)}]{CMS7Aem1}%
  \BibitemOpen
  \bibfield  {author} {\bibinfo {author} {\bibnamefont {{S. Chatrchyan \emph{et
  al.}}}} (\bibinfo {collaboration} {CMS Collaboration}),\ }\href {\doibase
  10.1007/JHEP04(2011)050} {\bibfield  {journal} {\bibinfo  {journal} {J. High
  Energy Phys.}\ }\bibinfo {volume} {2011} (\bibinfo {year} {2011})\ \bibinfo
  {pages} {50}}\BibitemShut {NoStop}%
\bibitem [{\citenamefont {{S. Chatrchyan \emph{et al.}}}(2012)}]{CMS7Ae1}%
  \BibitemOpen
  \bibfield  {author} {\bibinfo {author} {\bibnamefont {{S. Chatrchyan \emph{et
  al.}}}} (\bibinfo {collaboration} {CMS Collaboration}),\ }\href {\doibase
  10.1103/PhysRevLett.109.111806} {\bibfield  {journal} {\bibinfo  {journal}
  {Phys. Rev. Lett.}\ }\textbf {\bibinfo {volume} {109}},\ \bibinfo {pages}
  {111806} (\bibinfo {year} {2012})}\BibitemShut {NoStop}%
\bibitem [{\citenamefont {{S. Chatrchyan \emph{et al.}}}(2014)}]{CMS7Am1}%
  \BibitemOpen
  \bibfield  {author} {\bibinfo {author} {\bibnamefont {{S. Chatrchyan \emph{et
  al.}}}} (\bibinfo {collaboration} {CMS Collaboration}),\ }\href {\doibase
  10.1103/PhysRevD.90.032004} {\bibfield  {journal} {\bibinfo  {journal} {Phys.
  Rev. D}\ }\textbf {\bibinfo {volume} {90}},\ \bibinfo {pages} {032004}
  (\bibinfo {year} {2014})}\BibitemShut {NoStop}%
\bibitem [{\citenamefont {{V. Khachatryan \emph{et al.}}}(2016)}]{CMS8Am1}%
  \BibitemOpen
  \bibfield  {author} {\bibinfo {author} {\bibnamefont {{V. Khachatryan
  \emph{et al.}}}} (\bibinfo {collaboration} {CMS Collaboration}),\ }\href
  {\doibase 10.1140/epjc/s10052-016-4293-4} {\bibfield  {journal} {\bibinfo
  {journal} {Eur. Phys. J. C}\ }\textbf {\bibinfo {volume} {76}},\ \bibinfo
  {pages} {469} (\bibinfo {year} {2016})}\BibitemShut {NoStop}%
\bibitem [{\citenamefont {{R. Aaij \emph{et al.}}}(2014)}]{LHCbAmu7}%
  \BibitemOpen
  \bibfield  {author} {\bibinfo {author} {\bibnamefont {{R. Aaij \emph{et
  al.}}}} (\bibinfo {collaboration} {LHCb Collaboration}),\ }\href {\doibase
  10.1007/JHEP12(2014)079} {\bibfield  {journal} {\bibinfo  {journal} {J. High
  Energy Phys.}\ }\bibinfo {volume} {2014} (\bibinfo {year} {2014})\ \bibinfo
  {pages} {079}}\BibitemShut {NoStop}%
\bibitem [{\citenamefont {{R. Aaij \emph{et
  al.}}}(2016{\natexlab{a}})}]{LHCbAmel8}%
  \BibitemOpen
  \bibfield  {author} {\bibinfo {author} {\bibnamefont {{R. Aaij \emph{et
  al.}}}} (\bibinfo {collaboration} {LHCb Collaboration}),\ }\href {\doibase
  10.1007/JHEP10(2016)030} {\bibfield  {journal} {\bibinfo  {journal} {J. High
  Energy Phys.}\ }\bibinfo {volume} {2016} (\bibinfo {year}
  {2016}{\natexlab{a}})\ \bibinfo {pages} {030}}\BibitemShut {NoStop}%
\bibitem [{\citenamefont {{R. Aaij \emph{et
  al.}}}(2016{\natexlab{b}})}]{LHCbAmujet8}%
  \BibitemOpen
  \bibfield  {author} {\bibinfo {author} {\bibnamefont {{R. Aaij \emph{et
  al.}}}} (\bibinfo {collaboration} {LHCb Collaboration}),\ }\href {\doibase
  10.1007/JHEP05(2016)131} {\bibfield  {journal} {\bibinfo  {journal} {J. High
  Energy Phys.}\ }\bibinfo {volume} {2016} (\bibinfo {year}
  {2016}{\natexlab{b}})\ \bibinfo {pages} {131}}\BibitemShut {NoStop}%
\bibitem [{\citenamefont {Collins}\ and\ \citenamefont
  {Soper}(1977)}]{CollinsSoperFrame}%
  \BibitemOpen
  \bibfield  {author} {\bibinfo {author} {\bibfnamefont {J.~C.}\ \bibnamefont
  {Collins}}\ and\ \bibinfo {author} {\bibfnamefont {D.~E.}\ \bibnamefont
  {Soper}},\ }\href {\doibase 10.1103/PhysRevD.16.2219} {\bibfield  {journal}
  {\bibinfo  {journal} {Phys. Rev. D}\ }\textbf {\bibinfo {volume} {16}},\
  \bibinfo {pages} {2219} (\bibinfo {year} {1977})}\BibitemShut {NoStop}%
\bibitem [{\citenamefont {Mirkes}(1992)}]{MirkesA0to7a}%
  \BibitemOpen
  \bibfield  {author} {\bibinfo {author} {\bibfnamefont {E.}~\bibnamefont
  {Mirkes}},\ }\href {\doibase 10.1016/0550-3213(92)90046-E} {\bibfield
  {journal} {\bibinfo  {journal} {Nucl. Phys.}\ }\textbf {\bibinfo {volume}
  {B387}},\ \bibinfo {pages} {3} (\bibinfo {year} {1992})}\BibitemShut
  {NoStop}%
\bibitem [{\citenamefont {Mirkes}\ and\ \citenamefont
  {Ohnemus}(1994)}]{MirkesA0to7b}%
  \BibitemOpen
  \bibfield  {author} {\bibinfo {author} {\bibfnamefont {E.}~\bibnamefont
  {Mirkes}}\ and\ \bibinfo {author} {\bibfnamefont {J.}~\bibnamefont
  {Ohnemus}},\ }\href {\doibase 10.1103/PhysRevD.50.5692} {\bibfield  {journal}
  {\bibinfo  {journal} {Phys. Rev. D}\ }\textbf {\bibinfo {volume} {50}},\
  \bibinfo {pages} {5692} (\bibinfo {year} {1994})}\BibitemShut {NoStop}%
\bibitem [{\citenamefont {Frixione}\ \emph {et~al.}(2007)\citenamefont
  {Frixione}, \citenamefont {Nason},\ and\ \citenamefont
  {Oleari}}]{Powheg-Box}%
  \BibitemOpen
  \bibfield  {author} {\bibinfo {author} {\bibfnamefont {S.}~\bibnamefont
  {Frixione}}, \bibinfo {author} {\bibfnamefont {P.}~\bibnamefont {Nason}},
  and\ \bibinfo {author} {\bibfnamefont {C.}~\bibnamefont {Oleari}},\ }\href
  {\doibase 10.1088/1126-6708/2007/11/070} {\bibfield  {journal} {\bibinfo
  {journal} {J. High Energy Phys.}\ }\bibinfo {volume} {11} (\bibinfo {year}
  {2007})\ \bibinfo {pages} {070}}\BibitemShut {NoStop}%
\bibitem [{\citenamefont {Alioli}\ \emph {et~al.}(2008)\citenamefont {Alioli},
  \citenamefont {Nason}, \citenamefont {Oleari},\ and\ \citenamefont
  {Re}}]{PowhegBoxVBP}%
  \BibitemOpen
  \bibfield  {author} {\bibinfo {author} {\bibfnamefont {S.}~\bibnamefont
  {Alioli}}, \bibinfo {author} {\bibfnamefont {P.}~\bibnamefont {Nason}},
  \bibinfo {author} {\bibfnamefont {C.}~\bibnamefont {Oleari}}, and\ \bibinfo
  {author} {\bibfnamefont {E.}~\bibnamefont {Re}},\ }\href {\doibase
  10.1088/1126-6708/2008/07/060} {\bibfield  {journal} {\bibinfo  {journal} {J.
  High Energy Phys.}\ }\bibinfo {volume} {07} (\bibinfo {year} {2008})\
  \bibinfo {pages} {060}}\BibitemShut {NoStop}%
\bibitem [{\citenamefont {Sudakov}(1956{\natexlab{a}})}]{Sudakov-FFeng}%
  \BibitemOpen
  \bibfield  {author} {\bibinfo {author} {\bibfnamefont {V.~V.}\ \bibnamefont
  {Sudakov}},\ }\href@noop {} {\bibfield  {journal} {\bibinfo  {journal} {Sov.
  Phys. JETP}\ }\textbf {\bibinfo {volume} {3}},\ \bibinfo {pages} {65}
  (\bibinfo {year} {1956}{\natexlab{a}})}\BibitemShut {NoStop}%
\bibitem [{\citenamefont {Sudakov}(1956{\natexlab{b}})}]{Sudakov-FFrus}%
  \BibitemOpen
  \bibfield  {author} {\bibinfo {author} {\bibfnamefont {V.~V.}\ \bibnamefont
  {Sudakov}},\ }\href@noop {} {\bibfield  {journal} {\bibinfo  {journal} {Zh.
  Eksp. Teor. Fiz.}\ }\textbf {\bibinfo {volume} {30}},\ \bibinfo {pages} {87}
  (\bibinfo {year} {1956}{\natexlab{b}})}\BibitemShut {NoStop}%
\bibitem [{\citenamefont {Sj\"{o}strand}\ \emph {et~al.}(2006)\citenamefont
  {Sj\"{o}strand}, \citenamefont {Mrenna},\ and\ \citenamefont
  {Skands}}]{pythia64}%
  \BibitemOpen
  \bibfield  {author} {\bibinfo {author} {\bibfnamefont {T.}~\bibnamefont
  {Sj\"{o}strand}}, \bibinfo {author} {\bibfnamefont {S.}~\bibnamefont
  {Mrenna}}, and\ \bibinfo {author} {\bibfnamefont {P.~Z.}\ \bibnamefont
  {Skands}},\ }\href {\doibase 10.1088/1126-6708/2006/05/026} {\bibfield
  {journal} {\bibinfo  {journal} {J. High Energy Phys.}\ }\bibinfo {volume}
  {05} (\bibinfo {year} {2006})\ \bibinfo {pages} {026}}\BibitemShut {NoStop}%
\bibitem [{\citenamefont {{R. D. Ball \emph{et al.}}}(2015)}]{nnpdf301}%
  \BibitemOpen
  \bibfield  {author} {\bibinfo {author} {\bibnamefont {{R. D. Ball \emph{et
  al.}}}} (\bibinfo {collaboration} {NNPDF Collaboration}),\ }\href {\doibase
  10.1007/JHEP04(2015)040} {\bibfield  {journal} {\bibinfo  {journal} {J. High
  Energy Phys.}\ }\bibinfo {volume} {04} (\bibinfo {year} {2015})\ \bibinfo
  {pages} {040}}\BibitemShut {NoStop}%
\bibitem [{\citenamefont {{R. D. Ball \emph{et al.}}}(2013)}]{nnpdf302}%
  \BibitemOpen
  \bibfield  {author} {\bibinfo {author} {\bibnamefont {{R. D. Ball \emph{et
  al.}}}} (\bibinfo {collaboration} {NNPDF Collaboration}),\ }\href {\doibase
  10.1016/j.nuclphysb.2012.10.003} {\bibfield  {journal} {\bibinfo  {journal}
  {Nucl. Phys.}\ }\textbf {\bibinfo {volume} {B867}},\ \bibinfo {pages} {244}
  (\bibinfo {year} {2013})}\BibitemShut {NoStop}%
\bibitem [{\citenamefont {Ball}\ \emph {et~al.}(2012)\citenamefont {Ball},
  \citenamefont {Bertone}, \citenamefont {Cerutti}, \citenamefont {Debbio},
  \citenamefont {Forte}, \citenamefont {Guffanti}, \citenamefont {Latorre},
  \citenamefont {Rojo},\ and\ \citenamefont {Ubiali}}]{nnpdf303}%
  \BibitemOpen
  \bibfield  {author} {\bibinfo {author} {\bibfnamefont {R.~D.}\ \bibnamefont
  {Ball}}, \bibinfo {author} {\bibfnamefont {V.}~\bibnamefont {Bertone}},
  \bibinfo {author} {\bibfnamefont {F.}~\bibnamefont {Cerutti}}, \bibinfo
  {author} {\bibfnamefont {L.~D.}\ \bibnamefont {Debbio}}, \bibinfo {author}
  {\bibfnamefont {S.}~\bibnamefont {Forte}}, \bibinfo {author} {\bibfnamefont
  {A.}~\bibnamefont {Guffanti}}, \bibinfo {author} {\bibfnamefont {J.~I.}\
  \bibnamefont {Latorre}}, \bibinfo {author} {\bibfnamefont {J.}~\bibnamefont
  {Rojo}}, and\ \bibinfo {author} {\bibfnamefont {M.}~\bibnamefont {Ubiali}}
  (\bibinfo {collaboration} {NNPDF Collaboration}),\ }\href {\doibase
  10.1016/j.nuclphysb.2011.09.024} {\bibfield  {journal} {\bibinfo  {journal}
  {Nucl. Phys.}\ }\textbf {\bibinfo {volume} {B855}},\ \bibinfo {pages} {153}
  (\bibinfo {year} {2012})}\BibitemShut {NoStop}%
\bibitem [{\citenamefont {Ball}\ \emph {et~al.}(2011)\citenamefont {Ball},
  \citenamefont {Bertone}, \citenamefont {Cerutti}, \citenamefont {Debbio},
  \citenamefont {Forte}, \citenamefont {Guffanti}, \citenamefont {Latorre},
  \citenamefont {Rojo},\ and\ \citenamefont {Ubiali}}]{nnpdf304}%
  \BibitemOpen
  \bibfield  {author} {\bibinfo {author} {\bibfnamefont {R.~D.}\ \bibnamefont
  {Ball}}, \bibinfo {author} {\bibfnamefont {V.}~\bibnamefont {Bertone}},
  \bibinfo {author} {\bibfnamefont {F.}~\bibnamefont {Cerutti}}, \bibinfo
  {author} {\bibfnamefont {L.~D.}\ \bibnamefont {Debbio}}, \bibinfo {author}
  {\bibfnamefont {S.}~\bibnamefont {Forte}}, \bibinfo {author} {\bibfnamefont
  {A.}~\bibnamefont {Guffanti}}, \bibinfo {author} {\bibfnamefont {J.~I.}\
  \bibnamefont {Latorre}}, \bibinfo {author} {\bibfnamefont {J.}~\bibnamefont
  {Rojo}}, and\ \bibinfo {author} {\bibfnamefont {M.}~\bibnamefont {Ubiali}}
  (\bibinfo {collaboration} {NNPDF Collaboration}),\ }\href {\doibase
  10.1016/j.nuclphysb.2011.09.024} {\bibfield  {journal} {\bibinfo  {journal}
  {Nucl. Phys.}\ }\textbf {\bibinfo {volume} {B849}},\ \bibinfo {pages} {296}
  (\bibinfo {year} {2011})}\BibitemShut {NoStop}%
\bibitem [{\citenamefont {Ball}\ \emph {et~al.}(2010)\citenamefont {Ball},
  \citenamefont {Debbio}, \citenamefont {Forte}, \citenamefont {Guffanti},
  \citenamefont {Latorre}, \citenamefont {Rojo},\ and\ \citenamefont
  {Ubiali}}]{nnpdf305}%
  \BibitemOpen
  \bibfield  {author} {\bibinfo {author} {\bibfnamefont {R.~D.}\ \bibnamefont
  {Ball}}, \bibinfo {author} {\bibfnamefont {L.~D.}\ \bibnamefont {Debbio}},
  \bibinfo {author} {\bibfnamefont {S.}~\bibnamefont {Forte}}, \bibinfo
  {author} {\bibfnamefont {A.}~\bibnamefont {Guffanti}}, \bibinfo {author}
  {\bibfnamefont {J.~I.}\ \bibnamefont {Latorre}}, \bibinfo {author}
  {\bibfnamefont {J.}~\bibnamefont {Rojo}}, and\ \bibinfo {author}
  {\bibfnamefont {M.}~\bibnamefont {Ubiali}} (\bibinfo {collaboration} {NNPDF
  Collaboration}),\ }\href {\doibase 10.1016/j.nuclphysb.2010.05.008}
  {\bibfield  {journal} {\bibinfo  {journal} {Nucl. Phys.}\ }\textbf {\bibinfo
  {volume} {B838}},\ \bibinfo {pages} {136} (\bibinfo {year}
  {2010})}\BibitemShut {NoStop}%
\bibitem [{\citenamefont {Ball}\ \emph
  {et~al.}(2009{\natexlab{a}})\citenamefont {Ball}, \citenamefont {Debbio},
  \citenamefont {Forte}, \citenamefont {Guffanti}, \citenamefont {Latorre},
  \citenamefont {Piccione}, \citenamefont {Rojo},\ and\ \citenamefont
  {Ubiali}}]{nnpdf306}%
  \BibitemOpen
  \bibfield  {author} {\bibinfo {author} {\bibfnamefont {R.~D.}\ \bibnamefont
  {Ball}}, \bibinfo {author} {\bibfnamefont {L.~D.}\ \bibnamefont {Debbio}},
  \bibinfo {author} {\bibfnamefont {S.}~\bibnamefont {Forte}}, \bibinfo
  {author} {\bibfnamefont {A.}~\bibnamefont {Guffanti}}, \bibinfo {author}
  {\bibfnamefont {J.~I.}\ \bibnamefont {Latorre}}, \bibinfo {author}
  {\bibfnamefont {A.}~\bibnamefont {Piccione}}, \bibinfo {author}
  {\bibfnamefont {J.}~\bibnamefont {Rojo}}, and\ \bibinfo {author}
  {\bibfnamefont {M.}~\bibnamefont {Ubiali}} (\bibinfo {collaboration} {NNPDF
  Collaboration}),\ }\href {\doibase 10.1016/j.nuclphysb.2009.02.027}
  {\bibfield  {journal} {\bibinfo  {journal} {Nucl. Phys.}\ }\textbf {\bibinfo
  {volume} {B809}},\ \bibinfo {pages} {1} (\bibinfo {year}
  {2009}{\natexlab{a}})}\BibitemShut {NoStop}%
\bibitem [{\citenamefont {Ball}\ \emph
  {et~al.}(2009{\natexlab{b}})\citenamefont {Ball}, \citenamefont {Debbio},
  \citenamefont {Forte}, \citenamefont {Guffanti}, \citenamefont {Latorre},
  \citenamefont {Piccione}, \citenamefont {Rojo},\ and\ \citenamefont
  {Ubiali}}]{nnpdf306e}%
  \BibitemOpen
  \bibfield  {author} {\bibinfo {author} {\bibfnamefont {R.~D.}\ \bibnamefont
  {Ball}}, \bibinfo {author} {\bibfnamefont {L.~D.}\ \bibnamefont {Debbio}},
  \bibinfo {author} {\bibfnamefont {S.}~\bibnamefont {Forte}}, \bibinfo
  {author} {\bibfnamefont {A.}~\bibnamefont {Guffanti}}, \bibinfo {author}
  {\bibfnamefont {J.~I.}\ \bibnamefont {Latorre}}, \bibinfo {author}
  {\bibfnamefont {A.}~\bibnamefont {Piccione}}, \bibinfo {author}
  {\bibfnamefont {J.}~\bibnamefont {Rojo}}, and\ \bibinfo {author}
  {\bibfnamefont {M.}~\bibnamefont {Ubiali}} (\bibinfo {collaboration} {NNPDF
  Collaboration}),\ }\href {\doibase 10.1016/j.nuclphysb.2008.09.037}
  {\bibfield  {journal} {\bibinfo  {journal} {Nucl. Phys.}\ }\textbf {\bibinfo
  {volume} {B816}},\ \bibinfo {pages} {293} (\bibinfo {year}
  {2009}{\natexlab{b}})}\BibitemShut {NoStop}%
\bibitem [{\citenamefont {Forte}\ \emph {et~al.}(2002)\citenamefont {Forte},
  \citenamefont {Garrido}, \citenamefont {Latorre},\ and\ \citenamefont
  {Piccione}}]{nnpdf307}%
  \BibitemOpen
  \bibfield  {author} {\bibinfo {author} {\bibfnamefont {S.}~\bibnamefont
  {Forte}}, \bibinfo {author} {\bibfnamefont {L.}~\bibnamefont {Garrido}},
  \bibinfo {author} {\bibfnamefont {J.~I.}\ \bibnamefont {Latorre}}, and\
  \bibinfo {author} {\bibfnamefont {A.}~\bibnamefont {Piccione}},\ }\href
  {\doibase 10.1088/1126-6708/2002/05/062} {\bibfield  {journal} {\bibinfo
  {journal} {J. High Energy Phys.}\ }\bibinfo {volume} {05} (\bibinfo {year}
  {2002})\ \bibinfo {pages} {062}}\BibitemShut {NoStop}%
\bibitem [{\citenamefont {Ladinsky}\ and\ \citenamefont
  {Yuan}(1994)}]{ResBos1}%
  \BibitemOpen
  \bibfield  {author} {\bibinfo {author} {\bibfnamefont {G.~A.}\ \bibnamefont
  {Ladinsky}}\ and\ \bibinfo {author} {\bibfnamefont {C.-P.}\ \bibnamefont
  {Yuan}},\ }\href {\doibase 10.1103/PhysRevD.50.R4239} {\bibfield  {journal}
  {\bibinfo  {journal} {Phys. Rev. D}\ }\textbf {\bibinfo {volume} {50}},\
  \bibinfo {pages} {R4239} (\bibinfo {year} {1994})}\BibitemShut {NoStop}%
\bibitem [{\citenamefont {Bal\`{a}zs}\ and\ \citenamefont
  {Yuan}(1997)}]{ResBos2}%
  \BibitemOpen
  \bibfield  {author} {\bibinfo {author} {\bibfnamefont {C.}~\bibnamefont
  {Bal\`{a}zs}}\ and\ \bibinfo {author} {\bibfnamefont {C.-P.}\ \bibnamefont
  {Yuan}},\ }\href {\doibase 10.1103/PhysRevD.56.5558} {\bibfield  {journal}
  {\bibinfo  {journal} {Phys. Rev. D}\ }\textbf {\bibinfo {volume} {56}},\
  \bibinfo {pages} {5558} (\bibinfo {year} {1997})}\BibitemShut {NoStop}%
\bibitem [{\citenamefont {Landry}\ \emph {et~al.}(2003)\citenamefont {Landry},
  \citenamefont {Brock}, \citenamefont {Nadolsky},\ and\ \citenamefont
  {Yuan}}]{ResBos3}%
  \BibitemOpen
  \bibfield  {author} {\bibinfo {author} {\bibfnamefont {F.}~\bibnamefont
  {Landry}}, \bibinfo {author} {\bibfnamefont {R.}~\bibnamefont {Brock}},
  \bibinfo {author} {\bibfnamefont {P.~M.}\ \bibnamefont {Nadolsky}}, and\
  \bibinfo {author} {\bibfnamefont {C.-P.}\ \bibnamefont {Yuan}},\ }\href
  {\doibase 10.1103/PhysRevD.67.073016} {\bibfield  {journal} {\bibinfo
  {journal} {Phys. Rev. D}\ }\textbf {\bibinfo {volume} {67}},\ \bibinfo
  {pages} {073016} (\bibinfo {year} {2003})}\BibitemShut {NoStop}%
\bibitem [{\citenamefont {Konychev}\ and\ \citenamefont
  {Nadolsky}(2006)}]{ResBosc221}%
  \BibitemOpen
  \bibfield  {author} {\bibinfo {author} {\bibfnamefont {A.}~\bibnamefont
  {Konychev}}\ and\ \bibinfo {author} {\bibfnamefont {P.}~\bibnamefont
  {Nadolsky}},\ }\href {\doibase 10.1016/j.physletb.2005.12.063} {\bibfield
  {journal} {\bibinfo  {journal} {Phys. Lett. B}\ }\textbf {\bibinfo {volume}
  {633}},\ \bibinfo {pages} {710} (\bibinfo {year} {2006})}\BibitemShut
  {NoStop}%
\bibitem [{\citenamefont {{P. M. Nadolsky \emph{et al.}}}(2008)}]{Cteq66pdf}%
  \BibitemOpen
  \bibfield  {author} {\bibinfo {author} {\bibnamefont {{P. M. Nadolsky
  \emph{et al.}}}} (\bibinfo {collaboration} {CTEQ Collaboration}),\ }\href
  {\doibase 10.1103/PhysRevD.78.013004} {\bibfield  {journal} {\bibinfo
  {journal} {Phys. Rev. D}\ }\textbf {\bibinfo {volume} {78}},\ \bibinfo
  {pages} {013004} (\bibinfo {year} {2008})}\BibitemShut {NoStop}%
\bibitem [{\citenamefont {Collins}\ \emph {et~al.}(1985)\citenamefont
  {Collins}, \citenamefont {Soper},\ and\ \citenamefont {Sterman}}]{methodCSS}%
  \BibitemOpen
  \bibfield  {author} {\bibinfo {author} {\bibfnamefont {J.~C.}\ \bibnamefont
  {Collins}}, \bibinfo {author} {\bibfnamefont {D.~E.}\ \bibnamefont {Soper}},
  and\ \bibinfo {author} {\bibfnamefont {G.}~\bibnamefont {Sterman}},\ }\href
  {\doibase 10.1016/0550-3213(85)90479-1} {\bibfield  {journal} {\bibinfo
  {journal} {Nucl. Phys.}\ }\textbf {\bibinfo {volume} {B250}},\ \bibinfo
  {pages} {199} (\bibinfo {year} {1985})}\BibitemShut {NoStop}%
\bibitem [{\citenamefont {Collins}\ and\ \citenamefont
  {Soper}(1981)}]{wfactorCSS1}%
  \BibitemOpen
  \bibfield  {author} {\bibinfo {author} {\bibfnamefont {J.~C.}\ \bibnamefont
  {Collins}}\ and\ \bibinfo {author} {\bibfnamefont {D.~E.}\ \bibnamefont
  {Soper}},\ }\href {\doibase 10.1016/0550-3213(81)90339-4} {\bibfield
  {journal} {\bibinfo  {journal} {Nucl. Phys.}\ }\textbf {\bibinfo {volume}
  {B193}},\ \bibinfo {pages} {381} (\bibinfo {year} {1981})}\BibitemShut
  {NoStop}%
\bibitem [{\citenamefont {Collins}\ and\ \citenamefont
  {Soper}(1982)}]{wfactorCSS2}%
  \BibitemOpen
  \bibfield  {author} {\bibinfo {author} {\bibfnamefont {J.~C.}\ \bibnamefont
  {Collins}}\ and\ \bibinfo {author} {\bibfnamefont {D.~E.}\ \bibnamefont
  {Soper}},\ }\href {\doibase 10.1016/0550-3213(82)90453-9} {\bibfield
  {journal} {\bibinfo  {journal} {Nucl. Phys.}\ }\textbf {\bibinfo {volume}
  {B197}},\ \bibinfo {pages} {446} (\bibinfo {year} {1982})}\BibitemShut
  {NoStop}%
\bibitem [{\citenamefont {Collins}\ and\ \citenamefont
  {Soper}(1983)}]{wfactorCSS3}%
  \BibitemOpen
  \bibfield  {author} {\bibinfo {author} {\bibfnamefont {J.~C.}\ \bibnamefont
  {Collins}}\ and\ \bibinfo {author} {\bibfnamefont {D.~E.}\ \bibnamefont
  {Soper}},\ }\href {\doibase 10.1016/0550-3213(83)90235-3} {\bibfield
  {journal} {\bibinfo  {journal} {Nucl. Phys.}\ }\textbf {\bibinfo {volume}
  {B213}},\ \bibinfo {pages} {545(E)} (\bibinfo {year} {1983})}\BibitemShut
  {NoStop}%
\bibitem [{\citenamefont {{T. Aaltonen \emph{et al.}}}(2012)}]{zpt21}%
  \BibitemOpen
  \bibfield  {author} {\bibinfo {author} {\bibnamefont {{T. Aaltonen \emph{et
  al.}}}} (\bibinfo {collaboration} {CDF Collaboration}),\ }\href {\doibase
  10.1103/PhysRevD.86.052010} {\bibfield  {journal} {\bibinfo  {journal} {Phys.
  Rev. D}\ }\textbf {\bibinfo {volume} {86}},\ \bibinfo {pages} {052010}
  (\bibinfo {year} {2012})}\BibitemShut {NoStop}%
\bibitem [{\citenamefont {{C. Patrignani \emph{et al.}}}(2016)}]{pdg2016}%
  \BibitemOpen
  \bibfield  {author} {\bibinfo {author} {\bibnamefont {{C. Patrignani \emph{et
  al.}}}} (\bibinfo {collaboration} {Particle Data Group}),\ }\href {\doibase
  10.1088/1674-1137/40/10/100001} {\bibfield  {journal} {\bibinfo  {journal}
  {Chin. Phys. C}\ }\textbf {\bibinfo {volume} {40}},\ \bibinfo {pages}
  {100001} (\bibinfo {year} {2016})}\BibitemShut {NoStop}%
\bibitem [{\citenamefont {Cabibbo}(1963)}]{C-ckm}%
  \BibitemOpen
  \bibfield  {author} {\bibinfo {author} {\bibfnamefont {N.}~\bibnamefont
  {Cabibbo}},\ }\href {\doibase 10.1103/PhysRevLett.10.531} {\bibfield
  {journal} {\bibinfo  {journal} {Phys. Rev. Lett.}\ }\textbf {\bibinfo
  {volume} {10}},\ \bibinfo {pages} {531} (\bibinfo {year} {1963})}\BibitemShut
  {NoStop}%
\bibitem [{\citenamefont {Kobayashi}\ and\ \citenamefont
  {Maskawa}(1973)}]{KM-ckm}%
  \BibitemOpen
  \bibfield  {author} {\bibinfo {author} {\bibfnamefont {M.}~\bibnamefont
  {Kobayashi}}\ and\ \bibinfo {author} {\bibfnamefont {T.}~\bibnamefont
  {Maskawa}},\ }\href {\doibase 10.1143/PTP.49.652} {\bibfield  {journal}
  {\bibinfo  {journal} {Prog. Theor. Phys.}\ }\textbf {\bibinfo {volume}
  {49}},\ \bibinfo {pages} {652} (\bibinfo {year} {1973})}\BibitemShut
  {NoStop}%
\bibitem [{\citenamefont {{A. Abulencia \emph{et al.}}}(2007)}]{refCDFII}%
  \BibitemOpen
  \bibfield  {author} {\bibinfo {author} {\bibnamefont {{A. Abulencia \emph{et
  al.}}}} (\bibinfo {collaboration} {CDF Collaboration}),\ }\href {\doibase
  10.1088/0954-3899/34/12/001} {\bibfield  {journal} {\bibinfo  {journal} {J.
  Phys. G: Nucl. Part. Phys.}\ }\textbf {\bibinfo {volume} {34}},\ \bibinfo
  {pages} {2457} (\bibinfo {year} {2007})}\BibitemShut {NoStop}%
\bibitem [{\citenamefont {{T. Affolder \emph{et al.}}}(2004)}]{refCOT}%
  \BibitemOpen
  \bibfield  {author} {\bibinfo {author} {\bibnamefont {{T. Affolder \emph{et
  al.}}}},\ }\href {\doibase 10.1016/j.nima.2004.02.020} {\bibfield  {journal}
  {\bibinfo  {journal} {Nucl. Instrum. Methods Phys. Res., Sect. A}\ }\textbf
  {\bibinfo {volume} {526}},\ \bibinfo {pages} {249} (\bibinfo {year}
  {2004})}\BibitemShut {NoStop}%
\bibitem [{\citenamefont {{T. Aaltonen \emph{et al.}}}(2013)}]{refSVXII}%
  \BibitemOpen
  \bibfield  {author} {\bibinfo {author} {\bibnamefont {{T. Aaltonen \emph{et
  al.}}}},\ }\href {\doibase 10.1016/j.nima.2013.07.015} {\bibfield  {journal}
  {\bibinfo  {journal} {Nucl. Instrum. Methods Phys. Res., Sect. A}\ }\textbf
  {\bibinfo {volume} {729}},\ \bibinfo {pages} {153} (\bibinfo {year}
  {2013})}\BibitemShut {NoStop}%
\bibitem [{\citenamefont {{L. Balka \emph{et al.}}}(1988)}]{refCEM}%
  \BibitemOpen
  \bibfield  {author} {\bibinfo {author} {\bibnamefont {{L. Balka \emph{et
  al.}}}},\ }\href {\doibase 10.1016/0168-9002(88)90474-3} {\bibfield
  {journal} {\bibinfo  {journal} {Nucl. Instrum. Methods Phys. Res., Sect. A}\
  }\textbf {\bibinfo {volume} {267}},\ \bibinfo {pages} {272} (\bibinfo {year}
  {1988})}\BibitemShut {NoStop}%
\bibitem [{\citenamefont {{S. Bertolucci \emph{et al.}}}(1988)}]{refChad}%
  \BibitemOpen
  \bibfield  {author} {\bibinfo {author} {\bibnamefont {{S. Bertolucci \emph{et
  al.}}}},\ }\href {\doibase 10.1016/0168-9002(88)90476-7} {\bibfield
  {journal} {\bibinfo  {journal} {Nucl. Instrum. Methods Phys. Res., Sect. A}\
  }\textbf {\bibinfo {volume} {267}},\ \bibinfo {pages} {301} (\bibinfo {year}
  {1988})}\BibitemShut {NoStop}%
\bibitem [{\citenamefont {{M. Albrow \emph{et al.}}}(2002)}]{refPEM}%
  \BibitemOpen
  \bibfield  {author} {\bibinfo {author} {\bibnamefont {{M. Albrow \emph{et
  al.}}}},\ }\href {\doibase 10.1016/S0168-9002(01)01238-4} {\bibfield
  {journal} {\bibinfo  {journal} {Nucl. Instrum. Methods Phys. Res., Sect. A}\
  }\textbf {\bibinfo {volume} {480}},\ \bibinfo {pages} {524} (\bibinfo {year}
  {2002})}\BibitemShut {NoStop}%
\bibitem [{\citenamefont {{G. Apollinari \emph{et al.}}}(1998)}]{refPES}%
  \BibitemOpen
  \bibfield  {author} {\bibinfo {author} {\bibnamefont {{G. Apollinari \emph{et
  al.}}}},\ }\href {\doibase 10.1016/S0168-9002(98)00286-1} {\bibfield
  {journal} {\bibinfo  {journal} {Nucl. Instrum. Methods Phys. Res., Sect. A}\
  }\textbf {\bibinfo {volume} {412}},\ \bibinfo {pages} {515} (\bibinfo {year}
  {1998})}\BibitemShut {NoStop}%
\bibitem [{\citenamefont {{P. de Barbaro}}(1995)}]{refPHA}%
  \BibitemOpen
  \bibfield  {author} {\bibinfo {author} {\bibnamefont {{P. de Barbaro}}},\
  }\href {\doibase 10.1109/23.467920} {\bibfield  {journal} {\bibinfo
  {journal} {IEEE Trans. Nucl. Sci.}\ }\textbf {\bibinfo {volume} {42}},\
  \bibinfo {pages} {510} (\bibinfo {year} {1995})}\BibitemShut {NoStop}%
\bibitem [{\citenamefont {{T. Aaltonen \emph{et al.}}}(2016)}]{cdfAfb9eeprd}%
  \BibitemOpen
  \bibfield  {author} {\bibinfo {author} {\bibnamefont {{T. Aaltonen \emph{et
  al.}}}} (\bibinfo {collaboration} {CDF Collaboration}),\ }\href {\doibase
  10.1103/PhysRevD.93.11201} {\bibfield  {journal} {\bibinfo  {journal} {Phys.
  Rev. D}\ }\textbf {\bibinfo {volume} {93}},\ \bibinfo {pages} {112016}
  (\bibinfo {year} {2016})}\BibitemShut {NoStop}%
\bibitem [{\citenamefont {{T. Aaltonen \emph{et
  al.}}}(2017)}]{cdfAfb9eeprdErr}%
  \BibitemOpen
  \bibfield  {author} {\bibinfo {author} {\bibnamefont {{T. Aaltonen \emph{et
  al.}}}} (\bibinfo {collaboration} {CDF Collaboration}),\ }\href {\doibase
  10.1103/PhysRevD.95.119901} {\bibfield  {journal} {\bibinfo  {journal} {Phys.
  Rev. D}\ }\textbf {\bibinfo {volume} {95}},\ \bibinfo {pages} {119901}
  (\bibinfo {year} {2017})}\BibitemShut {NoStop}%
\bibitem [{\citenamefont {Sj\"{o}strand}\ \emph {et~al.}(2001)\citenamefont
  {Sj\"{o}strand}, \citenamefont {Ed\'{e}n}, \citenamefont {L\"{o}nnblad},
  \citenamefont {Miu}, \citenamefont {Mrenna},\ and\ \citenamefont
  {Norrbin}}]{Pythia621}%
  \BibitemOpen
  \bibfield  {author} {\bibinfo {author} {\bibfnamefont {T.}~\bibnamefont
  {Sj\"{o}strand}}, \bibinfo {author} {\bibfnamefont {P.}~\bibnamefont
  {Ed\'{e}n}}, \bibinfo {author} {\bibfnamefont {L.}~\bibnamefont
  {L\"{o}nnblad}}, \bibinfo {author} {\bibfnamefont {G.}~\bibnamefont {Miu}},
  \bibinfo {author} {\bibfnamefont {S.}~\bibnamefont {Mrenna}}, and\ \bibinfo
  {author} {\bibfnamefont {E.}~\bibnamefont {Norrbin}},\ }\href {\doibase
  10.1016/S0010-4655(00)00236-8} {\bibfield  {journal} {\bibinfo  {journal}
  {Comput. Phys. Commun.}\ }\textbf {\bibinfo {volume} {135}},\ \bibinfo
  {pages} {238} (\bibinfo {year} {2001})}\BibitemShut {NoStop}%
\bibitem [{\citenamefont {{H. L. Lai \emph{et al.}}}(2000)}]{Cteq5pdf}%
  \BibitemOpen
  \bibfield  {author} {\bibinfo {author} {\bibnamefont {{H. L. Lai \emph{et
  al.}}}} (\bibinfo {collaboration} {CTEQ Collaboration}),\ }\href {\doibase
  10.1007/s100529900196} {\bibfield  {journal} {\bibinfo  {journal} {Eur. Phys.
  J. C}\ }\textbf {\bibinfo {volume} {12}},\ \bibinfo {pages} {375} (\bibinfo
  {year} {2000})}\BibitemShut {NoStop}%
\bibitem [{\citenamefont {{T. Affolder \emph{et al.}}}(2000)}]{run1CDF-Z}%
  \BibitemOpen
  \bibfield  {author} {\bibinfo {author} {\bibnamefont {{T. Affolder \emph{et
  al.}}}} (\bibinfo {collaboration} {CDF Collaboration}),\ }\href {\doibase
  10.1103/PhysRevLett.84.845} {\bibfield  {journal} {\bibinfo  {journal} {Phys.
  Rev. Lett.}\ }\textbf {\bibinfo {volume} {84}},\ \bibinfo {pages} {845}
  (\bibinfo {year} {2000})}\BibitemShut {NoStop}%
\bibitem [{\citenamefont {{M. Albrow \emph{et al.}}}()}]{PyTuneAW}%
  \BibitemOpen
  \bibfield  {author} {\bibinfo {author} {\bibnamefont {{M. Albrow \emph{et
  al.}}}} (\bibinfo {collaboration} {Tev4LHC QCD Working Group}),\ }\href@noop
  {} {}\Eprint {http://arxiv.org/abs/hep-ph/0610012} {arXiv:hep-ph/0610012}
  \BibitemShut {NoStop}%
\bibitem [{\citenamefont {Barberio}\ and\ \citenamefont
  {Was}(1994)}]{Photos20a}%
  \BibitemOpen
  \bibfield  {author} {\bibinfo {author} {\bibfnamefont {E.}~\bibnamefont
  {Barberio}}\ and\ \bibinfo {author} {\bibfnamefont {Z.}~\bibnamefont {Was}},\
  }\href {\doibase 10.1016/0010-4655(94)90074-4} {\bibfield  {journal}
  {\bibinfo  {journal} {Computer Phys. Comm.}\ }\textbf {\bibinfo {volume}
  {79}},\ \bibinfo {pages} {291} (\bibinfo {year} {1994})}\BibitemShut
  {NoStop}%
\bibitem [{\citenamefont {Barberio}\ \emph {et~al.}(1991)\citenamefont
  {Barberio}, \citenamefont {van Eijk},\ and\ \citenamefont {Was}}]{Photos20b}%
  \BibitemOpen
  \bibfield  {author} {\bibinfo {author} {\bibfnamefont {E.}~\bibnamefont
  {Barberio}}, \bibinfo {author} {\bibfnamefont {B.}~\bibnamefont {van Eijk}},
  and\ \bibinfo {author} {\bibfnamefont {Z.}~\bibnamefont {Was}},\ }\href
  {\doibase 10.1016/0010-4655(91)90012-A} {\bibfield  {journal} {\bibinfo
  {journal} {Computer Phys. Comm.}\ }\textbf {\bibinfo {volume} {66}},\
  \bibinfo {pages} {115} (\bibinfo {year} {1991})}\BibitemShut {NoStop}%
\bibitem [{\citenamefont {Golonka}\ and\ \citenamefont
  {Was}(2006)}]{Photos20c}%
  \BibitemOpen
  \bibfield  {author} {\bibinfo {author} {\bibfnamefont {P.}~\bibnamefont
  {Golonka}}\ and\ \bibinfo {author} {\bibfnamefont {Z.}~\bibnamefont {Was}},\
  }\href {\doibase 10.1140/epjc/s2005-02396-4} {\bibfield  {journal} {\bibinfo
  {journal} {Eur. Phys. J. C}\ }\textbf {\bibinfo {volume} {45}},\ \bibinfo
  {pages} {97} (\bibinfo {year} {2006})}\BibitemShut {NoStop}%
\bibitem [{\citenamefont {Grindhammer}\ \emph {et~al.}(1990)\citenamefont
  {Grindhammer}, \citenamefont {Rudowicz},\ and\ \citenamefont
  {Peters}}]{nimGflash}%
  \BibitemOpen
  \bibfield  {author} {\bibinfo {author} {\bibfnamefont {G.}~\bibnamefont
  {Grindhammer}}, \bibinfo {author} {\bibfnamefont {M.}~\bibnamefont
  {Rudowicz}}, and\ \bibinfo {author} {\bibfnamefont {S.}~\bibnamefont
  {Peters}},\ }\href {\doibase 10.1016/0168-9002(90)90566-O} {\bibfield
  {journal} {\bibinfo  {journal} {Nucl. Instrum. Methods Phys. Res., Sect. A}\
  }\textbf {\bibinfo {volume} {290}},\ \bibinfo {pages} {469} (\bibinfo {year}
  {1990})}\BibitemShut {NoStop}%
\bibitem [{\citenamefont {Bodek}\ \emph {et~al.}(2012)\citenamefont {Bodek},
  \citenamefont {van Dyne}, \citenamefont {Han}, \citenamefont {Sakumoto},\
  and\ \citenamefont {Strelnikov}}]{muPcorrMethod}%
  \BibitemOpen
  \bibfield  {author} {\bibinfo {author} {\bibfnamefont {A.}~\bibnamefont
  {Bodek}}, \bibinfo {author} {\bibfnamefont {A.}~\bibnamefont {van Dyne}},
  \bibinfo {author} {\bibfnamefont {J.-Y.}\ \bibnamefont {Han}}, \bibinfo
  {author} {\bibfnamefont {W.}~\bibnamefont {Sakumoto}}, and\ \bibinfo {author}
  {\bibfnamefont {A.}~\bibnamefont {Strelnikov}},\ }\href {\doibase
  10.1140/epjc/s10052-012-2194-8} {\bibfield  {journal} {\bibinfo  {journal}
  {Eur. Phys. J. C}\ }\textbf {\bibinfo {volume} {72}},\ \bibinfo {pages}
  {2194} (\bibinfo {year} {2012})}\BibitemShut {NoStop}%
\bibitem [{\citenamefont {{A. Bhatti \emph{et al.}}}(2006)}]{refCdfJES}%
  \BibitemOpen
  \bibfield  {author} {\bibinfo {author} {\bibnamefont {{A. Bhatti \emph{et
  al.}}}},\ }\href {\doibase 10.1016/j.nima.2006.05.269} {\bibfield  {journal}
  {\bibinfo  {journal} {Nucl. Instrum. Methods Phys. Res., Sect. A}\ }\textbf
  {\bibinfo {volume} {566}},\ \bibinfo {pages} {375} (\bibinfo {year}
  {2006})}\BibitemShut {NoStop}%
\bibitem [{\citenamefont {Campbell}\ and\ \citenamefont
  {Ellis}(1999)}]{MCFM345}%
  \BibitemOpen
  \bibfield  {author} {\bibinfo {author} {\bibfnamefont {J.~M.}\ \bibnamefont
  {Campbell}}\ and\ \bibinfo {author} {\bibfnamefont {R.~K.}\ \bibnamefont
  {Ellis}},\ }\href {\doibase 10.1103/PhysRevD.60.113006} {\bibfield  {journal}
  {\bibinfo  {journal} {Phys. Rev. D}\ }\textbf {\bibinfo {volume} {60}},\
  \bibinfo {pages} {113006} (\bibinfo {year} {1999})}\BibitemShut {NoStop}%
\bibitem [{\citenamefont {Czakon}\ \emph {et~al.}(2013)\citenamefont {Czakon},
  \citenamefont {Fiedler},\ and\ \citenamefont {Mitov}}]{ttbarNNLO}%
  \BibitemOpen
  \bibfield  {author} {\bibinfo {author} {\bibfnamefont {M.}~\bibnamefont
  {Czakon}}, \bibinfo {author} {\bibfnamefont {P.}~\bibnamefont {Fiedler}},
  and\ \bibinfo {author} {\bibfnamefont {A.}~\bibnamefont {Mitov}},\ }\href
  {\doibase 10.1103/PhysRevLett.110.252004} {\bibfield  {journal} {\bibinfo
  {journal} {Phys. Rev. Lett.}\ }\textbf {\bibinfo {volume} {110}},\ \bibinfo
  {pages} {252004} (\bibinfo {year} {2013})}\BibitemShut {NoStop}%
\bibitem [{\citenamefont {{D. Acosta \emph{et al.}}}(2002)}]{cdfR2CLC}%
  \BibitemOpen
  \bibfield  {author} {\bibinfo {author} {\bibnamefont {{D. Acosta \emph{et
  al.}}}},\ }\href {\doibase 10.1016/S0168-9002(02)01445-6} {\bibfield
  {journal} {\bibinfo  {journal} {{Nucl. Instrum. Methods Phys. Res., Sect.
  A}}\ }\textbf {\bibinfo {volume} {494}},\ \bibinfo {pages} {57} (\bibinfo
  {year} {2002})}\BibitemShut {NoStop}%
\bibitem [{\citenamefont {{R. D. Ball \emph{et al.}}}(2017)}]{nnpdf301A}%
  \BibitemOpen
  \bibfield  {author} {\bibinfo {author} {\bibnamefont {{R. D. Ball \emph{et
  al.}}}} (\bibinfo {collaboration} {NNPDF Collaboration}),\ }\href {\doibase
  10.1140/epjc/s10052-017-5199-5} {\bibfield  {journal} {\bibinfo  {journal}
  {Eur. Phys. J. C}\ }\textbf {\bibinfo {volume} {77}},\ \bibinfo {pages} {663}
  (\bibinfo {year} {2017})}\BibitemShut {NoStop}%
\end{thebibliography}%

\end{document}